\documentclass{IEEEtran}
\usepackage[T1]{fontenc}
\usepackage[latin9]{inputenc}
\usepackage{color}
\usepackage{array}
\usepackage{float}
\usepackage{booktabs}
\usepackage{multirow}
\usepackage{amsmath}
\usepackage{amsthm}
\usepackage{amssymb}
\usepackage{graphicx}

\makeatletter

\providecommand{\tabularnewline}{\\}
\floatstyle{ruled}
\newfloat{algorithm}{tbp}{loa}
\providecommand{\algorithmname}{Algorithm}
\floatname{algorithm}{\protect\algorithmname}

\theoremstyle{plain}
\newtheorem{lem}{\protect\lemmaname}
\theoremstyle{definition}
\newtheorem{assumption}{Assumption}
\theoremstyle{remark}
\newtheorem{rem}{\protect\remarkname}
\theoremstyle{plain}
\newtheorem{thm}{\protect\theoremname}

\usepackage{cite}
\usepackage{multirow}
\usepackage{bbm}
\usepackage{tikz}
\usetikzlibrary{shapes.geometric, arrows, calc}
\usetikzlibrary{decorations.pathreplacing}
\newenvironment{cellvarwidth}[1][t]
    {\begin{varwidth}[#1]{\linewidth}}
    {\@finalstrut\@arstrutbox\end{varwidth}}
\usepackage{varwidth}

\makeatother

\providecommand{\lemmaname}{Lemma}
\providecommand{\remarkname}{Remark}
\providecommand{\theoremname}{Theorem}

\begin{document}
\global\long\def\norm#1{\left\Vert #1\right\Vert }%

\global\long\def\normsq#1{\left\Vert #1\right\Vert ^{2}}%

\global\long\def\abs#1{\left|#1\right|}%

\global\long\def\hatdot#1{\dot{\hat{#1}}}%

\global\long\def\hatddot#1{\ddot{\hat{#1}}}%

\global\long\def\tildedot#1{\dot{\tilde{#1}}}%

\global\long\def\tildeddot#1{\ddot{\tilde{#1}}}%

\global\long\def\dims#1{\in\mathbb{R}^{#1}}%

\global\long\def\posreals{\in\mathbb{R}_{\geq0}}%

\global\long\def\posints{\in\mathbb{Z}_{\geq0}}%

\global\long\def\bound{\in\mathcal{L}_{\infty}}%

\global\long\def\auc{\ \text{are uniformly continuous}}%

\global\long\def\iuc{\ \text{is uniformly continuous}}%

\global\long\def\vec#1{\text{vec}(#1)}%

\global\long\def\proj#1{\text{proj}(#1)}%

\global\long\def\dd#1#2{\frac{\partial#1}{\partial#2}}%

\global\long\def\deriv#1#2{\frac{{\rm d}#1}{{\rm d}#2}}%

\global\long\def\layer#1#2{\sigma^{(#1)}\left(#2\right)}%

\global\long\def\w#1#2{W_{#1}^{(#2)}}%

\global\long\def\wt#1#2{W_{#1}^{(#2)\top}}%

\global\long\def\m#1#2{\mathcal{M}_{#1}^{(#2)}}%

\global\long\def\mbar#1#2{\mathcal{\bar{M}}_{#1}^{(#2)}}%

\global\long\def\mk#1{m^{(#1)}}%

\global\long\def\bmj#1{\boldsymbol{{\rm m}}^{(#1)}}%

\global\long\def\mj#1{m^{(#1)}}%

\global\long\def\edge{i\in V}%

\global\long\def\inv{\in V}%

\global\long\def\d#1{d^{(#1)}}%

\title{Lyapunov-Based Graph Neural Networks for Adaptive Control of Multi-Agent
Systems}
\author{Brandon C. Fallin, Cristian F. Nino, Omkar Sudhir Patil, Zachary I.
Bell, and Warren E. Dixon\thanks{Brandon C. Fallin, Cristian F. Nino, Omkar Sudhir Patil, and Warren
E. Dixon are with the Department of Mechanical and Aerospace Engineering,
University of Florida, Gainesville, FL 32611, USA. Email: \{brandonfallin,
cristian1928, patilomkarsudhir, wdixon\}@ufl.edu.}\thanks{Zachary I. Bell is with the Munitions Directorate, Air Force Research
Laboratory, Eglin AFB, FL 32542 USA. Email: zachary.bell.10@us.af.mil.}\thanks{This research is supported in part by AFOSR grant FA9550-19-1-0169.
Any opinions, findings and conclusions or recommendations expressed
in this material are those of the author(s) and do not necessarily
reflect the views of the sponsoring agency.}}
\maketitle
\begin{abstract}
Graph neural networks (GNNs) have a message-passing framework in which
vector messages are exchanged between graph nodes and updated using
feedforward layers. The inclusion of distributed message-passing in
the GNN architecture makes them ideally suited for distributed control
and coordination tasks. Existing results develop GNN-based controllers
to address a variety of multi-agent control problems while compensating
for modeling uncertainties in the systems. However, these results
use GNNs that are pre-trained offline. This paper provides the first
result on GNNs with stability-driven online weight updates to address
the multi-agent target tracking problem. Specifically, new Lyapunov-based
distributed GNN and graph attention network (GAT)-based controllers
are developed to adaptively estimate unknown target dynamics and address
the second-order target tracking problem. A Lyapunov-based stability
analysis is provided to guarantee exponential convergence of the target
state estimates and agent states to a neighborhood of the target state.
Numerical simulations show a $20.8\%$ and $48.1\%$ position tracking
error performance improvement by the GNN and GAT architectures over
a baseline DNN architecture, respectively.
\end{abstract}


\section{Introduction\label{sec:intro}}

\IEEEPARstart{M}{ulti}-agent systems often operate in uncertain
environments which may include unknown, heterogeneous agent dynamics
or unknown disturbances. To address these uncertainties, traditional
machine learning approaches train neural networks (NNs) offline by
minimizing a loss function over a pre-collected dataset and deploying
the trained model to the system. However, the datasets required to
train these models can be difficult to obtain, may not match the operating
conditions of the environment, and do not adjust online to a mismatch
between the pretrained data and actual data. Adaptive control can
offer a powerful alternative by enabling real-time estimation of unknown
model parameters with stability guarantees for the system.

Lyapunov-based adaptive update laws for NN-based controllers are well
established for real-time learning for dynamical systems. Motivation
exists to generalize such update laws to online deep learning because
of the improved approximation performance of deep NNs (DNNs) \cite{LeCun.Bengio.ea2015}.
However, this generalization has been technically challenging because
the unknown DNN weights are embedded in a nested cascade of nonlinear
functions, preventing a structure that is suitable for Lyapunov-based
analysis. The authors of \cite{Patil.Le.ea2022} provided the first
breakthrough to enable real-time learning methods for uncertain weights
in all layers of DNNs by taking advantage of the compositional structure
of the architecture. The Lyapunov-based DNN (Lb-DNN) approach developed
in \cite{Patil.Le.ea2022} for fully connected feedforward networks
has been extended to incorporate a variety of deep learning architectures
\cite{Le.Patil.ea.2022a,Patil.Le.ea.2022,Vacchini.Sacchi.ea2023,Griffis.Patil.ea23_2,Hart.patil.ea2023,Ganie.Jagannathan2024,Le.Patil.ea2024},
enabling broad applications including: online approximation of unknown
dynamics \cite{Patil.Griffis.ea2023,Lu2024,Wu.Lu.ea2024,Mei2024},
herding agents with unknown interaction dynamics \cite{Makumi.Bell.ea23},
control input determination using historical sensor data \cite{Ryu.Choi2024},
safety constraint enforcement under uncertain dynamics \cite{Sweatland.Patil.ea2024},
and development of end-to-end learning methods \cite{Li.Cheah2023,Li.Nguyen.ea2023}. 

Much of the multi-agent adaptive NN literature relies on single-layer
NNs to compensate for unknown, heterogeneous agent dynamics \cite{Cheng.Hou.ea2010,Chen.Liu.ea2021,Wang.Tian.ea2024}.
Although NNs with a single hidden layer are capable of approximating
general nonlinear functions over a compact domain, DNNs provide improved
performance \cite{LeCun.Bengio.ea2015}. 

\begin{figure}[H]
\begin{centering}
\centering{}\includegraphics[width=1\columnwidth]{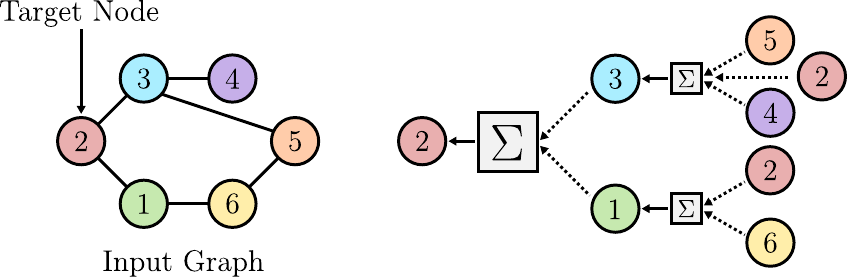}
\par\end{centering}
\caption{\label{fig:message-passing}Overview of the message-passing framework
employed at each GNN layer in which nodes aggregate messages from
their local neighborhoods. This figure shows a two-layer implementation
of the message-passing model \cite{Hamilton2020}. }
\end{figure}
Traditional deep learning architectures have specific structural limitations:
convolutional neural networks (CNNs) operate on grid-structured data
like images, while recurrent neural networks (RNNs), long short-term
memory (LSTM) blocks, and transformers process sequential data \cite{Hamilton2020,Hochreiter.Schmidhuber.1997,Vaswani.2017}.
Initially conceived as a spatial analogue to RNNs \cite{Scarselli.Gori.ea2008},
graph neural networks (GNNs) were developed to extend deep learning
to graph-valued data. 

As depicted in Fig. \ref{fig:message-passing}, GNNs operate through
a message-passing framework where nodes exchange vector messages that
are updated using feedforward NN layers. At every layer, the GNN aggregates
outputs from each agent's neighborhood, multiplies by the layer weights,
and produces an update using an elementwise activation function. The
traditional GNN architecture can also be augmented with an attention
mechanism to form the graph attention network (GAT) architecture \cite{Velickovic.Cucurull.ea2018}.
The attention mechanism enables each node to rank the importance of
messages from its neighbors. Leveraging attention is particularly
useful in applications where some nodes are more informative than
others \cite{Hamilton2020}. Unlike traditional NNs, GNNs efficiently
process and share information across networked agents, making them
ideally suited for distributed tasks among multiple agents. 

GNNs have been used extensively in the offline learning literature
for a variety of applications, including (i) node classification,
which aggregates information from neighboring nodes to sort graph
nodes into distinct categories \cite{Xiao.Wang.ea2022}, (ii) graph
classification, which combines local node features into global graph
representations \cite{Errica.Podda.ea2020}, and (iii) link prediction,
which uses graph-valued data to determine likely future connections
in settings like social media networks and online shopping platforms
\cite{Zhang.Chen2018}. Only recently have GNNs received attention
in the multi-agent literature, where they have been used to develop
decentralized controllers \cite{Gama.Tolstaya.ea2021}, perform coordinated
path planning \cite{Li.Gama.ea2020,Muthusamy.Owerko.ea2024} and perimeter
defense \cite{Lee.Zhou.ea2023}, facilitate deep reinforcement learning
\cite{Chen.Dong.ea2021}, and classify data from distributed sensor
networks \cite{Riess.Veveakis.ea2024}. Moreover, GNNs have been used
to solve the multi-robot decentralized tracking problem through an
imitation learning framework \cite{Zhou.Sharma.ea2022}. In \cite{Zhou.Sharma.ea2022},
the robots must jointly select actions to maximize target tracking
performance with local communications. The GNN in \cite{Zhou.Sharma.ea2022}
was trained on an action set of a state-of-the-art expert target tracking
algorithm offline, and its performance was shown to generalize for
large networks of robots. This method differs from the application
of GNNs in our work, in which we do not require offline training,
and update GNN parameters in real-time. These works show the promise
of the GNN architecture when controlling complex, distributed systems.
We build on these applications by providing stability guarantees for
a new GNN-based adaptive controller.

Many results have leveraged feedforward NNs to solve the distributed
target tracking problem, in which multiple agents are tasked with
monitoring, engaging with, or intercepting a target \cite{Peng.Wang.ea2013,Raj.Jagannathan.ea2020,Aryankia.Selmic2021,Nino.Patil.ea2024}.
However, none have used graph-theoretic learning methods to enhance
the NN's ability to approximate the target dynamics. Most notably,
\cite{Nino.Patil.ea2024} considered the use of deep neural networks
(DNNs) updated in real-time to achieve a target tracking objective
with unknown target dynamics and partial state feedback. This objective
was accomplished by developing an adaptive update law for each agent's
DNN and a distributed observer employing the DNN estimate to learn
the unknown state of the target. Despite its contribution to distributed
estimation capabilities of multi-agent systems, this work did not
leverage the underlying communication topology of the graph to inform
its DNN estimate. By not leveraging the underlying communication graph,
the target tracking performance can be limited because of the lack
of additional state information from network neighbors in the DNN
forward pass. 

In this work, we explore the application of a constructively developed
real-time, Lyapunov-based update law to the weights of the GNN architecture
and evaluate its performance when applied to the target tracking problem.
We note that the GNN can be applied to a variety of multi-agent control
problems including average consensus with unknown agent dynamics,
distributed state estimation of an uncertain target using networked
agents, or decentralized sensor fusion for multi-agent systems. This
work represents the first ever application of a GNN trained in real-time,
where its weight adaptation law is derived from Lyapunov-based methods.
We then augment the classical GNN architecture with attention, allowing
each layer of the GNN to rank the importance of its neighbors updates.
The attention augmentation results in the GAT architecture, and we
develop a Lyapunov-based update law for the attention weights in real-time.
These stability-driven adaptation laws allow for the developed architectures
to adapt to system uncertainties without offline training requirements;
however, offline training can be used to set the initial conditions
for each agent's weights. A Lyapunov-based stability analysis for
the target tracking problem is performed to prove exponential convergence
of the target state estimates and agent states to a neighborhood of
the target state. New bounds are developed for the eigenvalues of
the graph interaction matrix which allow gain conditions to be derived
in a decentralized manner, without knowledge of the global communication
topology. Additionally, a novel distributed weight adaptation law
is developed which ensures exponential convergence of the GNN weights
to a neighborhood of the ideal values. 

The analytical results were empirically validated by numerical simulations
which showed a $20.8\%$ and a $48.1\%$ position tracking error performance
improvement by the GNN and GAT architectures over a baseline DNN architecture,
respectively. This improvement was shown for a network of $6$ agents
with a path communication graph, where $3$ agents were linked to
the target. These results suggest that GNN and GAT architectures are
particularly valuable for real-world distributed systems where achieving
full network connectivity is challenging due to hardware constraints.

\section{Preliminaries\label{sec:prelims}}

\subsection{Notation}

Let $\mathbb{R}$ and $\mathbb{Z}$ denote the sets of reals and integers,
respectively. For $x\in\mathbb{R}$, let $\mathbb{R}_{\geq x}\triangleq[x,\infty)$
and $\mathbb{Z}_{\geq x}\triangleq\mathbb{R}_{\geq x}\cap\mathbb{Z}$.
Let ${\rm {\bf h}}_{i}$ represent the $i^{\text{th}}$ standard basis
in $\mathbb{R}^{N}$, where the standard basis is made up of vectors
that have one entry equal to $1$ and the remaining $N-1$ entries
equal to $0$. The $p\times p$ identity matrix, and the $p\times1$
column vector of ones are denoted by $I_{p}$ and $\mathbbm{1}_p$,
respectively. The $p\times1$ column vector of zeros is denoted by
$\boldsymbol{0}_{p}$. The enumeration operator $[\cdot]$ is defined
as $[N]\triangleq\{1,2,\ldots,N\}.$ A norm $\lVert\cdot\rVert$ is
submultiplicative if for any two matrices $A\in\mathbb{R}^{m\times n}$
and $B\in\mathbb{R}^{n\times p}$, it satisfies $\lVert AB\rVert\leq\lVert A\rVert\lVert B\rVert$.
The first partial derivative of a matrix $A\dims n$ with respect
to a vector $B\dims m$ is expressed in column convention as $\nabla_{B}A\dims{n\times m}$.
The Kronecker product of $A\in\mathbb{R}^{m\times n}$ and $B\in\mathbb{R}^{o\times p}$
is denoted by $A\otimes B\in\mathbb{R}^{mo\times np}$. The concatenation
operator of $A\dims{m\times n}$ and $B\dims{o\times n}$ is denoted
by $A\oplus B\dims{(m+o)\times n}$. Given a positive integer $N$
and collection $\{x_{i}\}_{i\in[N]}$, let $[x_{i}]_{i\in[N]}\triangleq[x_{1},x_{2},\ldots,x_{N}]\in\mathbb{R}^{m\times N}$
for $x_{i}\dims m$. For a set $\mathcal{S}\subset\mathbb{Z}$, let
$x_{j:j\in\mathcal{S}}$ denote the tuple $(x_{k_{1}},x_{k_{2}},\ldots,x_{k_{\lvert\mathcal{S}\rvert}})$,
where $k_{1}<k_{2}<\ldots<k_{\lvert\mathcal{S}\rvert}$ are the elements
of $\mathcal{S}$ in ascending order. The interior of a set $\mathcal{S}$
is denoted by $\text{int}(\mathcal{S})$. 

The Kronecker delta is denoted by $\delta_{a,b}$, where $\delta_{a,b}=1$
if $a=b$ and $\delta_{a,b}=0$ otherwise. For a set $A$ and an input
$x$, the indicator function is denoted by $\boldsymbol{1}_{A}(x)$,
where $\boldsymbol{1}_{A}(x)=1$ if $x\in A$, and $\boldsymbol{1}_{A}(x)=0$
otherwise. For $x\dims m$, let $\text{diag}(\cdot)$ denote the diagonalization
operator which assigns the $i^{\text{th}}$ value of $x$ to the $ii^{\text{th}}$
entry in an output $m\times m$ matrix for all $i\in[m]$. Similarly,
let $\text{blkdiag}(\cdot)$ denote the block diagonalization operator.
For $A\dims{m\times n}$ and $B\dims{o\times p}$, 
\[
\text{blkdiag}(A,B)\triangleq\begin{bmatrix}A & \boldsymbol{0}_{m\times p}\\
\boldsymbol{0}_{o\times n} & B
\end{bmatrix},
\]
where $\boldsymbol{0}_{m\times p}\dims{m\times p}$ is the $m\times p$
matrix of zeros and $\text{blkdiag}(A,B)\dims{(m+o)\times(n+p)}$.
Given $A\dims{m\times n}$ with columns $[a_{i}]_{i\in[n]}^{\top}\subset\mathbb{R}^{b}$,
$\vec A\triangleq[a_{i}^{\top},a_{2}^{\top},\ldots,a_{c}^{\top}]^{\top}\dims{mn}$.
Given any $A\in\mathbb{R}^{m\times n}$, $B\in\mathbb{R}^{n\times o}$,
$C\in\mathbb{R}^{o\times p}$, the vectorization operator satisfies
the property
\begin{equation}
\text{vec}(ABC)=\left(C^{\top}\otimes A\right)\text{vec}(B).\label{eq:vec-prop}
\end{equation}
Differentiating $\text{vec}(ABC)$ on both sides with respect to $\text{vec}(B)$
yields
\begin{equation}
\frac{\partial}{\partial\text{vec}(B)}\text{vec}(ABC)=\left(C^{\top}\otimes A\right).\label{eq:vec-diff-prop}
\end{equation}
Equations (\ref{eq:vec-prop}) and (\ref{eq:vec-diff-prop}) are proved
in \cite[Proposition 7.1.9]{Bernstein2009}. Additionally, for $A(x),B(x)\dims m$
such that $A(x),B(x)$ are differentiable, 
\begin{equation}
\dd{}x\left(A(x)\oplus B(x)\right)=\left(\dd{}xA(x)\oplus\dd{}xB(x)\right).\label{eq:concat-prop}
\end{equation}
Equation (\ref{eq:concat-prop}) is proved in the Appendix. Let $\text{proj}(a,b,c):\mathbb{R}^{n}\times\mathbb{R}^{n}\times\mathbb{R}\to\mathbb{R}^{n}$
denote the projection operator defined in \cite[Appendix E]{Krstic1995},
where for a smooth, convex function $\mathcal{P}:\mathbb{R}^{n}\to\mathbb{R}$,
a convex set $\Pi\triangleq\{b\dims n:\mathcal{P}(b)\leq c\}$, and
a user-defined symmetric and positive definite gain matrix $\Gamma\dims{n\times n}$,
\begin{equation}
\text{proj}(a,b,c)\triangleq\begin{cases}
a, & \begin{gathered}b\in\mathring{\Pi}\text{ or}\\
\nabla_{b}\mathcal{P}^{\top}a\leq0,
\end{gathered}
\\
\begin{gathered}\left(I_{n}-\min\left\{ 1,\frac{\mathcal{P}(b)}{c}\right\} \right.\\
\left.\cdot\Gamma\frac{\nabla_{b}\mathcal{P}\nabla_{b}\mathcal{P}^{\top}}{\nabla_{b}\mathcal{P}^{\top}\Gamma\nabla_{b}\mathcal{P}}\right)a,
\end{gathered}
 & \begin{gathered}b\in\Pi\backslash\mathring{\Pi}\text{ and}\\
\nabla_{b}\mathcal{P}^{\top}a>0.
\end{gathered}
\end{cases}\label{eq:proj}
\end{equation}

\subsection{Graphs}

For $N\in\mathbb{Z}_{\geq0}$, let $G\triangleq(V,E)$ be a static
and undirected graph with node set $V\triangleq[N]$ and edge set
$E\subseteq V\times V$. The edge $(i,j)\in E$ if and only if the
node $i$ can send information to node $j$. In this work, $G$ is
undirected, so $(i,j)\in E$ if and only if $(j,i)\in E$. Let $A\triangleq[a_{ij}]\dims{N\times N}$
denote the adjacency matrix of $G$, where $a_{ij}=\boldsymbol{1}_{E}(i,j)$
and $a_{ii}=0$ for all $i\in V$. An undirected graph is connected
if and only if there exists a sequence of edges in $E$ between any
two nodes in $V$. The neighborhood of node $i$ is denoted by $\mathcal{N}_{i}$,
where $\mathcal{N}_{i}\triangleq\left\{ j\in V:(j,i)\in E\right\} $.
The augmented neighborhood of node $i$ is denoted by $\bar{\mathcal{N}}_{i}$,
where $\bar{\mathcal{N}}_{i}\triangleq\mathcal{N}_{i}\cup\left\{ i\right\} $.
We denote node $i$'s $k$-hop neighborhood as $\mathcal{N}_{i}^{k}$.
Additionally, we denote the augmented $k$-hop neighborhood of node
$i$ as $\mathcal{\bar{N}}_{i}^{k}\triangleq\{i\}\cup\mathcal{N}_{i}^{k}$.
The degree matrix of $G$ is denoted by $D\triangleq\text{diag}(A\mathbbm{1})$.
The Laplacian matrix of $G$ is denoted by $\mathcal{L}_{G}\triangleq D-A\dims{N\times N}$.
In this work, we focus on a class of communication networks where
information flows through all nodes, which occurs when the graph $G$
is connected.

The set of permutations on $[N]$ is denoted by $S_{N}$. For a graph
$G$ and a permutation $S_{N}$, we define the graph permutation operation
as $p\ast G$, where $p\in S_{N}$. Two graphs are isomorphic if they
represent the exact same graph structure, differing only in the ordering
of their nodes in their corresponding adjacency matrices \cite{Hamilton2020}.
Formally, two graphs $G_{1}$ and $G_{2}$ are isomorphic if they
have the same number of nodes and there exists a permutation such
that $G_{1}=p\ast G_{2}$ \cite{Azizian.Lelarge2020}. 

\section{Graph Neural Network Architectures\label{sec:gnn-architectures}}

In this section, we explore two deep GNN architectures. We begin with
the traditional GNN. Then, we investigate the GAT architecture which
incorporates attention weights to compare the importance of neighboring
nodes' features at each layer. We derive closed-form expressions for
the first partial derivatives of $k$-layer GNN and GAT architectures
with respect to their weight vectors. Calculating the first partial
derivative of each architecture with respect to their layer weights
allows us to train the network in real-time using a novel Lyapunov-based
weight update law. 

\subsection{Graph Neural Network}

\begin{table*}[t]
\centering{}\caption{\label{tab:deep-jacobians}Partial Derivatives of the GNN and GAT
architectures with respect to $\text{vec}(\protect\w ij)$ and $a_{i}^{(j)}$,
respectively. Definitions of the terms $\varphi_{m^{(\ell)}}^{(\ell)},\Lambda_{m^{(\ell+1)}}^{(\ell)}$,
and $\varsigma_{m^{(\ell+1)}}^{(\ell)}$ are established in Table
\ref{tab:jacobian-terms}. }

\centering{}%
\begin{tabular}{ccc}
\toprule 
 & GNN & GAT\tabularnewline
\midrule
\midrule 
\begin{cellvarwidth}Partial derivative of $\phi_{i}$\\with respect
to $\text{vec}(\w ij)$ for $j=0,\ldots,k-1$\end{cellvarwidth} & {\scriptsize{}$\dd{\phi_{i}}{\vec{\w ij}}=\wt ik\varphi_{i}^{(k-1)}$} & {\scriptsize{}$\dd{\phi_{i}}{\vec{\w ij}}=\wt ik\Lambda_{i}^{(k-1)}$}\tabularnewline
\midrule 
\begin{cellvarwidth}Partial derivative of $\phi_{i}$\\with respect
to $a_{i}^{(j)}$ for $j=0,\ldots,k-1$\end{cellvarwidth} & {\scriptsize{}N/A} & {\scriptsize{}$\dd{\phi_{i}}{a_{i}^{(j)}}=\wt ik\varsigma_{i}^{(k-1)}$}\tabularnewline
\bottomrule
\end{tabular}
\end{table*}
The GNN architecture employs a message-passing framework in which
nodes exchange and update vector-valued messages using feedforward
layers. These messages are the outputs of each GNN layer calculated
at the node level. The GNN processes an input graph G with a set of
node features to generate node embeddings, which are the outputs of
the final GNN layer at each node. In this framework, each node's output
in a layer is informed by its previous layer output and messages received
from neighboring nodes. We use superscripts to denote layer-specific
elements. For example, $W_{i}^{(j)}$ denotes the $i^{\text{th}}$
node's weights for the $j^{\text{th}}$ GNN layer. Subscripts indicate
the specific node where an embedding or function is applied. 

Let the activation function for the $j^{\text{th}}$ GNN layer at
node $i$ be denoted by $\sigma^{(j)}(\cdot)$, where $\sigma(\cdot)$
is an element-wise smooth, bounded nonlinearity with bounded first
and second derivatives appended with $1$ to allow for biases such
that for $x\dims m$, $\sigma(x)\triangleq[\sigma(x_{0}),\ldots,\sigma(x_{m}),1]^{\top}$.
Let the output of layer $j$ for node $i$ of the GNN be denoted by
$\phi_{i}^{(j)}$. Let the aggregation function for the $j^{\text{th}}$
GNN layer at node $i$ be denoted by $\sum_{m\in\mathcal{\bar{N}}_{i}}\phi_{m}^{(j-1)}$.
Additionally, let $d^{(j)}$ represent the number of features at the
$j^{\text{th}}$ layer of the GNN for $j=0,\ldots,k$, where $j$
denotes the layer index. Let $d^{(in)}$ denote the dimension of the
GNN input at the base layer of each node. Let $d^{(out)}$ denote
the output dimension of the output layer of each node. Let $d^{(j)}=d^{(in)}+1$
when $j=-1$ and $d^{(j)}=d^{(out)}$ when $j=k$. Then, the message-passing
update for the $j^{\text{th}}$ layer of the $i^{\text{th}}$ node
is expressed as
\begin{equation}
\phi_{i}^{(j)}=\layer j{\wt ij\sum_{m\in\mathcal{\bar{N}}_{i}}\phi_{m}^{(j-1)}},\label{eq:message-passing}
\end{equation}
where $\w ij\dims{(\d{j-1}+1)\times\d j}$. We modify the framework
of the original GNN models proposed in \cite{Scarselli.Gori.ea2008}
and \cite{Merkwirth.Lengauer2005} by allowing each node to have distinct
layer weights. While synchronized layer weights at each node may be
desirable for some applications, allowing each node to have distinct
layer weights enables distributed weight updates at each node of the
GNN. Next, we define the GNN architecture for an arbitrary number
of layers. 

\begin{figure}[H]
\begin{centering}
\centering{}\includegraphics[width=0.85\columnwidth]{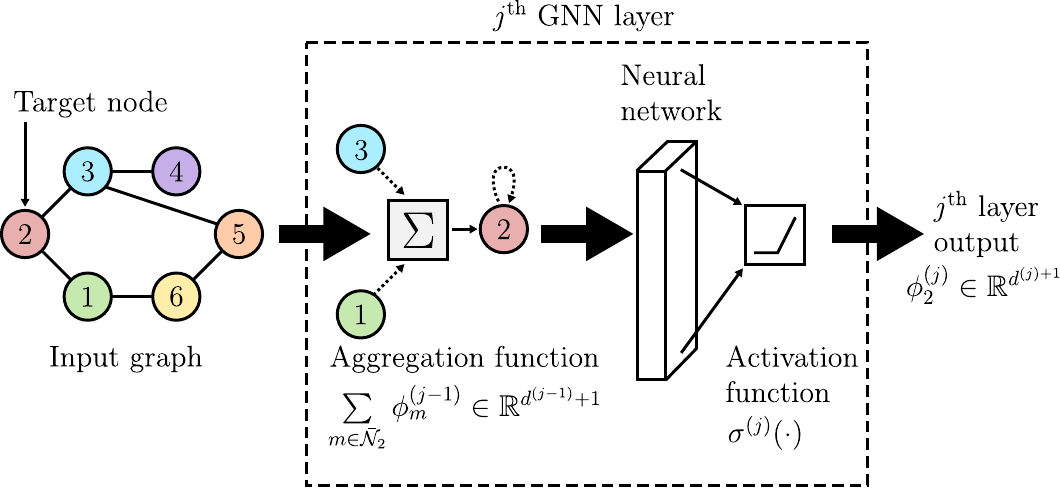}
\par\end{centering}
\caption{\label{fig:message-passing-1}Model of a single node's forward pass
for the $j^{\text{th}}$ layer of the GNN architecture. The output
of the previous layer is aggregated over the neighborhood of the target
node. The aggregated output is passed into a NN. The output of the
NN is the input to the $j^{\text{th}}$ layer activation function.
The activation function also appends a bias. The activated output
serves as an input for the next GNN layer \cite{Zeng.Tang2021}.}
\end{figure}
Let $\bar{\kappa}_{i}$ denote the $i^{\text{th}}$ node's input augmented
with a bias term such that $\bar{\kappa}_{i}\triangleq[\kappa_{i}^{\top},1]^{\top}\dims{d^{(in)}+1}$.
Let $\bar{\boldsymbol{\kappa}}\triangleq[\bar{\kappa}_{i}]_{i\in V}\dims{(\d{in}+1)\times N}$.
To distinguish between agent indices in cascading layers, we let $\mk j$
denote an arbitrary index corresponding to the $j^{\text{th}}$ layer,
where $\mk j\inv$. Then, we let $\boldsymbol{{\rm m}}^{(j)}$ denote
$\mk j\inv$. Let $\boldsymbol{\phi}^{(j)}\triangleq[\phi_{i}^{(j)},\ldots,\phi_{n}^{(j)}]=[\phi_{i}^{(j)}]_{\bmj j}\dims{(\d{j-1}+1)\times N}$.
Let $\bar{A}\dims{N\times N}$ represent the adjacency matrix with
self loops, and $\bar{A}_{i}\dims{1\times N}$ represent the $i^{\text{th}}$
row of the adjacency matrix. The message-passing GNN architecture
in (\ref{eq:message-passing}) with $k$ layers can be expressed as
\begin{equation}
\phi_{i}^{(j)}\triangleq\begin{cases}
\sigma^{(j)}(W_{i}^{(j)\top}\bar{\boldsymbol{\kappa}}\bar{A}_{i}^{\top}), & j=0,\\
\sigma^{(j)}(W_{i}^{(j)\top}\boldsymbol{\phi}^{(j-1)}\bar{A}_{i}^{\top}), & j=1,\ldots,k-1,\\
W_{i}^{(j)\top}\phi_{i}^{(j-1)}, & j=k.
\end{cases}\label{eq:deep-gnn-architecture}
\end{equation}
For an input $y^{(j)}\dims m$, the partial derivative $\dd{\sigma^{(j)}}{y^{(j)}}:\mathbb{R}^{d^{(j)}}\to\mathbb{R}^{(d^{(j)}+1)\times d^{(j)}}$
of the activation function vector at the $j^{\text{th}}$ layer with
respect to its input is given as $[\sigma^{\prime(j)}\left(y_{1}\right){\rm {\bf h}_{1}},\ldots,\sigma^{\prime(j)}\left(y_{d^{(j)}}\right){\rm {\bf h}}_{d^{(j)}},\boldsymbol{0}_{d^{(j)}}]^{\top}\dims{(d^{(j)}+1)\times d^{(j)}}$,
where ${\rm {\bf h}}_{i}$ is the $i^{\text{th}}$ standard basis
in $\mathbb{R}^{d^{(j)}}$ and $\boldsymbol{0}_{d^{(j)}}$ is the
zero vector in $\mathbb{R}^{d^{(j)}}$. The algorithm for the forward
pass of the message-passing GNN in (\ref{eq:deep-gnn-architecture})
is given in Algorithm \ref{alg:message-passing}. In Algorithm \ref{alg:message-passing},
the $\cup$ operator denotes dictionary merging, where unique values
are appended to the existing dictionary. 
\begin{algorithm}
\textbf{Input: }Graph $G=(V,E)$, number of layers $k$, agent input
$\kappa_{i}\dims{\d{in}}$ for all $i\in V$, adjacency matrix $A\dims{\lvert V\rvert\times\lvert V\rvert}$

\textbf{Output: }Agent GNN output $\phi_{i}^{(k)}\dims{\d{out}}$
for all $i\in V$

\textbf{Initialize:}

\qquad{}\textbf{for} each agent $i\in V$ \textbf{do} in parallel:

\qquad{}\qquad{}Set $\bar{\kappa}_{i}=[\kappa_{i}^{\top},1]^{\top}$

\qquad{}\qquad{}Initialize local dictionaries:

\qquad{}\qquad{}\qquad{}$D_{i}^{A}:$ Local adjacency vector

\qquad{}\qquad{}\qquad{}$D_{i}^{O}$: Activated layer outputs

\qquad{}\qquad{}\qquad{}$D_{i}^{W}$: Weight matrices

\qquad{}\textbf{end for}

1: \textbf{for} each layer $\ell=0,\ldots,k$ \textbf{do}:

2: \qquad{}\textbf{for} each agent $i\in V$ \textbf{do} in parallel:

3: \qquad{}\qquad{}\textbf{if }$\ell\neq k$\textbf{:}

4: \qquad{}\qquad{}\qquad{}Aggregate: $z_{i}^{(\ell)}=\boldsymbol{\phi}^{(\ell-1)}\bar{A}_{i}^{\top}$

5: \qquad{}\qquad{}\qquad{}Update: $\phi_{i}^{(\ell)}=\sigma^{(\ell)}(W_{i}^{(\ell)\top}z_{i}^{(\ell)})$

6: \qquad{}\qquad{}\textbf{else:}

7: \qquad{}\qquad{}\qquad{}Update: $\phi_{i}^{(\ell)}=W_{i}^{(\ell)\top}\phi_{i}^{(\ell-1)}$

8: \qquad{}\qquad{}\textbf{end if}

9: \qquad{}\qquad{}Store in local dictionaries:

10: \qquad{}\qquad{}\qquad{}$D_{i}^{O}[(i,\ell)]=\phi_{i}^{(\ell)}$

11: \qquad{}\qquad{}\qquad{}$D_{i}^{W}[(i,\ell)]=W_{i}^{(\ell)}$

12: \qquad{}\textbf{end for}

13: \qquad{}\textbf{for} each agent $i\in V$ \textbf{do} in parallel:

14: \qquad{}\qquad{}\textbf{for} each neighbor $j\in\mathcal{N}_{i}$
\textbf{do}:

15: \qquad{}\qquad{}\qquad{}Exchange and update dictionaries:

16: \qquad{}\qquad{}\qquad{}\qquad{}$D_{i}^{A}\leftarrow D_{i}^{A}\cup D_{j}^{A}$

17: \qquad{}\qquad{}\qquad{}\qquad{}$D_{i}^{O}\leftarrow D_{i}^{O}\cup D_{j}^{O}$

18: \qquad{}\qquad{}\qquad{}\qquad{}$D_{i}^{W}\leftarrow D_{i}^{W}\cup D_{j}^{W}$

19: \qquad{}\qquad{}\textbf{end for}

20: \qquad{}\textbf{end for}

21: \textbf{end for}

\caption{\label{alg:message-passing}GNN Message-Passing Algorithm}
\end{algorithm}

The vector of weights for the GNN at node $i$ is defined as $\theta_{i}\dims{p_{{\rm GNN}}}$,
where
\begin{equation}
\theta_{i}\triangleq\begin{bmatrix}\vec{\w i0}^{\top}, & \ldots{\color{blue},} & \vec{\w ik}^{\top}\end{bmatrix}^{\top},\label{eq:deep-gnn-weights}
\end{equation}
and $p_{{\rm GNN}}\triangleq\sum_{j=0}^{k}(\d j)(\d{j-1}+1)$. We
now compute the first partial derivative of the GNN architecture for
node $i$ in (\ref{eq:deep-gnn-architecture}) with respect to (\ref{eq:deep-gnn-weights}). 
\begin{lem}
\label{lem:deep-jacobian}For each node $i\in V$, the first partial
derivative of the GNN architecture at node $i$ in (\ref{eq:deep-gnn-architecture})
with respect to (\ref{eq:deep-gnn-weights}) is 
\begin{equation}
\nabla_{\theta_{i}}\phi_{i}\triangleq\begin{bmatrix}\dd{\phi_{i}}{\vec{\w i0}}^{\top}, & \ldots, & \dd{\phi_{i}}{\vec{\w ik}}^{\top}\end{bmatrix}^{\top},\label{eq:deep-gnn-jacobian}
\end{equation}
where $\nabla_{\theta_{i}}\phi_{i}\dims{d^{(out)}\times p_{{\rm GNN}}}$
and the partial derivatives of $\phi_{i}$ with respect to $\text{vec}(\w i{\ell})$
for all $\ell=0,\ldots,k$ is defined in Table \ref{tab:deep-jacobians}. 
\end{lem}
\begin{IEEEproof}
See the Appendix.
\end{IEEEproof}

\subsection{Graph Attention Network}

We refine the capabilities of the GNN architecture in (\ref{eq:deep-gnn-architecture})
by augmenting the input layer with an attention mechanism. Using attention,
each node ranks the importance of messages from its neighbors. Adding
attention to a GNN is a powerful strategy to increase its expressive
abilities. Attention is especially useful in applications where some
of a node's neighbors may be more informative than others \cite{Hamilton2020}.
The $i^{\text{th}}$ node's attention mechanism at layer $j$ is $\mathcal{M}_{i}^{(j)}:\mathbb{R}^{d^{(j)}}\times\mathbb{R}^{d^{(j)}}\to\mathbb{R}$.
The attention mechanism computes attention coefficients $c_{i,\ell}^{(j)}\triangleq\mathcal{M}_{i}^{(j)}(\wt ij\phi_{i}^{(j-1)},\wt ij\phi_{\ell}^{(j-1)})$
indicating the importance of node $\ell$'s features to node $i$.
These coefficients are calculated for all $\ell\in\mathcal{\bar{N}}_{i}$.
The attention coefficients are explicitly calculated as $c_{i,\ell}^{(j)}=a_{i}^{(j)\top}((\wt ij\phi_{i}^{(j-1)})\oplus(\wt ij\phi_{\ell}^{(j-1)}))$,
where $a_{i}^{(j)}\dims{2d^{(j)}}$ is the $i^{\text{th}}$ node's
vector of attention weights at layer $j$. Let $\boldsymbol{c}_{i}^{(j)}\triangleq[c_{i,\mj j}^{(j)}]_{\bmj j}$,
where $\boldsymbol{c}_{i}^{(j)}\dims{1\times N}$. The coefficients
are then normalized using the softmax function $\beta_{i,\ell}^{(j)}\dims{}$,
where
\begin{equation}
\beta_{i,\ell}^{(j)}\triangleq\frac{\exp(c_{i,\ell}^{(j)})}{\exp(\boldsymbol{c}_{i}^{(j)})\bar{A}_{i}^{\top}}.\label{eq:attention-coeff}
\end{equation}
Any activation that maps a vector $\boldsymbol{c}_{i}^{(j)}\dims{1\times N}$
to a vector $\boldsymbol{\beta}_{i}^{(j)}$ such that $\lVert\boldsymbol{\beta}_{i}^{(j)}\rVert_{1}=1$
is a valid activation function for the attention weights. The softmax
function is used because it is differentiable and maps attention weights
to a valid probability distribution. Let $\boldsymbol{B}_{i}^{(j)}\triangleq[\phi_{i}^{(j-1)}\beta_{i,1}^{(j)},\ldots,\phi_{N}^{(j-1)}\beta_{i,N}^{(j)}]=[\phi_{\mj j}^{(j-1)}\beta_{i,\mj j}^{(j)}]_{\bmj j}$,
where $\boldsymbol{B}_{i}^{(j)}\dims{(\d{j-1}+1)\times N}$. If $j=0$,
then $\boldsymbol{B}_{i}^{(0)}=[\bar{\kappa}_{\mj j}\beta_{i,\mj 0}^{(0)}]_{\bmj 0}$.
We combine the attention mechanism with the GNN architecture in (\ref{eq:deep-gnn-architecture})
to yield the GAT architecture with $k$ layers, expressed as
\begin{equation}
\phi_{i}^{(j)}\triangleq\begin{cases}
\sigma^{(j)}(\wt ij\boldsymbol{B}_{i}^{(j-1)}\bar{A}_{i}^{\top}), & j=0,\ldots,k-1,\\
W_{i}^{(j)\top}\phi_{i}^{(j-1)}, & j=k.
\end{cases}\label{eq:deep-gat-architecture}
\end{equation}
We define the vector of layer weights for the $i^{\text{th}}$ node
of the GAT as $\mathcal{W}_{i}\dims{p_{{\rm GNN}}}$, where $\mathcal{W}_{i}\triangleq[\text{vec}(\w i0)^{\top},\ldots,\text{vec}(\w ik)^{\top}]^{\top}.$
Additionally, we define the vector of attention weights for the $i^{\text{th}}$
node of the GAT as $\mathcal{Z}_{i}\dims{p_{{\rm ATT}}}$, where $\mathcal{Z}_{i}\triangleq[a_{i}^{(0)\top},\ldots,a_{i}^{(k-1)\top}]^{\top}$
and $p_{{\rm ATT}}\triangleq\sum_{j=0}^{k-1}2\d j$. Then, the total
vector of weights for node $i$ of the GAT is denoted by $\theta_{i}\dims{p_{{\rm GAT}}}$,
where
\begin{equation}
\theta_{i}\triangleq\begin{bmatrix}\mathcal{W}_{i}^{\top}, & \mathcal{Z}_{i}^{\top}\end{bmatrix}^{\top},\label{eq:deep-gat-weights}
\end{equation}
and $p_{{\rm GAT}}\triangleq p_{{\rm GNN}}+p_{{\rm ATT}}$. We now
compute the first partial derivative of the GAT architecture in (\ref{eq:deep-gnn-architecture})
with respect to $\theta_{i}$.
\begin{lem}
\label{lem:deep-gat-jacobian}For each node $i\in V$, the first partial
derivative of the GAT architecture at node $i$ in (\ref{eq:deep-gat-architecture})
with respect to (\ref{eq:deep-gat-weights}) is
\begin{equation}
\nabla_{\theta_{i}}\phi_{i}=\begin{bmatrix}\dd{\phi_{i}}{\mathcal{W}_{i}}^{\top}, & \dd{\phi_{i}}{\mathcal{Z}_{i}}^{\top}\end{bmatrix}^{\top},\label{eq:deep-gat-jacobian}
\end{equation}
where $\nabla_{\theta_{i}}\phi_{i}\dims{\d k\times p_{{\rm GAT}}}$.
The partial derivative of $\phi_{i}$ with respect to $\text{vec}(\w i{\ell})$
for $\ell=0,\ldots,k$ and the partial derivative of $\phi_{i}$ with
respect to $a_{i}^{(\ell)}$ for $\ell=0,\ldots,k-1$ are defined
in Table \ref{tab:deep-jacobians}. 
\end{lem}
\begin{IEEEproof}
See the Appendix.
\end{IEEEproof}

\subsection{Approximation Capabilities of Graph Neural Networks}

In this section, we establish the function approximation capabilities
of the GNN architectures in (\ref{eq:deep-gnn-architecture}) and
(\ref{eq:deep-gat-architecture}). Let $\mathcal{G}_{N}$ denote the
space of undirected graphs with $N$ nodes. Let $F\dims{Nm}$ denote
the ensemble feature vector of the graph, where each node has features
in $\mathbb{R}^{m}$. A function $f:\mathcal{G}_{N}\times\mathbb{R}^{Nm}\to\mathbb{R}^{N\ell}$
on a graph $G$ is invariant if $f(p\ast G,p\ast F)=f(G,F)$ for every
permutation $p\in S_{N}$, every $G\in\mathcal{G}_{N}$, and every
$F\dims{Nm}$. Practical examples of invariant functions on graphs
include graph-level statistics such as the number of nodes or the
graph diameter. A function $f:\mathcal{G}_{N}\times\mathbb{R}^{Nm}\to\mathbb{R}^{N\ell}$
is equivariant if $f(p\ast G,p\ast F)=p\ast f(G,F)$ for every permutation
$p\in S_{N}$, every $G\in\mathcal{G}_{N}$, and every $F\dims{Nm}$
\cite{Azizian.Lelarge2020}. Examples of equivariant functions on
graphs include message-passing updates or continuous functions of
a node's features evaluated at that node. In this work, we approximate
equivariant functions of the graph using a GNN. Next, we discuss limitations
of the GNN's discriminatory ability in relation to the Weisfeiler-Leman
test. 

The 1-Weisfeiler-Leman (1-WL) test is a graph isomorphism heuristic
that iteratively refines node labels based on neighborhood information.
The 1-WL's test ability to distinguish graph structures has been shown
to be equivalent to the discriminative ability of the GNN architecture.
In practice, the 1-WL test is carried out by assigning initial labels
to the nodes of the graph and iteratively updating each node's label
by aggregating and hashing the neighboring labels \cite{Hamilton2020}.
This process is repeated until the label assignments do not change,
or a fixed number of iterations are reached. The discriminatory ability
of the WL test is generalized through the $k$-WL test, with its discriminatory
ability increasing as $k$ grows. The message-passing GNN in (\ref{eq:deep-gnn-architecture})
universally approximates equivariant functions on a graph less separating
than the $2$-WL test \cite{Azizian.Lelarge2020}. Next, we establish
the universal function approximation ability of GNNs. 

Let $\mathcal{F}$ be a set of functions $f$ defined on a set $X$.
The equivalence relation $\text{eq}(\mathcal{F})$ defined by $\mathcal{F}$
on X states that for any $x,x^{\prime}\in X$, $(x,x^{\prime})\in\text{eq}(\mathcal{F})$
if and only if for all $f\in\mathcal{F}$, $f(x)=f(x^{\prime})$.
For a function $f$, we write $\text{eq}(f)$. Given two sets of functions
$\mathcal{F}$ and $\mathcal{E}$, we say that $\mathcal{F}$ is more
separating than $\mathcal{E}$ if $\text{eq}(\mathcal{F})\subseteq\text{eq}(\mathcal{E})$
\cite{Azizian.Lelarge2020}. The discriminatory power for the message-passing
GNN $\Phi$ in (\ref{eq:deep-gnn-architecture}) is expressed as $\text{eq}(\Phi)=\text{eq}(2\cdot\text{WL}_{\text{E}})$,
where $k\cdot\text{WL}_{\text{E}}$ denotes the $k$-WL test for equivariant
functions defined in \cite{Azizian.Lelarge2020}. 

Let $\mathcal{C}_{E}(X,Y)$ denote the set of continuous, equivariant
functions from $X$ to $Y$. The closure of a class of functions $\mathcal{F}$
by the uniform norm is denoted by $\text{cl(\ensuremath{\mathcal{F}})}$.
The universal function approximation theorem for GNNs is defined as
follows. 
\begin{lem}
\label{lem:univ-approx}\cite[Theorem 4]{Azizian.Lelarge2020} Let
$\Omega\subseteq G\times F$ be a compact set. Then, for the set of
GNNs $\mathcal{K}$, we have
\[
\text{cl}\left(\mathcal{K}\right)=\left\{ f\in\mathcal{C}_{E}(\Omega,\mathbb{R}^{N\ell})\ :\ \text{eq}\left(2\cdot\text{WL}_{\text{E}}\right)\subseteq\text{eq}\left(f\right)\right\} .
\]
\emph{Lemma \ref{lem:univ-approx} shows that $\mathcal{K}$ is dense
in $\mathcal{C}_{E}(\Omega,\mathbb{R}^{N\ell})$ by the Stone-Weierstrass
theorem, which states that if an algebra $\mathcal{A}$ of real continuous
function separates points, then $\mathcal{A}$ is dense in the set
of continuous functions on a compact set. This outcome allows us to
use the Lyapunov-based GNN (Lb-GNN) to universally approximate equivariant
functions on the graph of agents that are less separating than the
$2$-WL test. }
\end{lem}

\section{Problem Formulation\label{sec:prob-form}}

In this section, we apply the GNN architectures described in Section
\ref{sec:gnn-architectures} to the problem of target tracking with
unknown target dynamics and unknown agent interaction dynamics. While
we specifically address the target tracking problem, we note that
the GNN is a general architecture that can be applied to a variety
of multi-agent control problems, including average consensus with
unknown agent dynamics, distributed state estimation of an uncertain
target using networked agents, and sensor fusion for multi-agent systems. 

\subsection{System Dynamics and Network Topology}

Let $Q_{0}\triangleq[q_{0}^{\top},\dot{q}_{0}^{\top}]^{\top}\dims{2n}$,
where $q_{0},\dot{q}_{0}\dims n$ represent the target agent's unknown
position and velocity. Consider target dynamics denoted by 
\begin{equation}
\ddot{q}_{0}=f(Q_{0}),\label{eq:target-dynamics}
\end{equation}
where $f(\cdot):\mathbb{R}^{2n}\to\mathbb{R}^{n}$ is an unknown function
and $\ddot{q}_{0}\dims n$ represents the target agent's unknown acceleration. 
\begin{assumption}
\label{ass:target-bound}\cite[Assumption 1]{Nino.Patil.ea2024} The
target's position and velocity $q_{0}$ and $\dot{q}_{0}$ are bounded
such that $\norm{q_{0}(t)}\leq\bar{q}_{0}\in\mathbb{R}_{\geq0}$ and
$\norm{\dot{q}_{0}(t)}\leq\bar{\dot{q}}_{0}$ for all $t\in[t_{0},\infty)$,
where $\bar{q}_{0},\bar{\dot{q}}_{0}$ are known.
\end{assumption}
\begin{assumption}
\label{ass:lipschitz}\cite[Assumption 2]{Nino.Patil.ea2024} The
unknown drift dynamics of the target modeled by $f(Q_{0})$ in (\ref{eq:target-dynamics})
are Lipschitz continuous on the set $\mathcal{Y}_{1,i}$. The Lipschitz
constant defined on the set $\mathcal{Y}_{1,i}$ has a known upper
bound denoted by $L$. The set $\mathcal{Y}_{1,i}$ is is defined
as
\begin{equation}
\begin{gathered}\mathcal{Y}_{1,i}\triangleq\left\{ v_{1,i}\dims{2n}:\right.\\
\left.\norm{v_{1,i}}\leq2(\bar{q}_{0}+\bar{\dot{q}}_{0})+c_{\zeta}(2+\alpha_{1})\right\} ,
\end{gathered}
\label{eq:y1-i-1}
\end{equation}
where $\alpha_{1}\in\mathbb{R}_{>0}$ is a user-selected constant,
and $c_{\zeta}\in\mathbb{R}_{>0}$ is a known bounding constant.
\end{assumption}
Consider a network of $N$ agents indexed by $V$, modeled by the
connected and undirected graph $G=(V,E)$. Let $Q_{i}\triangleq[q_{i}^{\top},\dot{q}_{i}^{\top}]^{\top}\dims{2n}$,
where $q_{i},\dot{q}_{i}\dims n$ represent agent $i$'s known position
and velocity. Let $R_{i}\triangleq[1_{\mathcal{\bar{N}}_{i}}(m)Q_{m}^{\top}]_{m\in V}^{\top}\dims{2nN}$,
where $1_{\bar{\mathcal{N}}_{i}}(m)$ is the indicator function that
returns $1$ if $m\in\bar{\mathcal{N}}_{i}$ and $0$ otherwise. Agent
$i$'s dynamics are given by 
\begin{equation}
\ddot{q}_{i}=h\left(R_{i}\right)+u_{i},\label{eq:agent-dynamics}
\end{equation}
where $h(R_{i}):\mathbb{R}^{2nN}\to\mathbb{R}^{n}$ is an unknown,
continuous function, $\ddot{q}_{i}\dims n$ denote agent $i$'s unknown
acceleration and $u_{i}\dims n$ denotes agent $i$'s control input.
The function $h(\cdot)$ denotes unknown interaction dynamics between
agent $i$ and its neighboring agents. 

\subsection{Objective}

The primary objective of this work is to drive each agent to the target
by using only relative position and relative velocity measurements
made between agents in the network, and relative position and velocity
measurements with respect to the target if the agent and the target
are connected. 
\begin{assumption}
\label{ass:binary}\cite[Assumption 5]{Zegers.Deptula.ea.2022} Let
$b_{i}\in\left\{ 0,1\right\} $ denote a binary indicator of agent
$i$'s ability to take relative position and velocity measurements
with respect to the target. Each agent $i$ can measure (i) the target's
relative position, $q_{0}-q_{i}$, and relative velocity, $\dot{q}_{0}-\dot{q}_{i}$,
with respect to its own position and velocity if $b_{i}=1$, and (ii)
neighboring agents' relative position, $q_{j}-q_{i}$, and relative
velocity, $\dot{q}_{j}-\dot{q}_{i}$, with respect to its own position
and velocity if agent $j\in\mathcal{N}_{i}$. We assume that $b_{i}=1$
for at least one $i\in V$. 
\end{assumption}
To quantify the GNN target tracking objective, we define a position
tracking error as 
\begin{equation}
e_{i}\triangleq q_{0}-q_{i},\label{eq:pos-tracking-error}
\end{equation}
where $e_{i}\dims n$ for all $i\in V$. The signal $e_{i}$ is measurable
if $b_{i}=1$. The target dynamics in (\ref{eq:target-dynamics})
are unknown, so we develop a distributed observer which leverages
online learning techniques to estimate $f(\cdot)$. To this end, we
define the state estimation error as 
\begin{equation}
\tilde{q}_{i}\triangleq q_{0}-\hat{q}_{0,i},\label{eq:state-est-error}
\end{equation}
where $\tilde{q}_{i}\dims n$ for all $i\in V$, and $\hat{q}_{0,i},\hatdot q_{0,i}\dims n$
denote agent $i$'s estimate of the target agent's position and velocity,
respectively. Additionally, we define the state estimation regulation
error as
\begin{equation}
\hat{e}_{i}\triangleq\hat{q}_{0,i}-q_{i},\label{eq:state-est-reg-error}
\end{equation}
where $\hat{e}_{i}\dims n$ for all $i\in V$. The state estimation
regulation error signal $\hat{e}_{i}$ is measurable by all agents.
The state estimation error in (\ref{eq:state-est-error}) can be calculated
locally if $b_{i}=1$, where $\tilde{q}_{i}=e_{i}-\hat{e}_{i}$.
An auxiliary state estimation error is defined as
\begin{equation}
r_{1,i}\triangleq\tildedot q_{i}+\alpha_{1}\tilde{q}_{i},\label{eq:aux-state-est-error}
\end{equation}
where $r_{1,i}\dims n$ for all $i\in V$ and $\alpha_{1}\in\mathbb{R}_{\geq0}$
denotes a user-selected constant. Similarly, an auxiliary state estimation
regulation error is defined as
\begin{equation}
r_{2,i}\triangleq\hatdot e_{i}+\alpha_{2}\hat{e}_{i},\label{eq:aux-state-est-reg-error}
\end{equation}
where $r_{2,i}\dims n$ for all $i\in V$ and $\alpha_{2}\in\mathbb{R}_{>0}$
denotes a user-selected constant. The structure of (\ref{eq:aux-state-est-error})
and (\ref{eq:aux-state-est-reg-error}) are motivated by the subsequent
stability analysis in Theorem \ref{thm:stability-result} which indicates
that the boundedness and convergence of $\tilde{q}_{i}$ and $\hat{e}_{i}$
can be determined from the boundedness and convergence of $r_{1,i}$
and $r_{2,i}$, respectively. As a result, the following error system
development, control design, and stability analysis are focused on
the boundedness and convergence of $r_{1,i}$ and $r_{2,i}$. 

Taking the time derivative of (\ref{eq:aux-state-est-error}) yields
\begin{equation}
\dot{r}_{1,i}=f(Q_{0})-\hatddot q_{0,i}+\alpha_{1}\left(r_{1,i}-\alpha_{1}\tilde{q}_{i}\right).\label{eq:aux-deriv-1}
\end{equation}
In (\ref{eq:aux-deriv-1}), the input to the unknown function $f(\cdot)$
is $Q_{0}$, which is unknown to all agents. Therefore, directly learning
$f(Q_{0})$ is not feasible. To facilitate indirect learning, we inject
a reconstructible signal into (\ref{eq:aux-deriv-1}), which yields
\begin{equation}
\dot{r}_{1,i}=\tilde{f}_{i}+f\left(\hat{Q}_{0,i}\right)-\hatddot q_{0,i}+\alpha_{1}\left(r_{1,i}-\alpha_{1}\tilde{q}_{i}\right),\label{eq:aux-1-sub}
\end{equation}
where $\hat{Q}_{0,i}\triangleq[\hat{q}_{0,i}^{\top},\dot{\hat{q}}_{0,i}^{\top}]^{\top}\dims{2n}$
represents agent $i$'s estimate of the target's position and velocity,
respectively, and $\tilde{f}_{i}\triangleq f(Q_{0})-f(\hat{Q}_{0,i})$.
We use a deep Lb-GNN $\Phi_{1}$ to approximate $f(\hat{Q}_{0,i})$
at each node, where $\Phi_{1}\triangleq[\phi_{1,i}^{\top}]_{i\in V}^{\top}\dims{nN}$.
Similarly, the time derivative of (\ref{eq:aux-state-est-reg-error})
yields 
\begin{equation}
\dot{r}_{2,i}=\hatddot q_{0,i}-h(R_{i})-u_{i}+\alpha_{2}\left(r_{2,1}-\alpha_{2}\hat{e}_{i}\right).\label{eq:aux-deriv-2}
\end{equation}
We use a deep Lb-GNN $\Phi_{2}$ to approximate the unknown agent
interaction dynamics $h(R_{i})$ at each node, where $\Phi_{2}\triangleq[\phi_{2,i}^{\top}]_{i\in V}^{\top}\dims{nN}$.
The observer is designed based on (\ref{eq:aux-1-sub}) and the controller
is designed based on (\ref{eq:aux-deriv-2}). In the following section,
we design the distributed controller, observer, and adaptive update
laws.

\section{Control Synthesis\label{sec:control-synth}}

\subsection{Observer Design\label{subsec:observer-design}}

To develop the observer, we leverage the universal function approximation
properties of GNNs established in Lemma \ref{lem:univ-approx}, allowing
us to represent the unknown function $f(\hat{Q}_{0,i})$ at all nodes
with a deep Lb-GNN $\Phi_{1}$. Each node of a $k$-layer Lb-GNN is
a function of its $(k-1)$-hop neighbors' ideal weight values, $\theta_{j:j\in\mathcal{N}_{i}^{k-1}}^{\ast}$
and its $k$-hop neighbors' Lb-GNN inputs, $\hat{Q}_{0,j:j\in\mathcal{N}_{i}^{k}}$.
Node $i$'s component of the Lb-GNN $\Phi_{1}$ is denoted by $\phi_{1,i}(\hat{Q}_{0,i},\hat{Q}_{0,j:j\in\mathcal{N}_{i}^{k}},\theta_{1,i}^{\ast},\theta_{1,j:j\in\mathcal{N}_{i}^{k-1}}^{\ast})$.
For notational simplicity, we omit the arguments of the Lb-GNN and
let $\phi_{1,i}\triangleq\phi_{1,i}(\hat{Q}_{0,i},\hat{Q}_{0,j:j\in\mathcal{N}_{i}^{k}},\theta_{1,i},\theta_{1,j:j\in\mathcal{N}_{i}^{k-1}})$
and $\phi_{i,1}^{\ast}\triangleq\phi_{1,i}(\hat{Q}_{0,i},\hat{Q}_{0,j:j\in\mathcal{N}_{i}^{k}},\theta_{1,i}^{\ast},\theta_{1,j:j\in\mathcal{N}_{i}^{k-1}}^{\ast})$. 

Let $\hat{Q}_{0}\triangleq[\hat{Q}_{0,i}^{\top}]_{i\in V}^{\top}\dims{2nN}$.
We use a deep Lb-GNN to approximate the ensemble function $F(\hat{Q}_{0})\triangleq[f(\hat{Q}_{0,1})^{\top},\ldots,f(\hat{Q}_{0,N})^{\top}]^{\top}\dims{nN}.$
Lemma \ref{lem:univ-approx} holds if (i) the graph-valued function
being approximated by the Lb-GNN is less separating than the 2-WL
test, and (ii) the input space of the graph valued function $F(\hat{Q}_{0})$
is compact. The function $F(\hat{Q}_{0})$ is an equivariant function
of the graph $G$. The outputs of the function $F(\hat{Q}_{0})$ are
solely a function of each node's features. Therefore, this function
is less separating than the $2$-WL test. We subsequently prove that
the input space of $F(\hat{Q}_{0})$ is compact in the Lyapunov stability
analysis in Theorem \ref{thm:stability-result}. 

Next, we define the set $\Omega_{1}$ as
\begin{equation}
\Omega_{1}\triangleq G\times\mathcal{Y}_{1},\label{eq:omega-1-1}
\end{equation}
where
\begin{equation}
\mathcal{Y}_{1}\triangleq\left\{ v_{1}\dims{2nN}:v_{1,i}\in\mathcal{Y}_{1,i}\ \text{for all}\ i\in V\right\} ,\label{eq:y1-1}
\end{equation}
and $v_{1}\triangleq[v_{1,i}^{\top}]_{i\in V}^{\top}$. Let the loss
function for the GNN $\Phi_{1}$ be defined as $\mathcal{L}_{1}:\mathbb{R}^{pN}\to\mathbb{R}_{\geq0}${\footnotesize{}
\begin{equation}
\mathcal{L}_{1}(\theta_{1})\triangleq\int_{\Omega_{1}}\left(\norm{F(\hat{Q}_{0})-\Phi_{1}}^{2}+\sigma\left\Vert \theta_{1}\right\Vert ^{2}\right)d\mu(\hat{Q}_{0}),\label{eq:first-loss}
\end{equation}
}for all $i\in V$, where $\mu$ denotes the Lebesgue measure, $\sigma\in\mathbb{R}_{>0}$
denotes a regularizing constant, and the term $\sigma\lVert\theta_{1}\rVert^{2}$
represents $L_{2}$ regularization (also popularly known as ridge
regression in the machine learning community) \cite[Sec. 7.1.1]{Goodfellow2016}.
Let $\mathbb{\mathbb{f}}_{1}\subset\mathbb{R}^{pN}$ denote a user-selected
compact, convex parameter search space with a smooth boundary, satisfying
$\boldsymbol{0}_{pN}\in\text{int}(\mathbb{f}_{1})$. Additionally,
define $\bar{\theta}_{1}\triangleq\max_{\theta\in\mathbb{f}_{1}}\lVert\theta\rVert$. 

The objective is to identify the vector of ideal GNN parameters $\theta_{1}^{\ast}\in\mathbb{f}_{1}$
defined as\footnote{Although using a bounded search space can restrict the optimality
of the identified parameters to be local instead of global, it allows
the subsequent development to be analyzed from a convex optimization
perspective, which otherwise would be non-convex due to the nested
NIP structure of the GNN architecture in (\ref{eq:deep-gnn-architecture}).
Specifically, due to the strict convexity of the regularizing term
$\sigma\lVert\theta_{1}\rVert^{2}$ in (\ref{eq:first-loss}), there
exists $\sigma\in\mathbb{R}_{>0}$ which ensures $\mathcal{L}_{1}(\theta_{1})$
is convex for all $\theta_{1}\in\mathbb{f}_{1}$. Additionally, the
regularizing term has other advantages such as mitigation of overfitting
\cite[Sec. 7.1.1]{Goodfellow2016}. However, selecting very high values
of $\sigma$ can be counterproductive as it can obscure the contribution
of the $\lVert F(\hat{Q}_{0})-\Phi_{1}\rVert^{2}$ term to the loss
function while also causing underfitting \cite[Sec. 7.1.1]{Goodfellow2016};
therefore, there is a tradeoff between selecting low vs. high values
of $\sigma$.}
\begin{equation}
\theta_{1}^{\ast}\triangleq\underset{\theta\in\mathbb{f}_{1}}{\arg\min}\ \mathcal{L}_{1}(\theta_{1}),\label{eq:theta-star-1}
\end{equation}

\begin{rem}
\label{rem:univ-approx} Notice that the universal function approximation
property of GNNs was not invoked in the definition of $\theta_{1}^{*}$.
The universal function approximation theorem for GNNs established
in Lemma \ref{lem:univ-approx} states that the function space of
GNNs is dense in $\mathcal{C}_{E}(\Omega_{1},\mathbb{R}^{Nn})$. Let
$\varepsilon_{1}:\Omega_{1}\to\mathbb{R}^{nN}$ denote an unknown
function representing the function approximation error that can be
bounded as $\sup_{G\times\hat{Q}_{0}\in\Omega_{1}}\lVert\varepsilon_{1}\rVert\leq\bar{\varepsilon}_{1}$.
Therefore, $\int_{\Omega_{1}}\lVert F(\hat{Q}_{0})-\Phi_{1}\rVert^{2}d\mu(\hat{Q}_{0})<\varepsilon_{1}^{2}\mu(\Omega_{1})$.
As a result, for any prescribed $\bar{\varepsilon}_{1}\dims{}$, there
exists a GNN $\Phi_{1}$ such that there exist weights $\theta_{1}\dims{pN}$
which satisfy $\sup_{G\times\hat{Q}_{0}\in\Omega_{1}}\lVert F(\hat{Q}_{0})-\Phi_{1}(\hat{Q}_{0},\theta_{1})\rVert\leq\bar{\varepsilon}_{1}$.
However, determining a search space $\mathbb{f}_{1}$ for an arbitrary
$\bar{\varepsilon}_{1}$ remains an open challenge. Therefore, we
allow $\mathbb{f}_{1}$ to be arbitrarily selected in the above analysis,
at the expense of guarantees on the approximation accuracy. 
\end{rem}
Denote the $i^{\text{th}}$ component of $\varepsilon_{1}$ as $\varepsilon_{1,i}\in\mathbb{R}^{n}$
for all $i\in V$, where $\varepsilon_{1}=[\varepsilon_{1,i}^{\top}]_{i\in V}^{\top}$.
Each agent models the target agent's unknown drift dynamics as
\begin{equation}
f\left(\hat{Q}_{0,i}\right)=\phi_{1,i}^{\ast}+\varepsilon_{1,i}.\label{eq:lb-gnn-1}
\end{equation}
Substituting (\ref{eq:lb-gnn-1}) into (\ref{eq:aux-1-sub}) yields
\begin{equation}
\dot{r}_{1,i}=\tilde{f}_{i}+\phi_{1,i}^{\ast}+\varepsilon_{1,i}-\hatddot q_{0,i}+\alpha_{1}\left(r_{1,i}-\alpha_{1}\tilde{q}_{i}\right).\label{eq:taylor-approx-here}
\end{equation}
We perform a first-order Taylor approximation of $\phi_{1,i}^{\ast}$
noting that each node of a $k$-layer GNN is a function of its $(k-1)$-hop
neighbors' ideal weight values, $\theta_{1,j}^{\ast}$, where $j\in\mathcal{N}_{i}^{k-1}$.
Therefore, we perform a Taylor expansion about each of these ideal
weights. Define the parameter estimation error for each node of the
GNN, $\tilde{\theta}_{1,i}$, as
\begin{equation}
\tilde{\theta}_{1,i}\triangleq\theta_{1,i}^{\ast}-\hat{\theta}_{1,i}.\label{eq:param-est-error}
\end{equation}
Similarly to $\phi_{1,i}^{\ast}$, we let $\hat{\phi}_{1,i}\triangleq\phi_{1,i}(\hat{Q}_{0,i},\hat{Q}_{0,j:j\in\mathcal{N}_{i}^{k}},\hat{\theta}_{1,i},\hat{\theta}_{1,j:j\in\mathcal{N}_{i}^{k-1}})$.
Then, the first-order Taylor approximation of $\phi_{1,i}(\hat{Q}_{0,i},\hat{Q}_{0,j:j\in\mathcal{N}_{i}^{k}},\theta_{1,i}^{\ast},\theta_{1,j:j\in\mathcal{N}_{i}^{k-1}}^{\ast})$
about the point $(\hat{Q}_{0,i},\hat{Q}_{0,j},\hat{\theta}_{1,i},\hat{\theta}_{1,j:j\in\mathcal{N}_{i}^{k-1}})$
yields
\begin{equation}
\phi_{1,i}^{\ast}=\hat{\phi}_{1,i}+\sum_{j\in\mathcal{\bar{N}}_{i}^{k-1}}\nabla_{\hat{\theta}_{1,j}}\hat{\phi}_{1,i}\tilde{\theta}_{1,j}+\sum_{j\in\mathcal{\bar{N}}_{i}^{k-1}}T_{1,j},\label{eq:taylor-approx}
\end{equation}
where $T_{1,j}$ denotes the first Lagrange remainder, which accounts
for the error introduced by truncating the Taylor approximation after
the first-order term. Let $\overset{\circ}{\theta}_{1,j}\triangleq\hat{\theta}_{1,j}+s_{m}(\tilde{\theta}_{1,j})\tilde{\theta}_{1,j}$,
where $s_{m}(\tilde{\theta}_{1,j}):\mathbb{R}^{p}\to[0,1]$ for all
$m\in[n]$. By \cite[Theorem 15.29]{Fitzpatrick2009}, the remainder
term $T_{1,j}\in\mathbb{R}^{n}$ can be expressed as $T_{1,j}=\frac{1}{2}[T_{1,j}(s_{1}),\ldots,T_{1,j}(s_{n})]^{\top}$,
where $T_{1,j}(s_{m})\in\mathbb{R}$ is defined as $T_{1,j}(s_{m})\triangleq\tilde{\theta}_{1,j}^{\top}\nabla_{\overset{\circ}{\theta}_{1,j}}^{2}\phi_{1,i}(\hat{Q}_{0,i},\hat{Q}_{0,j:j\in\mathcal{N}_{i}^{k}},\overset{\circ}{\theta}_{1,j},\hat{\theta}_{1,\ell:\ell\in\mathcal{\bar{N}}_{i}^{k-1}\backslash j})\tilde{\theta}_{1,j}$
for all $m\in[n]$, and $\nabla_{\overset{\circ}{\theta}_{1,j}}^{2}\phi_{1,i}(\hat{Q}_{0,i},\hat{Q}_{0,j:j\in\mathcal{N}_{i}^{k}},\overset{\circ}{\theta}_{1,j},\hat{\theta}_{1,\ell:\ell\in\mathcal{\bar{N}}_{i}^{k-1}\backslash j})\dims{n\times p\times p}$
denotes the second partial derivative of $\hat{\phi}_{1,i}$ with
respect to $\overset{\circ}{\theta}_{1,j}$. 

Substituting (\ref{eq:taylor-approx}) into (\ref{eq:taylor-approx-here})
gives
\begin{equation}
\begin{aligned}\dot{r}_{1,i} & =\tilde{f}_{i}+\hat{\phi}_{1,i}+\sum_{j\in\mathcal{\bar{N}}_{i}^{k-1}}\nabla_{\hat{\theta}_{1,j}}\hat{\phi}_{1,i}\tilde{\theta}_{1,j}\\
 & +\sum_{j\in\mathcal{\bar{N}}_{i}^{k-1}}T_{1,j}+\varepsilon_{1,i}-\hatddot q_{0,i}+\alpha_{1}\left(r_{1,i}-\alpha_{1}\tilde{q}_{i}\right).
\end{aligned}
\label{eq:r1-deriv-nn}
\end{equation}
Based on (\ref{eq:r1-deriv-nn}), the distributed observer is designed
as 
\begin{equation}
\begin{gathered}\hatddot q_{0,i}=\hat{\phi}_{1,i}+\sum_{j\in\mathcal{N}_{i}^{k-1}}\nabla_{\hat{\theta}_{1,j}}\hat{\phi}_{1,i}\left(\hat{\theta}_{i}-\hat{\theta}_{j}\right)\\
+k_{1}\left(b_{i}\dot{\tilde{q}}_{i}+\sum_{j\in\mathcal{N}_{i}}\left(\hatdot q_{0,j}-\hatdot q_{0,i}\right)\right.\\
\left.+\alpha_{1}\sum_{j\in\mathcal{N}_{i}}\left(\hat{q}_{0,j}-\hat{q}_{0,i}\right)+\alpha_{1}b_{i}\tilde{q}_{i}\right),
\end{gathered}
\label{eq:observer}
\end{equation}
where $b_{i}$ ensures that the signals $\tilde{q}_{i}$ and $\tildedot q_{i}$
are only implemented by agents who are connected to the target. Substituting
(\ref{eq:observer}) into (\ref{eq:r1-deriv-nn}) yields 
\begin{equation}
\begin{aligned}\dot{r}_{1,i} & =\tilde{f}_{i}+\sum_{j\in\mathcal{\bar{N}}_{i}^{k-1}}\nabla_{\hat{\theta}_{1,j}}\hat{\phi}_{1,i}\tilde{\theta}_{1,i}+\alpha_{1}\left(r_{1,i}-\alpha_{1}\tilde{q}_{i}\right)\\
 & +\chi_{1,i}+\varepsilon_{1,i}-k_{1}\left(\sum_{j\in\mathcal{N}_{i}}\left(\hatdot q_{0,j}-\hatdot q_{0,i}\right)+b_{i}\tildedot q_{i}\right.\\
 & \left.+\alpha_{1}\sum_{j\in\mathcal{N}_{i}}\left(\hat{q}_{0,j}-\hat{q}_{0,i}\right)+\alpha_{1}b_{i}\tilde{q}_{i}\right),
\end{aligned}
\label{eq:observer-sub}
\end{equation}
where $\chi_{1,i}\triangleq\sum_{j\in\mathcal{\bar{N}}_{i}^{k-1}}T_{1,j}+\sum_{j\in\mathcal{N}_{i}^{k-1}}\nabla_{\hat{\theta}_{1,j}}\hat{\phi}_{1,i}(\theta_{1,j}^{\ast}-\theta_{1,i}^{\ast})$.
The graph interaction matrix is defined as 
\begin{equation}
\mathcal{H}\triangleq\left(\mathcal{L}_{G}+\mathcal{B}\right)\otimes I_{n},\label{eq:graph-interaction-matrix}
\end{equation}
where $\mathcal{B}\triangleq\text{diag}(b_{1},\ldots,b_{N})\dims{N\times N}$,
$\mathcal{H}\dims{nN\times nN}$, and $\mathcal{L}_{G}$ is the graph
Laplacian. 
\begin{rem}
\label{rem:pos-def}Assumption \ref{ass:binary} states that $b_{i}=1$
for at least one $b_{i}\in\{c_{1},\ldots,c_{N}\}$. Additionally,
the communication graph $G$ is connected. Thus, the graph interaction
matrix $\mathcal{H}$ defined in (\ref{eq:graph-interaction-matrix})
is symmetric and positive definite \cite[Corollary 4.33]{Qu2009}. 
\end{rem}
Define $\dot{r}_{1}\triangleq[\dot{r}_{1,i}^{\top}]_{i\inv}^{\top}\dims{nN}.$
Then, the ensemble auxiliary state estimation error dynamics are 
\begin{equation}
\dot{r}_{1}=\tilde{F}+\nabla_{\hat{\theta}_{1}}\hat{\Phi}_{1}\tilde{\theta}_{1}+\chi_{1}+\varepsilon_{1}-k_{1}\mathcal{H}r_{1}+\alpha_{1}\left(r_{1}-\alpha_{1}\tilde{q}\right),\label{eq:state-est-error-ensemble}
\end{equation}
where $\tilde{F}\triangleq[\tilde{f}_{i}^{\top}]_{i\inv}^{\top}\dims{nN}$,
$\tilde{q}\triangleq[\tilde{q}_{i}^{\top}]_{i\inv}^{\top}\dims{nN}$,
and $\nabla_{\hat{\theta}_{1}}\hat{\Phi}_{1}\triangleq\text{blkdiag}(\sum_{j\in\mathcal{\bar{N}}_{1}^{k-1}}\nabla_{\hat{\theta}_{1,j}}\hat{\phi}_{1,1},\ldots,\sum_{j\in\mathcal{\bar{N}}_{N}^{k-1}}\nabla_{\hat{\theta}_{1,j}}\hat{\phi}_{1,N})$.
We note that $\nabla_{\hat{\theta}_{1}}\hat{\Phi}_{1}\dims{nN\times pN}$.
Additionally, $\tilde{\theta}_{1}\triangleq[\tilde{\theta}_{1,i}^{\top}]_{i\inv}^{\top}\dims{pN}$,
$\chi_{1}\triangleq[\chi_{1,i}^{\top}]_{i\inv}^{\top}\dims{nN}$,
and $r_{1}\triangleq[r_{1,i}^{\top}]_{i\inv}^{\top}\dims{nN}$. In
the following subsection, we develop a distributed control law. 

\subsection{Controller Design}

Let $R\triangleq[R_{i}^{\top}]_{i\in V}^{\top}\dims{2nNN}$. We invoke
the universal function approximation theorem for GNNs to approximate
the ensemble function $H(R)\triangleq[h(R_{1})^{\top},\ldots,h(R_{N})^{\top}]^{\top}\dims{nN}.$
The function $H(R)$ is an equivariant function of the graph whose
outputs are solely a function of the $1$-hop neighborhood of each
node. This function is less separating than the $2$-WL test. We subsequently
prove that the input space of $H(R)$ is compact in the Lyapunov stability
analysis in Theorem \ref{thm:stability-result}. 

Define the set $\Omega_{2}$ as 
\begin{equation}
\Omega_{2}\triangleq G\times\mathcal{Y}_{2},\label{eq:omega-2-1}
\end{equation}
where $\mathcal{Y}_{2,i}\triangleq\{v_{2,i}\dims{2nN}:\lVert v_{2,i}\rVert\leq N(\bar{q}_{0}+\bar{\dot{q}}_{0}+c_{\zeta}(3+\alpha_{1}+\alpha_{2}))\}$,
and $\mathcal{Y}_{2}\triangleq\{v_{2}\dims{2nNN}:v_{2,i}\in\mathcal{Y}_{2,i}\ \text{for all}\ i\in V\}$,
where $v_{2}\triangleq[v_{2,i}^{\top}]_{i\in V}^{\top}$. Let $\theta_{2}\triangleq[\theta_{2,i}^{\top}]_{i\in V}^{\top}$.
As in Subsection \ref{subsec:observer-design}, let the loss function
for the GNN $\Phi_{2}$ at node $i$ be defined as $\mathcal{L}_{2}:\mathbb{R}^{pN}\to\mathbb{R}_{\geq0}${\small{}
\begin{equation}
\mathcal{L}_{2}(\theta_{2})\triangleq\int_{\Omega_{2}}\left(\left\Vert H(R)-\Phi_{2}\right\Vert ^{2}+\sigma\left\Vert \theta_{2}\right\Vert ^{2}\right)d\mu(R),\label{eq:first-loss-2}
\end{equation}
}where $\mu$ denotes the Lebesgue measure, $\sigma\in\mathbb{R}_{>0}$
denotes a regularizing constant, the term $\sigma\lVert\theta_{2}\rVert^{2}$
represents $L_{2}$ regularization. Let $\mathbb{\mathbb{f}}_{2}\subset\mathbb{R}^{pN}$
denote a user-selected compact, convex parameter search space with
a smooth boundary, satisfying $\boldsymbol{0}_{pN}\in\text{int}(\mathbb{f}_{2})$.
Additionally, define $\bar{\theta}_{2}\triangleq\max_{\theta\in\mathbb{f}_{2}}\lVert\theta\rVert$.
We denote the ideal parameters of the GNN $\Phi_{2}$ as $\theta_{2}^{\ast}\in\mathbb{f}_{2},$
where
\begin{equation}
\theta_{2}^{\ast}\triangleq\underset{\theta\in\mathbb{f}_{2}}{\arg\min}\ \mathcal{L}_{2}(\theta_{2}).\label{eq:theta-star-2}
\end{equation}
For clarity in the subsequent analysis, it is desirable that the solutions
$\theta_{1}^{*}$ and $\theta_{2}^{\ast}$ to (\ref{eq:theta-star-1})
and (\ref{eq:theta-star-2}), respectively, be unique. To this end,
the following assumption is made.
\begin{assumption}
\label{ass:convex-loss}The loss functions $\mathcal{L}_{1}$ and
$\mathcal{L}_{2}$ are strictly convex over the sets $\mathbb{f}_{1}$
and $\mathbb{f}_{2}$, respectively.
\end{assumption}
Let $\bar{\theta}\triangleq\max\{\bar{\theta}_{1},\bar{\theta}_{2}\}$.
Additionally, we let $\varepsilon_{2}:\Omega_{2}\to\mathbb{R}^{nN}$
denote an unknown function representing the function approximation
error that can be bounded as $\sup_{G\times R\in\Omega_{2}}\lVert\varepsilon_{2}\rVert\leq\bar{\varepsilon}_{2}$.
Denote the $i^{\text{th}}$ component of $\varepsilon_{2}$ as $\varepsilon_{2,i}\in\mathbb{R}^{n}$
for all $i\in V$, where $\varepsilon_{2}=[\varepsilon_{2,i}^{\top}]_{i\in V}^{\top}$.
We denote node $i$'s component of the deep Lb-GNN $\Phi_{2}$ as
$\phi_{2,i}^{\ast}\triangleq\phi_{2,i}(R_{i},R_{j:j\in\mathcal{N}_{i}^{k}},\theta_{2,i}^{\ast},\theta_{2,j:j\in\mathcal{N}_{i}^{k-1}}^{\ast})$.
Then, each agent models their unknown drift dynamics as
\begin{equation}
h\left(R_{i}\right)=\phi_{2,i}^{\ast}+\varepsilon_{2,i}.\label{eq:lb-gnn-2}
\end{equation}

Substituting (\ref{eq:lb-gnn-2}) into (\ref{eq:aux-deriv-2}) yields
\begin{equation}
\dot{r}_{2,i}=\hatddot q_{0,i}-\phi_{2,i}^{\ast}-\varepsilon_{2,i}-u_{i}+\alpha_{2}\left(r_{2,i}-\alpha_{2}\hat{e}_{i}\right).\label{eq:second-taylor-approx-here}
\end{equation}
We perform a first-order Taylor approximation of $\phi_{2,i}(R_{i},R_{j:j\in\mathcal{N}_{i}^{k}},\theta_{2,i}^{\ast},\theta_{2,j:j\in\mathcal{N}_{i}^{k-1}}^{\ast})$
about the point $(R_{i},R_{j:j\in\mathcal{N}_{i}^{k}},\hat{\theta}_{2,i},\hat{\theta}_{2,j:j\in\mathcal{N}_{i}^{k-1}})$
to find
\begin{equation}
\phi_{2,i}^{\ast}=\hat{\phi}_{2,i}+\sum_{j\in\mathcal{\bar{N}}_{i}^{k-1}}\nabla_{\hat{\theta}_{2,j}}\hat{\phi}_{2,i}\tilde{\theta}_{2,j}+\sum_{j\in\mathcal{\bar{N}}_{i}^{k-1}}T_{2,j},\label{eq:second-taylor-approx}
\end{equation}
where $T_{2,j}$ denotes the first Lagrange remainder. Substituting
(\ref{eq:second-taylor-approx}) into (\ref{eq:second-taylor-approx-here})
gives
\begin{equation}
\begin{gathered}\dot{r}_{2,i}=\hatddot q_{0,i}-\hat{\phi}_{2,i}-\sum_{j\in\mathcal{\bar{N}}_{i}^{k}}\nabla_{\hat{\theta}_{2,j}}\hat{\phi}_{2,i}\tilde{\theta}_{2,j}\\
-\sum_{j\in\bar{\mathcal{N}}_{i}^{k-1}}T_{2,j}-\varepsilon_{2,i}-u_{i}+\alpha_{2}\left(r_{2,i}-\alpha_{2}\hat{e}_{i}\right),
\end{gathered}
\label{eq:r2-deriv-nn}
\end{equation}
where the parameter estimation error for each node of the GNN, $\tilde{\theta}_{2,i}$,
is defined as
\begin{equation}
\tilde{\theta}_{2,i}\triangleq\theta_{2,i}^{\ast}-\hat{\theta}_{2,i}.\label{eq:param-est-error-2}
\end{equation}
Motivated by the structure of (\ref{eq:r2-deriv-nn}), the distributed
controller is designed as
\begin{equation}
\begin{gathered}u_{i}=\hatddot q_{0,i}-\hat{\phi}_{2,i}-\sum_{j\in\mathcal{\mathcal{N}}_{i}^{k-1}}\nabla_{\hat{\theta}_{2,j}}\hat{\phi}_{2,i}\left(\hat{\theta}_{2,i}-\hat{\theta}_{2,j}\right)\\
+k_{2}\left(\hatdot e_{i}+\alpha_{2}\hat{e}_{i}\right)
\end{gathered}
\label{eq:controller}
\end{equation}
where $k_{2}\in\mathbb{R}_{>0}$ is a user-defined constant. Substituting
(\ref{eq:controller}) into (\ref{eq:r2-deriv-nn}) gives  
\begin{equation}
\begin{gathered}\dot{r}_{2,i}=-\sum_{j\in\mathcal{\bar{N}}_{i}^{k-1}}\nabla_{\hat{\theta}_{2,j}}\hat{\phi}_{2,i}\tilde{\theta}_{2,i}-\chi_{2,i}\\
-\varepsilon_{2,i}-\left(k_{2}-\alpha_{2}\right)r_{2,i}-\alpha_{2}^{2}\hat{e}_{i},
\end{gathered}
\label{eq:controller-sub}
\end{equation}
and $\chi_{2,i}\triangleq\sum_{j\in\mathcal{\bar{N}}_{i}^{k-1}}T_{2,j}+\sum_{j\in\mathcal{N}_{i}^{k-1}}\nabla_{\hat{\theta}_{2,j}}\hat{\phi}_{2,i}\left(\theta_{2,j}^{\ast}-\theta_{1,j}^{\ast}\right)$.
Define $\dot{r}_{2}\triangleq[\dot{r}_{2,i}^{\top}]_{i\inv}^{\top}\dims{nN}$.
The ensemble auxiliary state estimation regulation error dynamics
are  
\begin{equation}
\dot{r}_{2}=-\nabla_{\hat{\theta}_{2}}\hat{\Phi}_{2}\tilde{\theta}_{2}-\chi_{2}-\varepsilon_{2}-\left(k_{2}-\alpha_{2}\right)r_{2}-\alpha_{2}^{2}\hat{e},\label{eq:state-est-reg-error-ensemble}
\end{equation}
where $\tilde{\theta}_{2}\triangleq[\tilde{\theta}_{2,i}^{\top}]_{i\inv}^{\top}\dims{pN}$,
$\chi_{2}\triangleq[\chi_{2,i}^{\top}]_{i\inv}^{\top}\dims{nN}$,
$r_{2}\triangleq[r_{2,i}^{\top}]_{i\inv}^{\top}\dims{nN}$, $\nabla_{\hat{\theta}_{2}}\hat{\Phi}_{2}\triangleq\text{blkdiag}(\sum_{j\in\mathcal{\bar{N}}_{1}^{k-1}}\nabla_{\hat{\theta}_{2,j}}\hat{\phi}_{2,1},\ldots,\sum_{j\in\mathcal{\bar{N}}_{N}^{k-1}}\nabla_{\hat{\theta}_{2,j}}\hat{\phi}_{2,N})\in\mathbb{R}^{nN\times pN}$,
and $\hat{e}\triangleq[\hat{e}_{i}^{\top}]_{i\inv}^{\top}\dims{nN}$.
In the following subsection, we develop Lyapunov-based adaptive update
laws for the deep Lb-GNNs $\Phi_{1}$ and $\Phi_{2}$. 

\subsection{Adaptive Update Law Design}

Based on the subsequent stability analysis, we first design the distributed
adaptive update law for the observer's deep Lb-GNN weights using the
projection operator defined in (\ref{eq:proj}), where 
\begin{equation}
\hatdot{\theta}_{1,i}=\text{proj}\left(\aleph_{1,i},\hat{\theta}_{1,i},\bar{\theta}\right),\label{eq:update-law-1}
\end{equation}
and
\begin{equation}
\begin{gathered}\aleph_{1,i}\triangleq\Gamma_{1,i}\left(-k_{3}\left(\sum_{j\in\mathcal{N}_{i}}\left(\hat{\theta}_{1,i}-\hat{\theta}_{1,j}\right)+\hat{\theta}_{1,i}\right)\right.\\
+\sum_{j\in\mathcal{\bar{N}}_{i}^{k-1}}\nabla_{\hat{\theta}_{1,j}}\hat{\phi}_{1,i}^{\top}\left(\sum_{j\in\mathcal{N}_{i}}\left(\hatdot q_{0,j}-\hatdot q_{0,i}\right)\right.\\
\left.\left.+b_{i}\tildedot q_{i}+\alpha_{1}\sum_{j\in\mathcal{N}_{i}}\left(\hat{q}_{0,j}-\hat{q}_{0,i}\right)+\alpha_{1}b_{i}\tilde{q}_{i}\right)\right),
\end{gathered}
\label{eq:aleph-1-i}
\end{equation}
where $k_{3}\in\mathbb{R}_{>0}$ is a user-defined gain, $\Gamma_{1,i}\dims{p\times p}$
is a symmetric, positive-definite, user-defined gain, $\nabla_{\hat{\theta}_{1,j}}\hat{\phi}_{1,i}$
denotes the first partial derivative of the selected architecture
$\hat{\phi}_{1,i}$ with respect to the vector of weights $\hat{\theta}_{1,j}$\footnote{The adaptive update law in (\ref{eq:update-law-1}) can be generalized
for any NN architecture for which a closed form of the first partial
derivative of $\hat{\phi}_{1,i}$ with respect to its weight vector
$\hat{\theta}_{1,i}$ has been calculated, including the GNN and GAT
architectures whose first partial derivatives with respect to their
respective vectors of weights were explicitly calculated in Section
\ref{sec:gnn-architectures}.}, $\bar{\theta}\dims p$ is the upper bound of the Lb-GNN weights,
and $p$ denotes the total number of weights for the selected GNN
architecture (e.g. for the GAT architecture, $p=p_{{\rm GAT}}$, where
$p_{{\rm GAT}}=\sum_{j=0}^{k-1}2\d j+\sum_{j=0}^{k}(\d j)(\d{j-1}+1)$).
The partial derivative of the selected architecture with respect to
the vector of weights in (\ref{eq:update-law-1}) is multiplied by
an implementable form of $r_{1,i}$ defined in (\ref{eq:aux-state-est-error}),
which each agent aims to minimize. Let $\hatdot{\theta}_{1}\triangleq[\dot{\hat{\theta}}_{1,i}^{\top}]_{i\inv}^{\top}\dims{pN}$.
The ensemble update law for the weights of the observer's deep Lb-GNN
is{\small{}
\begin{equation}
\hatdot{\theta}_{1}=\left[\text{proj}(\aleph_{1,1},\hat{\theta}_{1,1},\bar{\theta})^{\top},\ldots,\text{proj}(\aleph_{1,N},\hat{\theta}_{1,N},\bar{\theta})^{\top}\right]^{\top}.\label{eq:update-law-ensemble}
\end{equation}
}{\small\par}

Similarly, we design the distributed adaptive update law for the controller's
Lb-GNN using the projection operator defined in (\ref{eq:proj}),
where
\begin{equation}
\hatdot{\theta}_{2,i}=\text{proj}\left(\aleph_{2,i},\hat{\theta}_{2,i},\bar{\theta}\right),\label{eq:update-law-2}
\end{equation}
and
\begin{equation}
\begin{gathered}\aleph_{2,i}\triangleq\Gamma_{2,i}\left(-k_{4}\left(\sum_{j\in\mathcal{N}_{i}}\left(\hat{\theta}_{2,i}-\hat{\theta}_{2,j}\right)+\hat{\theta}_{2,i}\right)\right.\\
\left.-\sum_{j\in\mathcal{\bar{N}}_{i}^{k-1}}\nabla_{\hat{\theta}_{2,j}}\hat{\phi}_{2,i}^{\top}\left(\hatdot e_{i}+\alpha_{2}\hat{e}_{i}\right)\right),
\end{gathered}
\label{eq:aleph-2-i}
\end{equation}
where $k_{4}\in\mathbb{R}_{>0}$ is a user-defined gain, $\Gamma_{2,i}\dims{p\times p}$
is a symmetric, positive-definite, user-defined gain, and $\nabla_{\hat{\theta}_{2,j}}\hat{\phi}_{2,i}$
denotes the first partial derivative of the selected architecture
$\hat{\phi}_{2,i}$ with respect to the vector of weights $\hat{\theta}_{2,j}$.
The partial derivative of the selected architecture with respect to
the vector of weights in (\ref{eq:update-law-2}) is multiplied by
$r_{2,i}$ defined in (\ref{eq:aux-state-est-reg-error}), which each
agent seeks to minimize. Let $\hatdot{\theta}_{2}\triangleq[\dot{\hat{\theta}}_{2,i}^{\top}]_{i\inv}^{\top}\dims{pN}$.
The ensemble update law for the weights of the controller's Lb-GNN
is{\small{}
\begin{equation}
\hatdot{\theta}_{2}=\left[\text{proj}(\aleph_{2,1},\hat{\theta}_{2,1},\bar{\theta})^{\top},\ldots,\text{proj}(\aleph_{2,N},\hat{\theta}_{2,N},\bar{\theta})^{\top}\right]^{\top}.\label{eq:update-law-ensemble-2}
\end{equation}
}In the following subsection, we perform a stability analysis for
the ensemble system. 

\section{Stability Analysis\label{sec:stability}}

Define the concatenated state vector $\zeta:\mathbb{R}_{\geq0}\to\mathbb{R}^{N(4n+2p)}$
as $\zeta\triangleq[\tilde{q}^{\top},\hat{e}^{\top},r_{1}^{\top},r_{2}^{\top},\tilde{\theta}_{1}^{\top},\tilde{\theta}_{2}^{\top}]^{\top}$.
Let $\Gamma_{1}\triangleq\text{blkdiag}(\Gamma_{1,1},\ldots,\Gamma_{1,N})\dims{pN\times pN}$,
$\Gamma_{2}\triangleq\text{blkdiag}(\Gamma_{2,1},\ldots,\Gamma_{2,N})\dims{pN\times pN}$,
and $P\triangleq\text{blkdiag}(I_{2nN},\mathcal{H},I_{nN},\Gamma_{1}^{-1},\Gamma_{2}^{-1})\dims{N(4n+2p)\times N(4n+2p)}$.
By the definitions of $\Gamma_{1}$ and $\Gamma_{2}$ and Remark \ref{rem:pos-def},
the $P$ matrix is positive definite and symmetric. Using (\ref{eq:state-est-error-ensemble}),
(\ref{eq:state-est-reg-error-ensemble}), (\ref{eq:update-law-ensemble}),
and (\ref{eq:update-law-ensemble-2}) yields{\footnotesize{}
\begin{equation}
\dot{\zeta}=\begin{bmatrix}r_{1}-\alpha_{1}\tilde{q}\\
r_{2}-\alpha_{2}\hat{e}\\
\tilde{F}+\nabla_{\hat{\theta}_{1}}\hat{\Phi}_{1}\tilde{\theta}_{1}+\chi_{1}+\varepsilon_{1}-k_{1}\mathcal{H}r_{1}+\alpha_{1}\left(r_{1}-\alpha_{1}\tilde{q}\right)\\
-\nabla_{\hat{\theta}_{2}}\hat{\Phi}_{2}\tilde{\theta}_{2}-\chi_{2}-\varepsilon_{2}-\left(k_{2}-\alpha_{2}\right)r_{2}-\alpha_{2}^{2}\hat{e}\\
-\begin{bmatrix}\text{proj}(\aleph_{1,1},\bar{\theta})^{\top}, & \ldots, & \text{proj}(\aleph_{1,N},\bar{\theta})^{\top}\end{bmatrix}^{\top}\\
-\begin{bmatrix}\text{proj}(\aleph_{2,1},\bar{\theta})^{\top}, & \ldots, & \text{proj}(\aleph_{2,N},\bar{\theta})^{\top}\end{bmatrix}^{\top}
\end{bmatrix}.\label{eq:ensemble-closed-loop}
\end{equation}
}{\footnotesize\par}

We consider a candidate Lyapunov function defined as
\begin{equation}
V\triangleq\frac{1}{2}\zeta^{\top}P\zeta,\label{eq:lyap}
\end{equation}
where $V:\mathbb{R}^{N(4n+2p)}\to\mathbb{R}_{\geq0}$. Let 
\begin{equation}
\lambda_{1}\triangleq\frac{1}{2}\min\left\{ 1,\lambda_{\min}(\mathcal{H}),\frac{1}{\lambda_{\max}(\Gamma_{1})},\frac{1}{\lambda_{\max}(\Gamma_{2})}\right\} ,\label{eq:lambda1}
\end{equation}
 and
\begin{equation}
\lambda_{2}\triangleq\frac{1}{2}\max\left\{ 1,\lambda_{\max}(\mathcal{H}),\frac{1}{\lambda_{\min}(\Gamma_{1})},\frac{1}{\lambda_{\min}(\Gamma_{2})}\right\} .\label{eq:lambda2}
\end{equation}
 Then, by the Rayleigh-Ritz Theorem, (\ref{eq:lyap}) satisfies the
inequality
\begin{equation}
\lambda_{1}\norm{\zeta}^{2}\leq V(\zeta)\leq\lambda_{2}\norm{\zeta}^{2}.\label{eq:rayleigh-ritz}
\end{equation}

Let $\text{\ensuremath{\underbar{\ensuremath{\lambda}}}}_{\mathcal{H}}\triangleq\lambda_{\min}(\mathcal{H}),$
and $\bar{\lambda}_{\mathcal{H}}\triangleq\lambda_{\max}(\mathcal{H})$.
The quantities $\text{\ensuremath{\underbar{\ensuremath{\lambda}}}}_{\mathcal{H}}$
and $\bar{\lambda}_{\mathcal{H}}$ can only be computed if knowledge
of the global communication architecture is obtained \emph{a priori}.
Requiring knowledge of the network's inter-agent communication topology
is undesirable when implementing a decentralized control protocol
because the control gains must be sufficiently selected to ensure
stability in comparison to these values. To mitigate the need for
complete knowledge of the communication graph in the gain conditions
for each agent, we develop bounds on $\text{\ensuremath{\underbar{\ensuremath{\lambda}}}}_{\mathcal{H}}$
and $\bar{\lambda}_{\mathcal{H}}$ only in terms of the number of
agents in the network. The developed bounds allow agents to certify
that their gains are sufficiently large in a decentralized manner
using knowledge of the number of agents in the network. For large
networks, it may be difficult to know the exact number of agents,
but it is reasonable to use a conservative upper bound to satisfy
the sufficient gain conditions. 
\begin{lem}
\label{lem:lambda-bounds} The minimum eigenvalue of $\mathcal{H}$,
defined in (\ref{eq:graph-interaction-matrix}) and denoted by $\text{\ensuremath{\underbar{\ensuremath{\lambda}}}}_{\mathcal{H}}$
is equivalent to
\[
\text{\ensuremath{\underbar{\ensuremath{\lambda}}}}_{\mathcal{H}}=2\left(1+\cos\left(\frac{2N\pi}{2N+1}\right)\right),
\]
and the maximum eigenvalue of $\mathcal{H}$, denoted by $\bar{\lambda}_{\mathcal{H}}$,
is upper-bounded as $\bar{\lambda}_{\mathcal{H}}\leq N+1$.
\end{lem}
\begin{IEEEproof}
See the Appendix.
\end{IEEEproof}
For the subsequent stability analysis to hold, gain conditions must
be sequentially established to satisfy sufficient inequalities. Let
$\epsilon_{1},\epsilon_{2},\epsilon_{3}\in\mathbb{R}_{>0}$ denote
user-selected positive constants, and define $\lambda_{3}\in\mathbb{R}_{>0}$
as
\begin{equation}
\begin{gathered}\lambda_{3}\triangleq\min\left\{ \alpha_{1}-\frac{\left(1+\alpha_{1}^{2}\bar{\lambda}_{\mathcal{H}}\right)}{2\epsilon_{1}}-\frac{NL\bar{\lambda}_{\mathcal{H}}\left(1+\alpha_{1}\right)}{2\epsilon_{3}},\right.\\
\alpha_{2}-\frac{\left(1+\alpha_{2}^{2}\right)\epsilon_{2}}{2},\frac{k_{1}\text{\ensuremath{\underbar{\ensuremath{\lambda}}}}_{\mathcal{H}}^{2}}{2}-\alpha_{1}\bar{\lambda}_{\mathcal{H}}-\frac{\left(1+\alpha_{1}^{2}\bar{\lambda}_{\mathcal{H}}\right)\epsilon_{1}}{2}\\
-\frac{NL\bar{\lambda}_{\mathcal{H}}\left(1+\alpha_{1}\right)\epsilon_{3}}{2}-NL\bar{\lambda}_{\mathcal{H}},\\
\left.\frac{k_{2}}{2}-\alpha_{2}-\frac{\left(1+\alpha_{2}^{2}\right)}{2\epsilon_{2}},\frac{k_{3}}{2},\frac{k_{4}}{2}\right\} .
\end{gathered}
\label{eq:lambda-3}
\end{equation}
To ensure the positivity of $\lambda_{3}$, the following sufficient
gain conditions are established. First, select $\alpha_{1}>0$. Then,
select the positive constant $\epsilon_{3}$ as
\begin{equation}
\epsilon_{3}>NL\bar{\lambda}_{\mathcal{H}}\left(\frac{1}{2\alpha_{1}}+\frac{1}{2}\right).\label{eq:epsilon3}
\end{equation}
Next, let the positive constant $\epsilon_{1}$ be selected as
\begin{equation}
\epsilon_{1}>\frac{1+\alpha_{1}^{2}\bar{\lambda}_{\mathcal{H}}}{2\alpha_{1}-\frac{NL\bar{\lambda}_{\mathcal{H}}\left(1+\alpha_{1}\right)}{\epsilon_{3}}}.\label{eq:epsilon1}
\end{equation}
Selection of $\epsilon_{3}$ and $\epsilon_{1}$ according to the
sufficient conditions in (\ref{eq:epsilon3}) and (\ref{eq:epsilon1})
ensures that $\alpha_{1}-\frac{\left(1+\alpha_{1}^{2}\bar{\lambda}_{\mathcal{H}}\right)}{2\epsilon_{1}}-\frac{NL\bar{\lambda}_{\mathcal{H}}\left(1+\alpha_{1}\right)}{2\epsilon_{3}}>0$.
Next, select $\alpha_{2}>0$. Then, select the positive constant $\epsilon_{2}$
as 
\begin{equation}
\frac{2\alpha_{2}}{\left(1+\alpha_{2}^{2}\right)}>\epsilon_{2}.\label{eq:epsilon2}
\end{equation}
Selection of $\epsilon_{2}$ and $\alpha_{2}$ according to the sufficient
condition in (\ref{eq:epsilon2}) ensures that $\alpha_{2}-\frac{\left(1+\alpha_{2}^{2}\right)\epsilon_{2}}{2}>0$.
Lastly, sufficient gain conditions for the observer gain $k_{1}$
defined in (\ref{eq:observer}) and the controller gain $k_{2}$ defined
in (\ref{eq:controller-sub}) are
\begin{equation}
k_{1}>\frac{\bar{\lambda}_{\mathcal{H}}}{\text{\ensuremath{\underbar{\ensuremath{\lambda}}}}_{\mathcal{H}}^{2}}\left(2\alpha_{1}+\alpha_{1}^{2}\epsilon_{1}+NL\left(2+\epsilon_{3}+\alpha_{1}\epsilon_{3}\right)\right)+\frac{\epsilon_{1}}{\text{\ensuremath{\underbar{\ensuremath{\lambda}}}}_{\mathcal{H}}^{2}},\label{eq:k1}
\end{equation}
and
\begin{equation}
k_{2}>2\alpha_{2}+\frac{\left(1+\alpha_{2}^{2}\right)}{\epsilon_{2}},\label{eq:k2}
\end{equation}
ensuring that $\frac{k_{1}\text{\ensuremath{\underbar{\ensuremath{\lambda}}}}_{\mathcal{H}}^{2}}{2}-\alpha_{1}\bar{\lambda}_{\mathcal{H}}-\frac{\left(1+\alpha_{1}^{2}\bar{\lambda}_{\mathcal{H}}\right)\epsilon_{1}}{2}-\frac{NL\bar{\lambda}_{\mathcal{H}}\left(1+\alpha_{1}\right)\epsilon_{3}}{2}-NL\bar{\lambda}_{\mathcal{H}}>0$
and $\frac{k_{2}}{2}-\alpha_{2}-\frac{\left(1+\alpha_{2}^{2}\right)}{2\epsilon_{2}}>0$
hold, respectively. Next, we establish bounds for the first Lagrange
remainder of the GNNs $\Phi_{1}$ and $\Phi_{2}$.
\begin{lem}
\label{lem:lagrange-remainder-gnn}The $z^{\text{th}}$ component
of the $y^{\text{th}}$ GNN's first Lagrange remainder $T_{y,z}$
can be upper-bounded as
\[
\lVert T_{y,z}\rVert\leq\rho\left(\norm{\kappa}\right)\norm{\tilde{\theta}_{y,z}}^{2},
\]
where $\rho:\mathbb{R}_{\geq0}\to\mathbb{R}_{\geq0}$ is a strictly
increasing polynomial that is quadratic in the norm of the ensemble
GNN input $\lVert\kappa\rVert$.
\end{lem}
\begin{IEEEproof}
See the Appendix.
\end{IEEEproof}
\begin{lem}
\label{lem:lagrange-remainder-gat}The $z^{\text{th}}$ component
of the $y^{\text{th}}$ GAT's first Lagrange remainder $T_{y,z}$
can be upper-bounded as
\[
\lVert T_{y,z}\rVert\leq\rho\left(\norm{\kappa}\right)\norm{\tilde{\theta}_{y,z}}^{2},
\]
where $\rho:\mathbb{R}_{\geq0}\to\mathbb{R}_{\geq0}$ is a strictly
increasing polynomial of degree $2k$ in terms of the norm of the
ensemble GAT input $\lVert\kappa\rVert$, where $k$ is the number
of GAT layers. 
\end{lem}
\begin{IEEEproof}
See the Appendix.
\end{IEEEproof}
To facilitate the subsequent stability analysis, we define $\upsilon\in\mathbb{R}_{>0}$
as {\small{}
\begin{equation}
\begin{gathered}\upsilon\triangleq\frac{\bar{\lambda}_{\mathcal{H}}^{2}\bar{\varepsilon}_{1}^{2}}{2\text{\ensuremath{\underbar{\ensuremath{\lambda}}}}_{\mathcal{H}}^{2}k_{1}}+\frac{\bar{\varepsilon}_{2}^{2}}{2k_{2}}+\left(\frac{k_{3}}{2}+\frac{k_{4}}{2}\right)\bar{\theta}^{2}+\frac{N}{2\epsilon_{4}}\left(1+\bar{\lambda}_{\mathcal{H}}\right).\end{gathered}
\label{eq:upsilon-ball}
\end{equation}
}Lemma \ref{lem:univ-approx} established that the universal function
approximation property of GNNs for node-level tasks only holds over
a compact domain. Therefore, the inputs of the GNNs $\Phi_{1}$ and
$\Phi_{2}$ must lie on compact domains for all $t\in\mathbb{R}_{\geq0}$
such that Lemma \ref{lem:univ-approx} holds. We enforce this condition
by proving that $\hat{Q}_{0.i}\in\mathcal{Y}_{1,i}$ and $R_{i}\in\mathcal{Y}_{2,i}$
for all $t\in\mathbb{R}_{\geq0}$. This is achieved by yielding a
stability result which constrains $\zeta$ to a compact domain. Let
$\rho:\mathbb{R}_{\geq0}\to\mathbb{R}_{\geq0}$ denote a strictly
increasing function and define $\bar{\rho}(\cdot)\triangleq\rho(\cdot)-\rho(0)$,
where $\bar{\rho}$ is strictly increasing and invertible. Then, consider
the compact domain{\small{}
\begin{equation}
\mathcal{D}\triangleq\left\{ z\in\mathbb{R}^{N(4n+2p)}:\lVert z\rVert\leq\bar{\rho}^{-1}\left(\lambda_{3}-\lambda_{4}-\rho(0)\right)\right\} ,\label{eq:compact-domain}
\end{equation}
}where $\lambda_{4}\in\mathbb{R}_{>0}$ is a user-defined constant
and $\mathcal{D}\subset\mathbb{R}^{N(4n+2p)}$. Define $c_{\zeta}\in\mathbb{R}_{>0}$
as $c_{\zeta}\triangleq\bar{\rho}^{-1}\left(\lambda_{3}-\lambda_{4}-\rho(0)\right)$.
It follows that if $\zeta\in\mathcal{D}$, then the input to the GNN
$\Phi_{1}$ at node $i$ can be bounded as $\lVert\hat{Q}_{0,i}\rVert\leq(\alpha_{1}+2)c_{\zeta}+\bar{q}_{0}+\bar{\dot{q}}_{0}$,
and the input to the GNN $\Phi_{2}$ at node $i$ can be bounded as
$\lVert R_{i}\rVert\leq N((3+\alpha_{1}+\alpha_{2})c_{\zeta}+\bar{q}_{0}+\bar{\dot{q}}_{0})$.
Therefore, $\zeta\in\mathcal{D}$ implies $G\times\hat{Q}_{0}\in\Omega_{1}$
and $G\times R\in\Omega_{2}$. 

Since the solution $t\mapsto\zeta(t)$ is continuous\footnote{Continuous solutions exist over some time interval for systems satisfying
Caratheodory existence conditions. According to Caratheodory conditions
for the system $\dot{y}=f(y,t)$, $f$ should be locally bounded,
continuous in $y$ for each fixed $t$, and measurable in $t$ for
each fixed $y$ \cite[Ch. 2, Theorem 1.1]{Coddington.Levinson1955}.
The dynamics of $\dot{\zeta}$ in (\ref{eq:ensemble-closed-loop})
satisfy the Caratheodory conditions. }, there exists a time interval $\mathcal{I}\triangleq[t_{0},t_{1}]$
for $t_{1}>t_{0}$ such that $\zeta(t)\in\mathcal{D}$ for all $t\in\mathcal{I}$.
It follows that $G\times\hat{Q}_{0}(t)\in\Omega_{1}$ and $G\times R(t)\in\Omega_{2}$
for all $t\in\mathcal{I}$. Therefore, the universal function approximation
property of GNNs described in Lemma \ref{lem:univ-approx} holds for
the GNNs $\Phi_{1}$ and $\Phi_{2}$ on the interval $[t_{0},t_{1}]$.
In the subsequent stability analysis, we analyze the convergence properties
of the solutions and also establish that $\mathcal{I}$ can be extended
to $[t_{0},\infty)$. 

To this end, we define the set of initial conditions as {\tiny{}
\begin{equation}
\mathcal{S}\triangleq\left\{ z\in\mathbb{R}^{N(4n+2p)}:\lVert z\rVert<\sqrt{\frac{\lambda_{1}}{\lambda_{2}}}\bar{\rho}^{-1}\left(\lambda_{3}-\lambda_{4}-\rho(0)\right)-\sqrt{\frac{\upsilon}{\lambda_{4}}}\right\} ,\label{eq:initial-conditions}
\end{equation}
}where $\mathcal{S}\subset\mathcal{D}$. Based on the use of a bounded
search space in (\ref{eq:theta-star-1}) and (\ref{eq:theta-star-2})
and the use of the projection operator in (\ref{eq:update-law-1})
and (\ref{eq:update-law-2}), $\tilde{\theta}_{1,i}$ and $\tilde{\theta}_{2,i}$
can each be upper-bounded as $2\bar{\theta}$ for all $i\in V$, allowing
the user to verify that an initial condition $\zeta(t_{0})$ is in
$\mathcal{S}$ given $\lambda_{1},\lambda_{2},\lambda_{3},\lambda_{4}$,
and $\upsilon$. The uniform ultimate bound is defined as\footnote{The radius of the set $\mathcal{U}$ can be made smaller by setting
the observer gain $k_{1}$, controller gain $k_{2}$, and constant
$\epsilon_{4}$ arbitrarily large, thus diminishing the impact of
the GNN function approximation error. However, the set $\mathcal{U}$
cannot be made arbitrarily small because it is limited by the size
of the $\sigma$-modification gains $k_{3}$ and $k_{4}$, which also
dictate the rate of convergence of $\lVert\zeta(t)\rVert$. } 
\begin{equation}
\mathcal{U}\triangleq\left\{ z\in\mathbb{R}^{N(4n+2p)}:z\leq\sqrt{\frac{\lambda_{2}\upsilon}{\lambda_{1}\lambda_{4}}}\right\} .\label{eq:final-set}
\end{equation}
The following theorem presents the main result.
\begin{thm}
\label{thm:stability-result}For target and agent dynamics in (\ref{eq:target-dynamics})
and (\ref{eq:agent-dynamics}), respectively, the observer in (\ref{eq:observer}),
controller in (\ref{eq:controller}), and the adaptive update law
in (\ref{eq:update-law-1}) and (\ref{eq:update-law-2}) guarantee
that for any initial condition of the states $\zeta(t_{0})\in\mathcal{S}$,
$\zeta$ exponentially converges to $\mathcal{U}$, where{\small{}
\begin{equation}
\norm{\zeta(t)}\leq\sqrt{\frac{\lambda_{2}}{\lambda_{1}}\left(\frac{\upsilon}{\lambda_{4}}+e^{-\frac{\lambda_{4}}{\lambda_{2}}\left(t-t_{0}\right)}\left(\normsq{\zeta(t_{0})}-\frac{\upsilon}{\lambda_{4}}\right)\right)},\label{eq:norm-zeta}
\end{equation}
}for all $t\in[t_{0},\infty)$ given that the constants and control
gains $\epsilon_{3},\epsilon_{1},\epsilon_{2},k_{1},$ and $k_{2}$
are selected according to the sufficient conditions in (\ref{eq:epsilon3})-(\ref{eq:k2}),
respectively and the sufficient gain condition $\lambda_{3}>\lambda_{4}+\rho\left(\sqrt{\frac{\lambda_{2}\upsilon}{\lambda_{1}\lambda_{4}}}\right)$
is satisfied.
\end{thm}
\begin{IEEEproof}
Substituting (\ref{eq:state-est-error-ensemble}), (\ref{eq:state-est-reg-error-ensemble}),
(\ref{eq:update-law-ensemble}), and (\ref{eq:update-law-ensemble-2})
into the derivative of (\ref{eq:lyap}) yields{\small{}
\begin{equation}
\begin{gathered}\dot{V}=\alpha_{1}\tilde{q}^{\top}\tilde{q}-\alpha_{2}\hat{e}^{\top}\hat{e}-k_{1}r_{1}^{\top}\mathcal{H}^{\top}\mathcal{H}r_{1}+\tilde{q}^{\top}r_{1}\\
-\left(k_{2}-\alpha_{2}\right)r_{2}^{\top}r_{2}+r_{1}^{\top}\mathcal{H}^{\top}\nabla_{\hat{\theta}_{1}}\hat{\Phi}_{1}\tilde{\theta}_{1}+\hat{e}^{\top}r_{2}\\
-r_{2}^{\top}\nabla_{\hat{\theta}_{2}}\hat{\Phi}_{2}\tilde{\theta}_{2}+\alpha_{1}r_{1}^{\top}\mathcal{H}^{\top}r_{1}-\alpha_{1}^{2}r_{1}^{\top}\mathcal{H}^{\top}\tilde{q}\\
-\alpha_{2}^{2}r_{2}^{\top}\hat{e}+r_{1}^{\top}\mathcal{H}^{\top}\tilde{F}+r_{1}^{\top}\mathcal{H}^{\top}\chi_{1}+r_{1}^{\top}\mathcal{H}\varepsilon_{1}-r_{2}\chi_{2}-r_{2}^{\top}\varepsilon_{2}\\
-\tilde{\theta}_{1}^{\top}\Gamma_{1}^{-1}\left[\text{proj}(\aleph_{1,1},\hat{\theta}_{1,1},\bar{\theta})^{\top},\ldots,\text{proj}(\aleph_{1,N},\hat{\theta}_{1,N},\bar{\theta})^{\top}\right]^{\top}\\
-\tilde{\theta}_{2}^{\top}\Gamma_{2}^{-1}\left[\text{proj}(\aleph_{2,1},\hat{\theta}_{2,1},\bar{\theta})^{\top},\ldots,\text{proj}(\aleph_{2,N},\hat{\theta}_{2,N},\bar{\theta})^{\top}\right]^{\top}.
\end{gathered}
\label{eq:lyap-dot}
\end{equation}
}Performing algebraic manipulation and applying \cite[Lemma E.1.IV]{Krstic1995}
yields the bound
\begin{equation}
\begin{gathered}-\tilde{\theta}_{1}^{\top}\Gamma_{1}^{-1}\left[\text{proj}(\aleph_{1,1},\hat{\theta}_{1,1},\bar{\theta})^{\top},\right.\\
\left.\ldots,\text{proj}(\aleph_{1,N},\hat{\theta}_{1,N},\bar{\theta})^{\top}\right]^{\top}\leq-\tilde{\theta}_{1}^{\top}\Gamma_{1}^{-1}\aleph_{1},
\end{gathered}
\label{eq:proj-bound-1}
\end{equation}
where $\aleph_{1}\triangleq[\aleph_{1,1}^{\top},\ldots,\aleph_{1,N}^{\top}]^{\top}\dims{pN}$
and $\aleph_{1,i}$ is defined in (\ref{eq:aleph-1-i}). Similar to
the graph interaction matrix in (\ref{eq:graph-interaction-matrix}),
we define the weight interaction matrix, $\mathcal{J}$ as $\mathcal{J}\triangleq\mathcal{L}_{G}\otimes I_{p}$,
where $\mathcal{L}_{G}$ is the graph Laplacian and $\mathcal{J}\dims{pN\times pN}$.
The ensemble form of $\aleph_{1,i}$ is\footnote{The terms featuring the weight interaction matrix $\mathcal{J}$ in
(\ref{eq:aleph-ensemble-1}) and (\ref{eq:aleph-ensemble-2}) function
as a consensus protocol with respect to the GNN weights. Specifically,
agents leverage their neighbors' estimates of the ideal GNN weights
to refine their own estimates.}
\begin{align}
\aleph_{1} & =\Gamma_{1}\left(k_{3}\mathcal{J}\tilde{\theta}_{1}-k_{3}\hat{\theta}_{1}+\nabla_{\hat{\theta}_{1}}\hat{\Phi}_{1}^{\top}\mathcal{H}r_{1}\right).\label{eq:aleph-ensemble-1}
\end{align}
 Once again, we apply \cite[Lemma E.1.IV]{Krstic1995} which yields
the bound
\begin{equation}
\begin{gathered}-\tilde{\theta}_{2}^{\top}\Gamma_{2}^{-1}\left[\text{proj}(\aleph_{2,1},\hat{\theta}_{2,1},\bar{\theta})^{\top},\right.\\
\left.\ldots,\text{proj}(\aleph_{2,N},\hat{\theta}_{2,N},\bar{\theta})^{\top}\right]^{\top}\leq-\tilde{\theta}_{2}^{\top}\Gamma_{2}^{-1}\aleph_{2},
\end{gathered}
\label{eq:proj-bound-2}
\end{equation}
 where $\aleph_{2}\triangleq[\aleph_{2,1}^{\top},\ldots,\aleph_{2,N}^{\top}]^{\top}\dims{pN}$
and $\aleph_{2,i}$ is defined in (\ref{eq:aleph-2-i}). The ensemble
form of $\aleph_{2,i}$ is
\begin{equation}
\aleph_{2}=\Gamma_{2}\left(k_{4}\mathcal{J}\tilde{\theta}_{2}-k_{4}\hat{\theta}_{2}-\nabla_{\hat{\theta}_{2}}\hat{\Phi}_{2}^{\top}r_{2}\right).\label{eq:aleph-ensemble-2}
\end{equation}
Applying the bounds in (\ref{eq:proj-bound-1}) and (\ref{eq:proj-bound-2})
and substituting (\ref{eq:aleph-ensemble-1}) and (\ref{eq:aleph-ensemble-2})
into (\ref{eq:lyap-dot}) gives
\begin{equation}
\begin{gathered}\dot{V}\leq\alpha_{1}\tilde{q}^{\top}\tilde{q}-\alpha_{2}\hat{e}^{\top}\hat{e}-k_{1}r_{1}^{\top}\mathcal{H}^{\top}\mathcal{H}r_{1}\\
-\left(k_{2}-\alpha_{2}\right)r_{2}^{\top}r_{2}-k_{3}\tilde{\theta}_{1}^{\top}\tilde{\theta}_{1}^{\top}-k_{4}\tilde{\theta}_{2}^{\top}\tilde{\theta}_{2}^{\top}\\
+\tilde{q}^{\top}r_{1}+\hat{e}^{\top}r_{2}+\alpha_{1}r_{1}^{\top}\mathcal{H}^{\top}r_{1}-\alpha_{1}^{2}r_{1}^{\top}\mathcal{H}^{\top}\tilde{q}\\
-\alpha_{2}^{2}r_{2}^{\top}\hat{e}+k_{3}\tilde{\theta}_{1}^{\top}\theta_{1}^{\ast}-k_{3}\tilde{\theta}_{1}^{\top}\mathcal{J}\tilde{\theta}_{1}+k_{4}\tilde{\theta}_{2}^{\top}\theta_{2}^{\ast}+r_{1}^{\top}\mathcal{H}\varepsilon_{1}\\
-k_{4}\tilde{\theta}_{2}^{\top}\mathcal{J}\tilde{\theta}_{2}+r_{1}^{\top}\mathcal{H}^{\top}\tilde{F}+r_{1}^{\top}\mathcal{H}^{\top}\chi_{1}-r_{2}\chi_{2}-r_{2}^{\top}\varepsilon_{2}.
\end{gathered}
\label{eq:lyap-dot-1}
\end{equation}
We note that $\lambda_{\min}(\mathcal{J})=\lambda_{\min}(\mathcal{L}_{G})=0$.
Then, $-k_{3}\tilde{\theta}_{1}^{\top}\mathcal{J}\tilde{\theta}_{1}\leq0$
and $-k_{4}\tilde{\theta}_{2}^{\top}\mathcal{J}\tilde{\theta}_{2}\leq0$.
Using Young's inequality yields the bounds
\begin{equation}
\begin{gathered}\tilde{q}^{\top}r_{1}-\alpha_{1}^{2}r_{1}^{\top}\mathcal{H}^{\top}\tilde{q}\leq\frac{\left(1+\alpha_{1}^{2}\bar{\lambda}_{\mathcal{H}}\right)\epsilon_{1}}{2}\norm{r_{1}}^{2}\\
+\frac{\left(1+\alpha_{1}^{2}\bar{\lambda}_{\mathcal{H}}\right)}{2\epsilon_{1}}\norm{\tilde{q}}^{2},
\end{gathered}
\label{eq:alpha-youngs-1}
\end{equation}
and
\begin{equation}
\hat{e}^{\top}r_{2}-\alpha_{2}^{2}r_{2}^{\top}\hat{e}\leq\frac{\left(1+\alpha_{2}^{2}\right)\epsilon_{2}}{2}\norm{\hat{e}}^{2}+\frac{\left(1+\alpha_{2}^{2}\right)}{2\epsilon_{2}}\norm{r_{2}}^{2},\label{eq:alpha-youngs-2}
\end{equation}
where $\epsilon_{1},\epsilon_{2}$ were introduced in (\ref{eq:epsilon3})
and (\ref{eq:epsilon2}), respectively. Using Young's inequality,
Assumption \ref{ass:target-bound}, and Assumption \ref{ass:lipschitz}
to upper-bound $r_{1}^{\top}\mathcal{H}^{\top}\tilde{F}$ yields
\begin{align}
r_{1}^{\top}\mathcal{H}\tilde{F} & \leq NL\bar{\lambda}_{\mathcal{H}}\norm{r_{1}}^{2}+NL\bar{\lambda}_{\mathcal{H}}\left(1+\alpha_{1}\right)\norm{r_{1}}\norm{\tilde{q}}.\label{eq:tilde-youngs}
\end{align}
Using Young's inequality and nonlinear damping yields
\begin{align}
-k_{1}\text{\ensuremath{\underbar{\ensuremath{\lambda}}}}_{\mathcal{H}}^{2}\norm{r_{1}}^{2}-r_{1}^{\top}\mathcal{H}^{\top}\varepsilon_{1} & \leq-\frac{k_{1}}{2}\text{\ensuremath{\underbar{\ensuremath{\lambda}}}}_{\mathcal{H}}^{2}\norm{r_{1}}^{2}+\frac{\bar{\lambda}_{\mathcal{H}}^{2}\bar{\varepsilon}_{1}^{2}}{2\text{\ensuremath{\underbar{\ensuremath{\lambda}}}}_{\mathcal{H}}^{2}k_{1}},\label{eq:eps-youngs-1}
\end{align}
and
\begin{equation}
-k_{2}\normsq{r_{2}}-r_{2}\varepsilon_{2}\leq-\frac{k_{2}}{2}\norm{r_{2}}^{2}+\frac{\bar{\varepsilon}_{2}^{2}}{2k_{2}}.\label{eq:eps-youngs-2}
\end{equation}
By Young's inequality, Lemma \ref{lem:lagrange-remainder-gnn}, and
Lemma \ref{lem:lagrange-remainder-gat}, there exists some $\rho_{G,1}:\mathbb{R}_{\geq0}\to\mathbb{R}_{\geq0}$,
$\rho_{G,2}:\mathbb{R}_{\geq0}\to\mathbb{R}_{\geq0}$, $\rho_{T,1}:\mathbb{R}_{\geq0}\to\mathbb{R}_{\geq0}$
and $\rho_{T,2}:\mathbb{R}_{\geq0}\to\mathbb{R}_{\geq0}$, such that
$\rho_{G,1},\rho_{G,2},\rho_{T,1}$ and $\rho_{T,2}$ are strictly
increasing functions. Then,
\begin{equation}
\begin{gathered}r_{1}^{\top}\mathcal{H}\chi_{1}\leq\frac{N\bar{\lambda}_{\mathcal{H}}}{2\epsilon_{4}}\\
+N\bar{\lambda}_{\mathcal{H}}\left(\rho_{T,1}\left(\norm{\zeta}\right)\norm{\zeta}+\frac{\epsilon_{4}}{2}\left(\rho_{G,1}\left(\norm{\zeta}\right)\right)^{2}\right)\norm{\zeta}^{2},
\end{gathered}
\label{eq:chi-lagrange-1}
\end{equation}
and
\begin{equation}
\begin{gathered}r_{2}^{\top}\chi_{2}\leq\frac{N}{2\epsilon_{4}}\\
+N\left(\rho_{T,2}\left(\norm{\zeta}\right)\norm{\zeta}+\frac{\epsilon_{4}}{2}\left(\rho_{G,2}\left(\norm{\zeta}\right)\right)^{2}\right)\norm{\zeta}^{2},
\end{gathered}
\label{eq:chi-lagrange-2}
\end{equation}
where $\epsilon_{4}\in\mathbb{R}_{>0}$ is a user-defined constant.
Since $\rho_{G,1},\rho_{G,2},\rho_{T,1}$ and $\rho_{T,2}$ are strictly
increasing functions, there exist some strictly increasing functions
$\rho_{1}:\mathbb{R}_{\geq0}\to\mathbb{R}_{\geq0}$ and $\rho_{2}:\mathbb{R}_{\geq0}\to\mathbb{R}_{\geq0}$
such that{\small{}
\begin{equation}
N\bar{\lambda}_{\mathcal{H}}\left(\rho_{T,1}\left(\norm{\zeta}\right)\norm{\zeta}+\frac{\epsilon_{4}}{2}\left(\rho_{G,1}\left(\norm{\zeta}\right)^{2}\right)\right)\leq\rho_{1}\left(\norm{\zeta}\right),\label{eq:r1-chi-bound}
\end{equation}
}and
\begin{equation}
N\left(\rho_{T,2}\left(\norm{\zeta}\right)\norm{\zeta}+\frac{\epsilon_{4}}{2}\left(\rho_{G,2}\left(\norm{\zeta}\right)\right)^{2}\right)\leq\rho_{2}\left(\norm{\zeta}\right).\label{eq:r2-chi-bound}
\end{equation}
Lastly, since $\rho_{1}$ and $\rho_{2}$ are strictly increasing
functions, there exists some $\rho:\mathbb{R}_{\geq0}\to\mathbb{R}_{\geq0}$
such that $\rho(\lVert\zeta\rVert)>\rho_{1}(\lVert\zeta\rVert)+\rho_{2}(\lVert\zeta\rVert)$.
Using (\ref{eq:alpha-youngs-1})-(\ref{eq:r2-chi-bound}) to upper-bound
(\ref{eq:lyap-dot-1}) yields
\begin{equation}
\begin{aligned}\dot{V} & \leq-\left(\alpha_{1}-\frac{\left(1+\alpha_{1}^{2}\bar{\lambda}_{\mathcal{H}}\right)}{2\epsilon_{1}}-\frac{NL\bar{\lambda}_{\mathcal{H}}\left(1+\alpha_{1}\right)}{2\epsilon_{3}}\right)\normsq{\tilde{q}}\\
 & -\left(\alpha_{2}-\frac{\left(1+\alpha_{2}^{2}\right)\epsilon_{2}}{2}\right)\normsq{\hat{e}}-\left(\frac{k_{1}\text{\ensuremath{\underbar{\ensuremath{\lambda}}}}_{\mathcal{H}}^{2}}{2}-\alpha_{1}\bar{\lambda}_{\mathcal{H}}\right.\\
 & -\frac{\left(1+\alpha_{1}^{2}\bar{\lambda}_{\mathcal{H}}\right)\epsilon_{1}}{2}-\frac{NL\bar{\lambda}_{\mathcal{H}}\left(1+\alpha_{1}\right)\epsilon_{3}}{2}\\
 & -NL\bar{\lambda}_{\mathcal{H}}\biggr)\normsq{r_{1}}-\left(\frac{k_{2}}{2}-\alpha_{2}-\frac{\left(1+\alpha_{2}^{2}\right)}{2\epsilon_{2}}\right)\normsq{r_{2}}\\
 & -\left(\frac{k_{3}}{2}\right)\normsq{\tilde{\theta}_{1}}-\left(\frac{k_{4}}{2}\right)\norm{\tilde{\theta}_{2}}^{2}+\left(\frac{k_{3}}{2}+\frac{k_{4}}{2}\right)\bar{\theta}^{2}\\
 & +\frac{\bar{\lambda}_{\mathcal{H}}^{2}\bar{\varepsilon}_{1}^{2}}{2\text{\ensuremath{\underbar{\ensuremath{\lambda}}}}_{\mathcal{H}}^{2}k_{1}}+\frac{\bar{\varepsilon}_{2}^{2}}{2k_{2}}+\frac{N}{2\epsilon_{4}}\left(1+\bar{\lambda}_{\mathcal{H}}\right)+\rho\left(\lVert\zeta\rVert\right)\lVert\zeta\rVert^{2}.
\end{aligned}
\label{eq:lyap-dot-simplified}
\end{equation}
Provided constants and control gains $\epsilon_{3},\epsilon_{1},\epsilon_{2},k_{1}$,
and $k_{2}$ are selected according to the sufficient conditions in
(\ref{eq:epsilon3})-(\ref{eq:k2}), respectively, then (\ref{eq:rayleigh-ritz})
can be used to upper-bound (\ref{eq:lyap-dot-simplified}) as 
\begin{equation}
\dot{V}\leq-\left(\lambda_{3}-\rho\left(\lVert\zeta\rVert\right)\right)\norm{\zeta}^{2}+\upsilon,\label{eq:lyap-dot-lambda-3}
\end{equation}
 for all $t\in\mathcal{I}$, where $\upsilon$ is defined in (\ref{eq:upsilon-ball})
and $\lambda_{3}$ is defined in (\ref{eq:lambda-3}). Since the solution
$t\mapsto\zeta(t)$ is continuous, we have $\zeta(t)\in\mathcal{D}$
for all $t\in\mathcal{I}$. Consequently, applying (\ref{eq:rayleigh-ritz})
to (\ref{eq:lyap-dot-lambda-3}) yields 
\begin{equation}
\dot{V}\leq-\frac{\lambda_{4}}{\lambda_{2}}V(\zeta)+\upsilon,\label{eq:lyap-dot-small}
\end{equation}
for all $t\in\mathcal{I}$. Solving the differential inequality in
(\ref{eq:lyap-dot-small}) over $\mathcal{I}$ gives {\small{}
\begin{equation}
V\left(\zeta(t)\right)\leq V\left(\zeta(t_{0})\right)e^{-\frac{\lambda_{4}}{\lambda_{2}}\left(t-t_{0}\right)}+\frac{\lambda_{2}\upsilon}{\lambda_{4}}\left(1-e^{-\frac{\lambda_{4}}{\lambda_{2}}\left(t-t_{0}\right)}\right).\label{eq:lyap-almost-there}
\end{equation}
}Applying (\ref{eq:rayleigh-ritz}) and (\ref{eq:lyap-almost-there})
yields

{\small{}
\begin{equation}
\lVert\zeta(t)\rVert\leq\sqrt{e^{-\frac{\lambda_{4}}{\lambda_{2}}\left(t-t_{0}\right)}\left(\frac{\lambda_{2}}{\lambda_{1}}\normsq{\zeta(t_{0})}-\frac{\lambda_{2}\upsilon}{\lambda_{1}\lambda_{4}}\right)+\frac{\lambda_{2}\upsilon}{\lambda_{1}\lambda_{4}}},\label{eq:norm-zeta-proof}
\end{equation}
}for all $t\in\mathcal{I}$. Next, we must show that $\mathcal{I}$
can be extended to $[t_{0},\infty)$. 

Let $t\mapsto\zeta(t)$ be a solution to the ordinary differential
equation (\ref{eq:ensemble-closed-loop}) with initial condition $\zeta(t_{0})\in\mathcal{S}$.
By \cite[Lemma E.1]{Krstic1995}, the projection operator defined
in (\ref{eq:proj}) is locally Lipschitz in its arguments. Then, the
right hand side of (\ref{eq:ensemble-closed-loop}) is piecewise continuous
in $t$ and locally Lipschitz in $\zeta$ for all $t\geq t_{0}$ and
$\zeta\in\mathbb{R}^{N(4n+2p)}$. Taking the upper-bound of (\ref{eq:norm-zeta-proof})
yields
\begin{equation}
\lVert\zeta(t)\rVert<\sqrt{\frac{\lambda_{2}}{\lambda_{1}}}\norm{\zeta(t_{0})}+\sqrt{\frac{\lambda_{2}\upsilon}{\lambda_{1}\lambda_{4}}},\label{eq:norm-zeta-upper-bound}
\end{equation}
for all $t\in\mathcal{\mathcal{I}}$. Since $\zeta(t_{0})\in\mathcal{S}$,
we have that 
\begin{equation}
\lVert\zeta(t_{0})\rVert<\sqrt{\frac{\lambda_{1}}{\lambda_{2}}}\bar{\rho}^{-1}\left(\lambda_{3}-\lambda_{4}-\rho(0)\right)-\sqrt{\frac{\upsilon}{\lambda_{4}}}.\label{eq:s-def}
\end{equation}
Applying (\ref{eq:s-def}) to (\ref{eq:norm-zeta-upper-bound}) gives
$\lVert\zeta(t)\rVert<\bar{\rho}^{-1}\left(\lambda_{3}-\lambda_{4}-\rho(0)\right)$
for all $t\in\mathcal{I}$, which by the definition of (\ref{eq:compact-domain}),
yields $\zeta(t)\in\mathcal{D}$ for all $t\in\mathcal{\mathcal{I}}.$
Since $\zeta(t)$ remains in the compact set $\mathcal{D}\subset\mathbb{R}^{N(4n+2p)}$
for all $t\in\mathcal{\mathcal{I}}$, by \cite[Theorem 3.3]{Khalil2002},
a unique solution exists for all $t\geq t_{0}$. Therefore, $\mathcal{I}=[t_{0},\infty)$. 

Thus, for $\zeta(t_{0})\in\mathcal{S}$, we have{\small{}
\begin{equation}
\lVert\zeta(t)\rVert\leq\sqrt{e^{-\frac{\lambda_{4}}{\lambda_{2}}\left(t-t_{0}\right)}\left(\frac{\lambda_{2}}{\lambda_{1}}\normsq{\zeta(t_{0})}-\frac{\lambda_{2}\upsilon}{\lambda_{1}\lambda_{4}}\right)+\frac{\lambda_{2}\upsilon}{\lambda_{1}\lambda_{4}}},\label{eq:norm-zeta-all-time}
\end{equation}
}and $\zeta(t)\in\mathcal{D}$ for all $t\in[t_{0},\infty)$. The
limit of (\ref{eq:norm-zeta-all-time}) as $t\to\infty$ yields $\lVert\zeta\rVert\leq\sqrt{\frac{\lambda_{2}\upsilon}{\lambda_{1}\lambda_{4}}}$,
which indicates that $\zeta(t)$ converges to the set $\mathcal{U}$.
For any bounded set $\mathcal{C}\subset\mathbb{R}^{N(4n+2p)}$ with
radius ${\tt R}_{\mathcal{C}}>0$, selecting $\lambda_{3}>\lambda_{4}+\rho(\sqrt{\frac{\lambda_{2}}{\lambda_{1}}}{\tt R}_{\mathcal{C}}+\sqrt{\frac{\lambda_{2}\upsilon}{\lambda_{1}\lambda_{4}}})$
ensures that $\lVert\zeta(t)\rVert\leq\sqrt{e^{-\frac{\lambda_{4}}{\lambda_{2}}\left(t-t_{0}\right)}\left(\frac{\lambda_{2}}{\lambda_{1}}\normsq{\zeta(t_{0})}-\frac{\lambda_{2}\upsilon}{\lambda_{1}\lambda_{4}}\right)+\frac{\lambda_{2}\upsilon}{\lambda_{1}\lambda_{4}}}$.
This implies the stability result is semi-global \cite[Remark 2]{Pettersen2017},
as the set of stabilizing initial conditions can be made arbitrarily
large by appropriately adjusting $\lambda_{3}$ to encompass any bounded
subset of $\mathbb{R}^{N(4n+2p)}$.

Next, we prove that Assumption \ref{ass:lipschitz} holds for all
$t\in\mathbb{R}_{\geq0}$, which states that the unknown drift dynamics
modeled by $f(Q_{0})$ are Lipschitz on the set $\mathcal{Y}_{1,i}$,
where the Lipschitz constant has a known upper bound $L$. Since $\zeta(t)\in\mathcal{D}$
for all $t\in[t_{0},\infty)$, it follows that $\lVert\zeta(t)\rVert\leq c_{\zeta}$
for all $t\in[t_{0},\infty)$. Applying the triangle inequality, (\ref{eq:state-est-error}),
and (\ref{eq:aux-state-est-reg-error}) yields $\lVert Q_{0}-\hat{Q}_{0,i}\rVert\leq c_{\zeta}(2+\alpha_{1})$
for $\zeta\in\mathcal{D}$. Thus, $Q_{0}-\hat{Q}_{0,i}\in\mathcal{Y}_{1,i}$
for all $i\in V$ and Assumption \ref{ass:lipschitz} holds everywhere.

We prove that $\hat{Q}_{0,i}\in\mathcal{Y}_{1,i}$ for all $i\in V$
such that Lemma \ref{lem:univ-approx} holds by applying (\ref{eq:state-est-error}),
(\ref{eq:aux-state-est-error}), and Assumption \ref{ass:target-bound}
as above to find $\lVert\hat{Q}_{0,i}\rVert\leq\bar{q}_{0}+\bar{\dot{q}}_{0}+c_{\zeta}(2+\alpha_{1})$
for $\zeta\in\mathcal{D}$. Thus, $\hat{Q}_{0,i}\in\mathcal{Y}_{1,i}$
for all $i\in V$, $\hat{Q}_{0}\in\mathcal{Y}_{1}$, and the universal
function approximation property of the deep Lb-GNN $\Phi_{1}$ described
in Lemma \ref{lem:univ-approx} holds everywhere. We repeat this process
to show that Lemma \ref{lem:univ-approx} holds for the deep Lb-GNN
$\Phi_{2}$. We upper-bound $\norm{Q_{i}}$ using Assumption \ref{ass:target-bound}
and the triangle inequality which yields
\begin{equation}
\norm{Q_{i}}\leq\bar{q}_{0}+\bar{\dot{q}}_{0}+c_{\zeta}\left(3+\alpha_{1}+\alpha_{2}\right),\label{eq:q-upper-bound}
\end{equation}
for $\zeta\in\mathcal{D}$. Next, we upper-bound $\lVert R_{i}\rVert$.
We apply the triangle inequality, (\ref{eq:q-upper-bound}), and the
fact that $\lVert R_{i}\rVert$ is the largest for a fully connected
graph, where $1_{\mathcal{\bar{N}}_{i}}(m)=1$ for all $m\in V$,
which yields $\lVert R_{i}\rVert\leq N(\bar{q}_{0}+\bar{\dot{q}}_{0}+c_{\zeta}(3+\alpha_{1}+\alpha_{2}))$
for $\zeta\in\mathcal{D}$. Therefore, for $\zeta\in\mathcal{D}$,
$R_{i}\in\mathcal{Y}_{2,i}$ holds for all $i\in V$, $R\in\mathcal{Y}_{2}$,
and the universal function approximation property of the deep GNN
$\Phi_{2}$ described in Lemma \ref{lem:univ-approx} holds everywhere.

Since $\lVert\zeta\rVert\leq c_{\zeta}$, $\lVert\tilde{q}\rVert,\lVert\hat{e}\rVert,\lVert r_{1}\rVert,\lVert r_{2}\rVert,\lVert\tilde{\theta}_{1}\rVert,\lVert\tilde{\theta}_{2}\rVert\leq c_{\zeta}$.
Therefore, $\lVert\tilde{q}\rVert,\lVert\hat{e}\rVert,\lVert r_{1}\rVert,\lVert r_{2}\rVert,\lVert\tilde{\theta}_{1}\rVert,\lVert\tilde{\theta}_{2}\rVert\in\mathcal{L}_{\infty}$.
Since $\lVert\tilde{q}\rVert,\lVert r_{1}\rVert\in\mathcal{L}_{\infty}$,
then $\lVert\hat{Q}_{0}\rVert\in\mathcal{L}_{\infty}$. Based on the
fact that $\lVert\hat{Q}_{0}\rVert\in\mathcal{L}_{\infty}$ and the
use of the projection operator in (\ref{eq:update-law-1}), $\lVert\hat{\Phi}_{1}\rVert$
is bounded. Since $\lVert\hat{e}\rVert,\lVert r_{2}\rVert\in\mathcal{L}_{\infty}$,
then $\lVert R\rVert\in\mathcal{L}_{\infty}$. Based on the fact that
$\lVert R\rVert\in\mathcal{L}_{\infty}$ and the use of the projection
operator in (\ref{eq:update-law-2}), $\lVert\hat{\Phi}_{2}\rVert$
is bounded. Based on the fact that $\lVert\tilde{\theta}_{1}\rVert,\lVert\tilde{\theta}_{2}\rVert\in\mathcal{L}_{\infty}$,
the use of the projection operator, and the use of a bounded search
space in (\ref{eq:theta-star-1}) and (\ref{eq:theta-star-2}), $\lVert\hat{\theta}_{1}\rVert,\lVert\hat{\theta}_{2}\rVert\in\mathcal{L}_{\infty}$.
Based on the use of activation functions with bounded derivatives
and the fact that $\lVert\hat{Q}_{0}\rVert,\lVert R\rVert,\lVert\hat{\theta}_{1}\rVert,\lVert\hat{\theta}_{2}\rVert\in\mathcal{L}_{\infty}$,
$\lVert\nabla_{\hat{\theta}_{1}}\hat{\Phi}_{1}\rVert$ and $\lVert\nabla_{\hat{\theta}_{2}}\hat{\Phi}_{2}\rVert$
are bounded. Let $\ddot{\hat{q}}_{0}\triangleq[\ddot{\hat{q}}_{0,i}^{\top}]_{i\in V}^{\top}\dims{nN}$.
Based on the fact that $\lVert r_{1}\rVert,\lVert\hat{\theta}\rVert_{1}\in\mathcal{L}_{\infty}$
and the boundedness of $\lVert\hat{\Phi}_{1}\rVert$ and $\lVert\nabla_{\hat{\theta}_{1}}\hat{\Phi}_{1}\rVert$,
$\lVert\ddot{\hat{q}}_{0}\rVert$ is bounded. Let $u\triangleq[u_{i}^{\top}]_{i\in V}^{\top}\dims{nN}$.
Based on the fact that $\lVert r_{2}\rVert,\lVert\hat{\theta}_{2}\rVert\in\mathcal{L}_{\infty}$
and the boundedness of $\lVert\hat{\Phi}_{2}\rVert$, $\lVert\nabla_{\hat{\theta}_{2}}\hat{\Phi}_{2}\rVert$,
and $\lVert\ddot{\hat{q}}_{0}\rVert$, $\lVert u\rVert$ is bounded.
Based on the fact that $\lVert\tilde{\theta}_{1}\rVert,\lVert\hat{\theta}_{1}\rVert,\lVert r_{1}\rVert\in\mathcal{L}_{\infty}$
and the boundedness of $\lVert\nabla_{\hat{\theta}_{1}}\hat{\Phi}_{1}\rVert$,
$\lVert\hatdot{\theta}_{1}\rVert$ is bounded. Based on the fact that
$\lVert\tilde{\theta}_{2}\rVert,\lVert\hat{\theta}_{2}\rVert,\lVert r_{2}\rVert\in\mathcal{L}_{\infty}$
and the boundedness of $\lVert\nabla_{\hat{\theta}_{2}}\hat{\Phi}_{2}\rVert$,
$\lVert\hatdot{\theta}_{2}\rVert$ is bounded. 

\end{IEEEproof}

\section{Simulation Results\label{sec:simulation}}

\begin{table*}[t]
\centering{}\caption{\label{tab:compare-results}Performance comparison results for various
NN architectures and communication topologies. For a network of $N$
agents and a signal $s_{i}({\tt k})\in\mathbb{R}^{n}$ with ${\tt K}\in\mathbb{Z}_{>0}$
samples, the average root mean square (RMS) value of the signal $s_{RMS}$
is given as $s_{{\rm RMS}}\triangleq1/N\cdot\sum_{i\in V}(1/{\tt K}\cdot\sum_{{\tt k}=1}^{{\tt K}}s_{i}({\tt k})^{\top}s_{i}({\tt k}))$.}

\centering{}%
\begin{tabular}{ccccccccc}
\toprule 
 & \multirow{10}{*}{} & \multicolumn{7}{c}{Path Graph $(P_{6})$}\tabularnewline
\cmidrule{1-1} \cmidrule{3-9} \cmidrule{4-9} \cmidrule{5-9} \cmidrule{6-9} \cmidrule{7-9} \cmidrule{8-9} \cmidrule{9-9} 
NN Architecture &  & $e_{{\rm RMS}}$ & $\dot{e}_{{\rm RMS}}$ & $\tilde{\Phi}_{1,{\rm RMS}}[0:10]$ & $\tilde{\Phi}_{1,{\rm RMS}}[10:60]$ & $\tilde{\Phi}_{2,{\rm RMS}}[0:10]$ & $\tilde{\Phi}_{2,{\rm RMS}}[10:60]$ & $u_{{\rm RMS}}$\tabularnewline
\cmidrule{1-1} \cmidrule{3-9} \cmidrule{4-9} \cmidrule{5-9} \cmidrule{6-9} \cmidrule{7-9} \cmidrule{8-9} \cmidrule{9-9} 
DNN$+$DNN &  & $0.4615$ & $0.4207$ & $0.8241$ & $0.5414$ & $0.2755$ & $0.0801$ & $1.088$\tabularnewline
\cmidrule{1-1} \cmidrule{3-9} \cmidrule{4-9} \cmidrule{5-9} \cmidrule{6-9} \cmidrule{7-9} \cmidrule{8-9} \cmidrule{9-9} 
GNN$+$GNN &  & $0.3745$ & $0.4008$ & $0.8921$ & $0.4809$ & $2.468$ & $0.0855$ & $1.420$\tabularnewline
\cmidrule{1-1} \cmidrule{3-9} \cmidrule{4-9} \cmidrule{5-9} \cmidrule{6-9} \cmidrule{7-9} \cmidrule{8-9} \cmidrule{9-9} 
GAT$+$GNN &  & $0.2826$ & $0.3614$ & $2.289$ & $0.4017$ & $2.698$ & $0.0688$ & $1.621$\tabularnewline
\cmidrule{1-1} \cmidrule{3-9} \cmidrule{4-9} \cmidrule{5-9} \cmidrule{6-9} \cmidrule{7-9} \cmidrule{8-9} \cmidrule{9-9} 
 &  & \multicolumn{7}{c}{Ring Graph $(R_{6})$}\tabularnewline
\cmidrule{1-1} \cmidrule{3-9} \cmidrule{4-9} \cmidrule{5-9} \cmidrule{6-9} \cmidrule{7-9} \cmidrule{8-9} \cmidrule{9-9} 
NN Architecture &  & $e_{{\rm RMS}}$ & $\dot{e}_{{\rm RMS}}$ & $\tilde{\Phi}_{1,{\rm RMS}}[0:10]$ & $\tilde{\Phi}_{1,{\rm RMS}}[10:60]$ & $\tilde{\Phi}_{2,{\rm RMS}}[0:10]$ & $\tilde{\Phi}_{2,{\rm RMS}}[10:60]$ & $u_{{\rm RMS}}$\tabularnewline
\cmidrule{1-1} \cmidrule{3-9} \cmidrule{4-9} \cmidrule{5-9} \cmidrule{6-9} \cmidrule{7-9} \cmidrule{8-9} \cmidrule{9-9} 
DNN$+$DNN &  & $0.4389$ & $0.4027$ & $0.8243$ & $0.5390$ & $0.2773$ & $0.0800$ & $1.086$\tabularnewline
\cmidrule{1-1} \cmidrule{3-9} \cmidrule{4-9} \cmidrule{5-9} \cmidrule{6-9} \cmidrule{7-9} \cmidrule{8-9} \cmidrule{9-9} 
GNN$+$GNN &  & $0.3144$ & $0.3609$ & $0.9347$ & $0.4346$ & $2.131$ & $0.0799$ & $1.399$\tabularnewline
\cmidrule{1-1} \cmidrule{3-9} \cmidrule{4-9} \cmidrule{5-9} \cmidrule{6-9} \cmidrule{7-9} \cmidrule{8-9} \cmidrule{9-9} 
GAT$+$GNN &  & $0.2730$ & $0.3415$ & $2.304$ & $0.3737$ & $2.549$ & $0.0632$ & $1.616$\tabularnewline
\cmidrule{1-1} \cmidrule{3-9} \cmidrule{4-9} \cmidrule{5-9} \cmidrule{6-9} \cmidrule{7-9} \cmidrule{8-9} \cmidrule{9-9} 
 &  & \multicolumn{7}{c}{Star Graph $(S_{6})$}\tabularnewline
\cmidrule{1-1} \cmidrule{3-9} \cmidrule{4-9} \cmidrule{5-9} \cmidrule{6-9} \cmidrule{7-9} \cmidrule{8-9} \cmidrule{9-9} 
NN Architecture &  & $e_{{\rm RMS}}$ & $\dot{e}_{{\rm RMS}}$ & $\tilde{\Phi}_{1,{\rm RMS}}[0:10]$ & $\tilde{\Phi}_{1,{\rm RMS}}[10:60]$ & $\tilde{\Phi}_{2,{\rm RMS}}[0:10]$ & $\tilde{\Phi}_{2,{\rm RMS}}[10:60]$ & $u_{{\rm RMS}}$\tabularnewline
\cmidrule{1-1} \cmidrule{3-9} \cmidrule{4-9} \cmidrule{5-9} \cmidrule{6-9} \cmidrule{7-9} \cmidrule{8-9} \cmidrule{9-9} 
DNN$+$DNN &  & $0.5198$ & $0.4676$ & $0.8230$ & $0.5473$ & $0.3605$ & $0.0831$ & $1.105$\tabularnewline
\cmidrule{1-1} \cmidrule{3-9} \cmidrule{4-9} \cmidrule{5-9} \cmidrule{6-9} \cmidrule{7-9} \cmidrule{8-9} \cmidrule{9-9} 
GNN$+$GNN &  & $0.4120$ & $0.4584$ & $0.9971$ & $0.5087$ & $2.549$ & $0.0828$ & $1.556$\tabularnewline
\cmidrule{1-1} \cmidrule{3-9} \cmidrule{4-9} \cmidrule{5-9} \cmidrule{6-9} \cmidrule{7-9} \cmidrule{8-9} \cmidrule{9-9} 
GAT$+$GNN &  & $0.2942$ & $0.3968$ & $2.069$ & $0.3879$ & $2.481$ & $0.0645$ & $1.792$\tabularnewline
\cmidrule{1-1} \cmidrule{3-9} \cmidrule{4-9} \cmidrule{5-9} \cmidrule{6-9} \cmidrule{7-9} \cmidrule{8-9} \cmidrule{9-9} 
 &  & \multicolumn{7}{c}{Complete Graph $(K_{6})$}\tabularnewline
\cmidrule{1-1} \cmidrule{3-9} \cmidrule{4-9} \cmidrule{5-9} \cmidrule{6-9} \cmidrule{7-9} \cmidrule{8-9} \cmidrule{9-9} 
NN Architecture &  & $e_{{\rm RMS}}$ & $\dot{e}_{{\rm RMS}}$ & $\tilde{\Phi}_{1,{\rm RMS}}[0:10]$ & $\tilde{\Phi}_{1,{\rm RMS}}[10:60]$ & $\tilde{\Phi}_{2,{\rm RMS}}[0:10]$ & $\tilde{\Phi}_{2,{\rm RMS}}[10:60]$ & $u_{{\rm RMS}}$\tabularnewline
\cmidrule{1-1} \cmidrule{3-9} \cmidrule{4-9} \cmidrule{5-9} \cmidrule{6-9} \cmidrule{7-9} \cmidrule{8-9} \cmidrule{9-9} 
DNN$+$DNN &  & $0.4267$ & $0.3922$ & $0.8241$ & $0.5378$ & $0.8817$ & $0.0793$ & $1.131$\tabularnewline
\cmidrule{1-1} \cmidrule{3-9} \cmidrule{4-9} \cmidrule{5-9} \cmidrule{6-9} \cmidrule{7-9} \cmidrule{8-9} \cmidrule{9-9} 
GNN$+$GNN &  & $0.2680$ & $0.3105$ & $2.070$ & $0.3364$ & $2.220$ & $0.0494$ & $1.430$\tabularnewline
\cmidrule{1-1} \cmidrule{3-9} \cmidrule{4-9} \cmidrule{5-9} \cmidrule{6-9} \cmidrule{7-9} \cmidrule{8-9} \cmidrule{9-9} 
GAT$+$GNN &  & $0.2607$ & $0.3161$ & $2.434$ & $0.3281$ & $2.181$ & $0.0614$ & $1.482$\tabularnewline
\cmidrule{1-1} \cmidrule{3-9} \cmidrule{4-9} \cmidrule{5-9} \cmidrule{6-9} \cmidrule{7-9} \cmidrule{8-9} \cmidrule{9-9} 
 &  & \multicolumn{7}{c}{Acyclic Graph}\tabularnewline
\cmidrule{1-1} \cmidrule{3-9} \cmidrule{4-9} \cmidrule{5-9} \cmidrule{6-9} \cmidrule{7-9} \cmidrule{8-9} \cmidrule{9-9} 
NN Architecture &  & $e_{{\rm RMS}}$ & $\dot{e}_{{\rm RMS}}$ & $\tilde{\Phi}_{1,{\rm RMS}}[0:10]$ & $\tilde{\Phi}_{1,{\rm RMS}}[10:60]$ & $\tilde{\Phi}_{2,{\rm RMS}}[0:10]$ & $\tilde{\Phi}_{2,{\rm RMS}}[10:60]$ & $u_{{\rm RMS}}$\tabularnewline
\cmidrule{1-1} \cmidrule{3-9} \cmidrule{4-9} \cmidrule{5-9} \cmidrule{6-9} \cmidrule{7-9} \cmidrule{8-9} \cmidrule{9-9} 
DNN$+$DNN &  & $0.4836$ & $0.4394$ & $0.8240$ & $0.5436$ & $0.4391$ & $0.0812$ & $1.102$\tabularnewline
\cmidrule{1-1} \cmidrule{3-9} \cmidrule{4-9} \cmidrule{5-9} \cmidrule{6-9} \cmidrule{7-9} \cmidrule{8-9} \cmidrule{9-9} 
GNN$+$GNN &  & $0.3952$ & $0.4250$ & $0.9185$ & $0.4919$ & $2.174$ & $0.0834$ & $1.411$\tabularnewline
\cmidrule{1-1} \cmidrule{3-9} \cmidrule{4-9} \cmidrule{5-9} \cmidrule{6-9} \cmidrule{7-9} \cmidrule{8-9} \cmidrule{9-9} 
GAT$+$GNN &  & $0.2912$ & $0.3900$ & $2.256$ & $0.4110$ & $2.242$ & $0.0662$ & $1.568$\tabularnewline
\cmidrule{1-1} \cmidrule{3-9} \cmidrule{4-9} \cmidrule{5-9} \cmidrule{6-9} \cmidrule{7-9} \cmidrule{8-9} \cmidrule{9-9} 
\end{tabular}
\end{table*}
To demonstrate the performance of the developed GNN and GAT control
designs, simulations were performed with a network of $N=6$ agents
tasked with tracking a moving target. Each simulation was performed
for a duration of $60$ seconds. An all-layer first partial derivative
of a DNN with respect to its weights as developed in \cite{Patil.Le.ea2022}
for DNNs with smooth activation functions was used as a baseline for
the adaptive update law in (\ref{eq:update-law-1}). The DNN architectures
listed in Table \ref{tab:compare-results} correspond to $\Phi_{1}$
and $\Phi_{2}$ for each simulation, e.g., the $\text{GAT}+\text{GNN}$
configuration indicates that $\Phi_{1}$ uses the GAT architecture
and $\Phi_{2}$ uses the GNN architecture. Let $\tilde{\Phi}_{1}\triangleq\hat{\Phi}_{1}-F(Q_{0})$
denote the function approximation error of the DNN $\Phi_{1}$ and
let $\tilde{\Phi}_{2}\triangleq\hat{\Phi}_{2}-H(R)$ denote the function
approximation error of the DNN $\Phi_{2}$. The quantities listed
in Table \ref{tab:compare-results} are as follows: $e_{{\rm RMS}}$
denotes the average RMS position tracking between agents and target,
$\dot{e}_{{\rm RMS}}$ denotes the average RMS velocity tracking between
agents and target, $\tilde{\Phi}_{1,{\rm RMS}}[X:Y]$ denotes the
average RMS function approximation error of $\Phi_{1}$ from $X$
seconds to $Y$ seconds, $\tilde{\Phi}_{2,{\rm RMS}}[X:Y]$ denotes
the average RMS function approximation error of $\Phi_{2}$ from $X$
seconds to $Y$ seconds, and $u_{{\rm RMS}}$ denotes the average
RMS control effort of the agents. 

The target has drift dynamics
\[
\begin{bmatrix}\ddot{x}_{0}\\
\ddot{y}_{0}\\
\ddot{z}_{0}
\end{bmatrix}=\begin{bmatrix}\cos\left(\dot{x}_{0}\right)-\sin\left(\dot{y}_{0}\right)+\cos\left(2\dot{z}_{0}\right)\\
\dot{x}_{0}-\dot{y}_{0}+\dot{z}_{0}+\frac{y_{0}}{\sqrt{1+\lvert y_{0}\rvert}}\\
\sin\left(\dot{y}_{0}\right)-\dot{x}_{0}\dot{z}_{0}
\end{bmatrix},
\]
where $q_{0}\triangleq[x_{0},y_{0},z_{0}]^{\top}$ denotes the target's
spatial coordinates in $\mathbb{R}^{3}$ and $\dot{q}_{0}\triangleq[x_{0},y_{0},z_{0}]^{\top}$
denotes the target's velocity in $\mathbb{R}^{3}$. The dynamics of
the agents are given by
\begin{figure}
\begin{centering}
\centering{}\includegraphics[width=1\columnwidth]{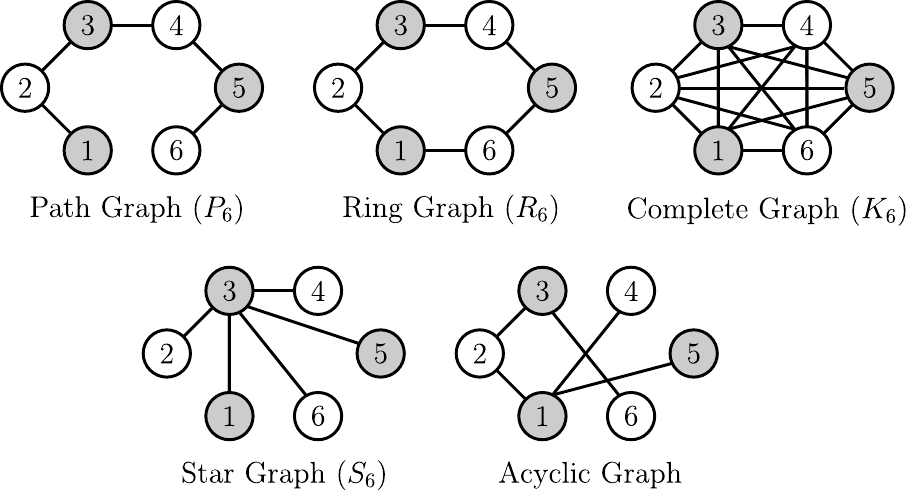}
\par\end{centering}
\caption{\label{fig:graph-topology}Visualization of communication topologies
examined in Table \ref{tab:compare-results}. Shaded agents are connected
to the target, where $b_{i}=1$.}
\end{figure}
\[
\begin{bmatrix}\ddot{x}_{i}\\
\ddot{y}_{i}\\
\ddot{z}_{i}
\end{bmatrix}=\begin{bmatrix}\sum_{j\in\mathcal{N}_{i}}\frac{1}{20,000\left(y_{i}-y_{j}\right)^{2}}\\
\sum_{j\in\mathcal{N}_{i}}\left(\dot{z}_{i}-\dot{z}_{j}\right)\cos\left(\dot{x}_{i}\right)\\
\sum_{j\in\mathcal{N}_{i}}\cos\left(\dot{z}_{i}\dot{z}_{j}\right)\frac{\dot{x}_{i}-\dot{x}_{j}}{\sqrt{1+\lvert\dot{x}_{i}-\dot{x}_{j}\rvert}}
\end{bmatrix}+u_{i},
\]
where $q_{i}\triangleq[x_{i},y_{i},z_{i}]^{\top}$ denotes the agent's
spatial coordinates in $\mathbb{R}^{3}$ and $\dot{q}_{i}\triangleq[\dot{x}_{i},\dot{y}_{i},\dot{z}_{i}]^{\top}$
denotes the agent's velocity in $\mathbb{R}^{3}$. The initial conditions
of the agents were selected to form an equilateral $N$-gon around
the origin with radius of $10$ $m$. The initial position of the
target was selected such that $q_{0}(t_{0})=[-3,2,10]^{\top}$ $m$,
and the initial velocity of the target was selected such that $\dot{q}_{0}(t_{0})=[-1,0,-2]^{\top}$
$m/s$. The following communication topologies were considered: path,
ring, star, complete, and acyclic. In the star configuration, all
agents are connected to a pinned agent where $b_{i}=1$. For all communication
topologies, half of the agents are connected to the target such that
$b_{i}=1$. 
\begin{figure}
\begin{centering}
\centering{}\includegraphics[width=0.9\columnwidth]{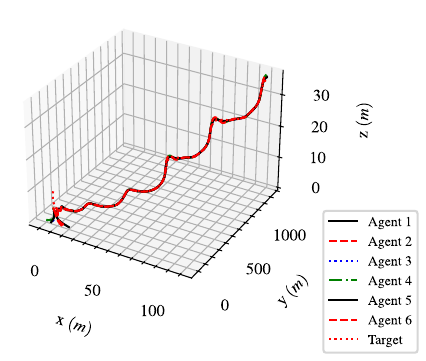} 
\par\end{centering}
\caption{\label{fig:3d-path}Trajectory visualization for the $\text{GNN}+\text{GNN}$
configuration and acyclic graph communication topology.}

\end{figure}

The GAT was used only for the approximation of the unknown target
dynamics because the updates from agents not connected to the target
are less informative than the agents who are connected to the target.
The use of attention helps to discern between these updates. However,
the updates from all neighbors should be weighted equally to approximate
agent interaction dynamics, motivating the use of the standard GNN
rather than the GAT to approximate the unknown interaction dynamics.

In all simulations, each agent was initialized using the same set
of weights. For all $i\in V$ and $j\in\{0,\ldots,k\}$, the weights
of the DNN and GNN were initialized from the distribution $\w ij\sim\mathcal{N}(0,0.03)$
and the weights of the GAT were initialized from the distribution
$\w ij\sim\mathcal{N}(0,0.3)$, where $\w ij\dims{(\d{j-1}+1)\times\d j}$.
For all $i\in V$ and $j\in\{0,\ldots,k-1\}$, the attention weights
of the $j^{\text{th}}$ layer of the GAT were initialized from the
distribution $a_{i}^{(j)}\sim\mathcal{N}(0,0.3)$, where $a_{i}^{(j)}\dims{1\times2\d j}$.
These distributions were empirically found to yield the best results
for each architecture. The $\tanh(\cdot)$ activation function was
used for all hidden layers in all architectures. 
\begin{figure}
\begin{centering}
\centering{}\includegraphics[width=0.9\columnwidth]{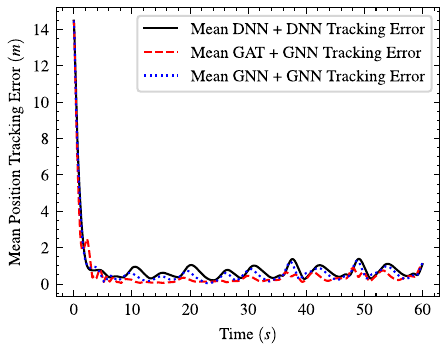}
\par\end{centering}
\caption{\label{fig:position-tracking-error}Mean position tracking error response
for each NN architecture with the acyclic graph communication topology.}
\end{figure}

Gain values were empirically selected and held constant for all NN
configurations. For all $i\in V$, the chosen gains are as follows:
$\alpha_{1}=0.85,\alpha_{2}=2.45,k_{1}=6.5,k_{2}=3.85,k_{3}=0.01,k_{4}=0.08$,
and $\Gamma_{1}=\Gamma_{2}=0.875\cdot I_{p\times p}$. For every agent,
the DNN architecture has $6$ layers and $24$ neurons at each layer
while the GNN and GAT architectures have $2$ layers and $24$ neurons
at each layer. The number of GNN layers must be carefully selected
for decentralized applications. Due to the message-passing framework
of the GNN, the number of required successive communications between
neighbors is equivalent to the number of message-passing layers in
the network. In the multi-agent control literature, the number of
GNN layers is typically limited to between $1$ and $4$ layers \cite{Li.Gama.ea2020,Gama.Tolstaya.ea2021,Zhou.Sharma.ea2022}. 

\begin{figure}
\begin{centering}
\centering{}\includegraphics[width=0.9\columnwidth]{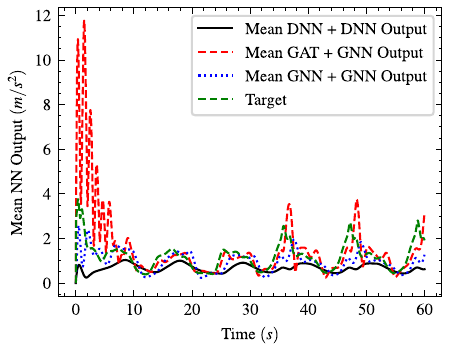}
\par\end{centering}
\caption{\label{fig:function-approx-error}Mean NN output with the acyclic
graph communication topology.}
\end{figure}

A three-dimensional visualization of the agents and target for the
GNN architecture and the acyclic graph communication topology is given
in Fig. \ref{fig:3d-path}. As seen in Fig. \ref{fig:position-tracking-error},
which shows the mean position tracking error performance of all NN
configurations, the transient performance in the first $3$ seconds
is the same for all NN architectures because the robust state-feedback
terms are predominant in the control input, and the NNs have just
started to learn the target dynamics. However, after $3$ seconds,
the differences in precision of the target tracking are more prevalent.
As evidenced in Fig. \ref{fig:function-approx-error}, which shows
the mean function approximation capabilities of all NN configurations
with respect to the mean drift dynamics of the target agent, the $\text{GNN}+\text{GNN}$
and $\text{GAT}+\text{GNN}$ configurations are able to achieve more
accurate estimates of the target dynamics than the $\text{DNN}+\text{DNN}$
configuration. 

The initial spike in the function approximation error in Fig. \ref{fig:function-approx-error}
is a consequence of the message-passing framework of the GNN architecture.
When the network is initialized, the estimate of the target dynamics
is inaccurate for all agents. When these estimates are exchanged among
nodes in the forward pass, the initial function approximation error
is amplified. The function approximation overshoot is quickly corrected
as the GNN learns. The same transient performance is not seen for
the $\text{DNN}+\text{DNN}$ configuration. The lack of compensation
for the unknown target and interaction dynamics by the DNN leads to
the oscillatory tracking error performance seen in Fig. \ref{fig:position-tracking-error}. 

 The performance improvement offered by the GNN and GAT architectures
is most apparent in communication topologies that are less connected.
In the path graph communication topology, the $\text{GNN}+\text{GNN}$
and $\text{GAT}+\text{GNN}$ configurations yielded a $20.8\%$ and
$48.1\%$ improvement over the baseline RMS position tracking error
performance of the $\text{DNN}+\text{DNN}$ configuration, respectively.
The improvement in RMS position tracking error performance can be
attributed to the GNN and GAT's ability to leverage information from
neighboring nodes during the forward pass. Incorporating neighborhood
messages in each GNN update allows the network to accurately learn
the target dynamics even when few agents are connected to the target
and not all agents can communicate. The developed GNN and GAT-based
controllers are impactful for implementation in real-world distributed
systems, where achieving fully connected communication topologies
can be challenging due to onboard hardware constraints. 

\section{Conclusion\label{sec:conclusion}}

In this work, a new Lyapunov-based GNN adaptive controller and observer
are developed to address the second-order target tracking problem
with unknown target dynamics and unknown inter-agent interaction dynamics.
Real-time training methods for the GNN and GAT architectures are introduced.
Additionally, a new distributed weight adaptation law is developed
to ensure exponential convergence of each agent's GNN weights to a
neighborhood of the ideal values. A Lyapunov-based stability analysis
is conducted to guarantee exponential convergence of the target state
estimates and agent states to a neighborhood of the target state.
Numerical simulations validate the theoretical findings, revealing
that the GNN and GAT architectures achieve $20.8\%$ and $48.1\%$
improvements in position tracking error performance, respectively,
compared to a baseline DNN architecture. These results were demonstrated
using a network of 6 agents with a path communication topology, indicating
that GNN and GAT architectures offer particular advantages for practical
distributed systems where complete network connectivity is often constrained
by hardware limitations. Future applications of the GNN architecture
include distributed state estimation with partial feedback for a target
with unknown dynamics as well as consensus and formation control for
multi-agent systems with unknown dynamics. 

\appendix{}

\begin{table*}
\caption{\label{tab:jacobian-terms}Definitions of terms used in computing
the first partial derivatives of the GNN and GAT architectures with
respect to their weight vectors.}

\centering{}%
\begin{tabular}{cc}
\toprule 
Term Definition & Term Dimensions\tabularnewline
\midrule
\midrule 
{\scriptsize{}$\pi_{i}^{(\ell)}\triangleq\dd{\layer{\ell}{x_{i}^{(\ell)}}}{x_{i}^{(\ell)}}$} & {\scriptsize{}$(d^{(\ell)}+1)\times d^{(\ell)}$}\tabularnewline
\midrule 
{\scriptsize{}$\xi_{i}^{(\ell)}\triangleq I_{\d{\ell}}\otimes\phi_{i}^{(\ell-1)\top}$} & {\scriptsize{}$\d{\ell}\times\d{\ell}(\d{\ell-1}+1)$}\tabularnewline
\midrule 
{\scriptsize{}$\iota_{i}^{(\ell)}\triangleq I_{\d{\ell}}\otimes\left(\bar{A}_{i}\boldsymbol{\phi}^{(\ell-1)\top}\right)$} & {\scriptsize{}$\d{\ell}\times\d{\ell}(\d{\ell-1}+1)$}\tabularnewline
\midrule 
{\scriptsize{}$\Delta_{i}^{(\ell)}\triangleq\pi_{i}^{(\ell)}\wt i{\ell}\tau_{i}^{(\ell)}$} & {\scriptsize{}$(\d{\ell}+1)\times N(\d{\ell-1}+1)$}\tabularnewline
\midrule 
{\scriptsize{}$\gamma_{i,m}^{(\ell)}\triangleq\exp(c_{i,m}^{(\ell)})$} & {\scriptsize{}$1\times1$}\tabularnewline
\midrule 
{\scriptsize{}$\epsilon_{i}^{(\ell)}\triangleq\exp(\boldsymbol{c}_{i}^{(\ell)})\bar{A}_{i}^{\top}$} & {\scriptsize{}$1\times1$}\tabularnewline
\midrule 
{\scriptsize{}$\eta_{i}^{(\ell)}\triangleq\frac{1}{\left(\exp\left(\boldsymbol{c}_{i}^{(\ell)}\right)\bar{A}_{i}^{\top}\right)^{2}}$} & {\scriptsize{}$1\times1$}\tabularnewline
\midrule 
{\scriptsize{}$\tau_{i}^{(\ell)}\triangleq\bar{A}_{i}\otimes I_{(\d{\ell-1}+1)}$} & {\scriptsize{}$(\d{\ell-1}+1)\times N(\d{\ell-1}+1)$}\tabularnewline
\midrule 
{\scriptsize{}$\varphi_{\mj{\ell+1}}^{(\ell)}\triangleq\begin{cases}
\Delta_{\mj{\ell+1}}^{(\ell)}[(\varphi_{\mj{\ell}}^{(\ell-1)})^{\top}]_{\bmj{\ell}}^{\top}, & \ell=k-1,\ldots,j+1\\
\delta_{i,\mj{\ell+1}}\pi_{\mj{\ell+1}}^{(\ell)}\iota_{\mj{\ell+1}}^{(\ell)}, & \ell=j.
\end{cases}$} & {\scriptsize{}$(\d{\ell}+1)\times\d j(\d{j-1}+1)$}\tabularnewline
\midrule 
{\scriptsize{}$\Lambda_{\mj{\ell+1}}^{(\ell)}\triangleq\begin{cases}
\Delta_{\mj{\ell+1}}^{(\ell)}\left[\left(\Lambda_{\mj{\ell}}^{(\ell-1)}\beta_{\mj{\ell+1},\mj{\ell}}^{(\ell)}+\phi_{\mj{\ell}}^{(\ell-1)}\Upsilon_{\mj{\ell+1},\mj{\ell}}^{(\ell)}\right)^{\top}\right]_{\bmj{\ell}}^{\top} & \begin{gathered}\ell=k-1,\\
\ldots,j+1
\end{gathered}
\\
\begin{gathered}\pi_{\mj{\ell+1}}^{(\ell)}\delta_{i,\mj{\ell+1}}\iota_{\mj{\ell+1}}^{(\ell)}+\pi_{\mj{\ell+1}}^{(\ell)}\wt{\mj{\ell+1}}{\ell}\tau_{\mj{\ell+1}}^{(\ell)}\left[\left(\phi_{\mj{\ell}}^{(\ell-1)}\Upsilon_{\mj{\ell+1},\mj{\ell}}^{(\ell)}\right)^{\top}\right]_{\bmj{\ell}}^{\top}\end{gathered}
 & \ell=j
\end{cases}$} & {\scriptsize{}$(\d{\ell}+1)\times\d j(\d{j-1}+1)$}\tabularnewline
\midrule 
{\scriptsize{}$\begin{gathered}\Upsilon_{\mj{\ell+1},\mj{\ell}}^{(\ell)}\triangleq\eta_{\mj{\ell+1}}^{(\ell)}\left(\epsilon_{\mj{\ell+1}}^{(\ell)}\varkappa_{\mj{\ell+1},\mj{\ell}}^{(\ell)}-\gamma_{\mj{\ell+1},\mj{\ell}}^{(\ell)}\bar{A}_{\mj{\ell+1}}\left[\left(\varkappa_{\mj{\ell+1},\mj{\ell}}^{(\ell)}\right)^{\top}\right]_{\bmj{\ell}}^{\top}\right)\end{gathered}
$} & {\scriptsize{}$1\times\d j(\d{j-1}+1)$}\tabularnewline
\midrule 
{\scriptsize{}$\varkappa_{\mj{\ell+1},\mj{\ell}}^{(\ell)}\triangleq\begin{cases}
\gamma_{\mj{\ell+1},\mj{\ell}}^{(\ell)}a_{\mj{\ell+1}}^{(\ell)}\left(\left(\wt{\mj{\ell+1}}{\ell}\Lambda_{\mj{\ell+1}}^{(\ell-1)}\right)\oplus\left(\wt{\mj{\ell+1}}{\ell}\Lambda_{\mj{\ell}}^{(\ell-1)}\right)\right), & \begin{gathered}\ell=k-1,\\
\ldots,j+1
\end{gathered}
\\
\gamma_{\mj{\ell+1},\mj{\ell}}^{(\ell)}a_{\mj{\ell+1}}^{(\ell)\top}\delta_{i,\mj{\ell+1}}\left(\left(\xi_{\mj{\ell+1}}^{(\ell)}\right)\oplus\left(\xi_{\mj{\ell}}^{(\ell)}\right)\right), & \ell=j
\end{cases}$} & {\scriptsize{}$1\times\d j(\d{j-1}+1)$}\tabularnewline
\midrule 
{\scriptsize{}$\varsigma_{\mj{\ell+1}}^{(\ell)}\triangleq\begin{cases}
\Delta_{\mj{\ell+1}}^{(\ell)}\left[\left(\varsigma_{\mj{\ell}}^{(\ell-1)}\beta_{\mj{\ell+1},\mj{\ell}}^{(\ell)}+\phi_{\mj{\ell}}^{(\ell-1)}\vartheta_{\mj{\ell+1},\mj{\ell}}^{(\ell)}\right)^{\top}\right]_{\bmj{\ell}}^{\top}, & \begin{gathered}\begin{gathered}\ell=k-1,\\
\ldots,j+1
\end{gathered}
\end{gathered}
\\
\pi_{\mj{\ell+1}}^{(\ell)}\wt{\mj{\ell+1}}{\ell}\tau_{\mj{\ell+1}}^{(\ell)}\left[\left(\phi_{\mj{\ell}}^{(\ell-1)}\vartheta_{\mj{\ell+1},\mj{\ell}}^{(\ell)}\right)^{\top}\right]_{\bmj{\ell}}^{\top}, & \ell=j
\end{cases}$} & {\scriptsize{}$(\d{\ell}+1)\times2\d j$}\tabularnewline
\midrule 
{\scriptsize{}$\vartheta_{\mj{\ell+1},\mj{\ell}}^{(\ell)}\triangleq\eta_{\mj{\ell+1}}^{(\ell)}\left(\epsilon_{\mj{\ell+1}}^{(\ell)}\varrho_{\mj{\ell+1},\mj{\ell}}^{(\ell)}-\gamma_{\mj{\ell+1},\mj{\ell}}^{(\ell)}\bar{A}_{\mj{\ell+1}}\left[\left(\varrho_{\mj{\ell+1},\mj{\ell}}^{(\ell)}\right)^{\top}\right]_{\bmj{\ell}}^{\top}\right)$} & {\scriptsize{}$1\times2\d j$}\tabularnewline
\midrule 
{\scriptsize{}$\varrho_{\mj{\ell+1},\mj{\ell}}^{(\ell)}\triangleq\begin{cases}
\gamma_{\mj{\ell+1},\mj{\ell}}^{(\ell)}a_{\mj{\ell+1}}^{(\ell)}\left(\left(\wt{\mj{\ell+1}}{\ell}\varsigma_{\mj{\ell+1}}^{(\ell-1)}\right)\oplus\left(\wt{\mj{\ell+1}}{\ell}\varsigma_{\mj{\ell}}^{(\ell-1)}\right)\right), & \begin{gathered}\ell=k-1,\\
\ldots,j+1
\end{gathered}
\\
\gamma_{\mj{\ell+1},\mj{\ell}}^{(\ell)}\delta_{i,\mj{\ell+1}}\left(\left(\wt{\mj{\ell+1}}{\ell}\phi_{\mj{\ell+1}}^{(\ell-1)}\right)\oplus\left(\wt{\mj{\ell+1}}{\ell}\phi_{\mj{\ell}}^{(\ell-1)}\right)\right)^{\top}, & \ell=j
\end{cases}$} & {\scriptsize{}$1\times2\d j$}\tabularnewline
\bottomrule
\end{tabular}
\end{table*}

\begin{IEEEproof}[Proof of Lemma \ref{lem:deep-jacobian}]
The recursive model of the GNN architecture in (\ref{eq:deep-gnn-architecture})
is given as
\begin{equation}
\phi_{i}=\wt ik\sigma^{(k-1)}\left(\wt i{k-1}\left[\phi_{\mj{k-1}}^{(k-2)}\right]_{\bmj{k-1}}\bar{A}_{i}^{\top}\right).\label{eq:deep-gnn-recursive}
\end{equation}
We begin by taking the partial derivative of (\ref{eq:deep-gnn-recursive})
with respect to $\vec{\w ij}$, which yields 
\begin{equation}
\begin{gathered}\dd{\phi_{i}}{\vec{\w ij}}=\wt ik\pi_{i}^{(k-1)}\wt i{k-1}\\
\cdot\tau_{i}^{(k-1)}\left[\left(\dd{}{\vec{\w ij}}\phi_{\mj{k-1}}^{(k-2)}\right)^{\top}\right]_{\bmj{k-1}}^{\top}.
\end{gathered}
\label{eq:deep-deriv-struct}
\end{equation}
Next, we evaluate the partial derivative of $\phi_{\mj{k-1}}^{(k-2)}$
with respect to $\vec{\w ij}$. This operation yields 
\begin{equation}
\begin{gathered}\dd{\phi_{\mj{k-1}}^{(k-2)}}{\vec{\w ij}}=\pi_{\mj{k-1}}^{(k-2)}\wt{\mj{k-1}}{k-2}\\
\cdot\tau_{\mj{k-1}}^{(k-2)}\left[\left(\dd{}{\vec{\w ij}}\phi_{\mj{k-2}}^{(k-3)}\right)^{\top}\right]_{\bmj{k-2}}^{\top}.
\end{gathered}
\label{eq:deep-deriv-struct-1}
\end{equation}
Equation (\ref{eq:deep-deriv-struct-1}) has the same derivative structure
seen in (\ref{eq:deep-deriv-struct}), but with a deeper layer index.
We continue executing the partial derivatives of $\phi_{\mj{\ell}}^{(\ell-1)}$
from $\ell=k-1,k-2,\ldots,j+1$ until $\ell=j$, where 
\[
\dd{\phi_{\mj{j+1}}^{(j)}}{\vec{\w ij}}=\pi_{\mj{j+1}}^{(j)}\dd{}{\vec{\w ij}}\iota_{\mj{j+1}}^{(j)}\vec{\w{\mj{j+1}}j}.
\]
We note that $\dd{}{\vec{\w ij}}\vec{\w{\mj{j+1}}j}=\boldsymbol{0}_{\d j\times(\d{j-1}+1)}$
if $\mj{j+1}\neq i$. Therefore,
\begin{equation}
\dd{\phi_{\mj{j+1}}^{(j)}}{\vec{\w ij}}=\pi_{\mj{j+1}}^{(j)}\delta_{i,\mj{j+1}}\iota_{\mj{j+1}}^{(j)}.\label{eq:deep-deriv-struct-2}
\end{equation}
Next, we leverage a recursive term to capture the nested structure
of the partial derivative of $\phi_{i}$ with respect to $\vec{\w ij}$.
Recall the definitions of $\Delta_{i}^{(j)}\dims{(\d j+1)\times N(\d{j-1}+1)}$
and $\varphi_{\mj{\ell+1}}^{(\ell)}\dims{(\d{\ell}+1)\times\d{\ell}(\d{\ell-1}+1)}$
from Table \ref{tab:jacobian-terms}. We note that $\varphi_{\mj{\ell+1}}^{(\ell)}$
for $\ell=k-1,\ldots,j+1$ captures the structure of the partial derivative
in (\ref{eq:deep-deriv-struct}) and (\ref{eq:deep-deriv-struct-1}),
while $\varphi_{\mj{\ell+1}}^{(\ell)}$ for $\ell=j$ captures the
root partial derivative in (\ref{eq:deep-deriv-struct-2}). Therefore,
the partial derivative of $\phi_{i}$ with respect to $\vec{\w ij}$
is equal to
\[
\dd{\phi_{i}}{\vec{\w ij}}=\wt ik\varphi_{i}^{(k-1)},
\]
where $\dd{\phi_{i}}{\vec{\w ij}}\dims{\d k\times\d j(\d{in}+1)}$.
Then, the first partial derivative of the GNN architecture in (\ref{eq:deep-gnn-architecture})
with respect to (\ref{eq:deep-gnn-weights}) is equal to (\ref{eq:deep-gnn-jacobian}).
\end{IEEEproof}
\begin{IEEEproof}[Proof of Lemma \ref{lem:deep-gat-jacobian}]
We begin by taking the partial derivative of the outer layer of the
GAT architecture in (\ref{eq:deep-gat-architecture}) with respect
to $\vec{\w ij}$ as
\begin{equation}
\begin{gathered}\dd{\phi_{i}}{\vec{\w ij}}=\wt ik\Delta_{i}^{(k-1)}\\
\cdot\left[\left(\dd{}{\vec{\w ij}}\phi_{\mj{k-1}}^{(k-2)}\beta_{i,\mj{k-1}}^{(k-1)}\right)^{\top}\right]_{\bmj{k-1}}^{\top}.
\end{gathered}
\label{eq:apply-product-rule}
\end{equation}
Note that both $\phi_{\mj{k-1}}^{(k-2)}$ and $\beta_{i,\mj{k-1}}^{(k-1)}$
are functions of $\w ij$. Therefore, we apply the product rule to
(\ref{eq:apply-product-rule}) which gives{\small{}
\begin{equation}
\begin{gathered}\dd{\phi_{i}}{\vec{\w ij}}=\wt ik\Delta_{i}^{(k-1)}\left[\left(\dd{\phi_{\mj{k-1}}^{(k-2)}}{\vec{\w ij}}\beta_{i,\mj{k-1}}^{(k-1)}\right.\right.\\
\left.\left.+\phi_{\mj{k-1}}^{(k-2)}\dd{\beta_{i,\mj{k-1}}^{(k-1)}}{\vec{\w ij}}\right)^{\top}\right]_{\bmj{k-1}}^{\top}.
\end{gathered}
\label{eq:deep-product-rule}
\end{equation}
}Evaluating the partial derivative of $\phi_{\mj{k-1}}^{(k-2)}$ with
respect to $\vec{\w ij}$ yields
\[
\begin{gathered}\Lambda_{m^{(k-1)}}^{(k-2)}\triangleq\dd{\phi_{\mj{k-1}}^{(k-2)}}{\vec{\w ij}}=\Delta_{\mj{k-1}}^{(k-2)}\\
\cdot\left[\left(\dd{}{\vec{\w ij}}\phi_{\mj{k-2}}^{(k-3)}\beta_{\mj{k-1},\mj{k-2}}^{(k-2)}\right)^{\top}\right]_{\bmj{k-2}}^{\top}.
\end{gathered}
\]
Evaluating the partial derivative of $\beta_{i,\mj{k-2}}^{(k-1)}$
with respect to $\vec{\w ij}$ yields
\begin{equation}
\dd{\beta_{i,\mj{k-1}}^{(k-1)}}{\vec{\w ij}}=\dd{}{\vec{\w ij}}\frac{\exp\left(c_{i,\mj{k-1}}^{(k-1)}\right)}{\exp\left(\boldsymbol{c}_{i}^{(k-1)}\right)\bar{A}_{i}^{\top}}.\label{eq:apply-quotient-rule}
\end{equation}
We apply the quotient rule to (\ref{eq:apply-quotient-rule}) such
that 
\begin{equation}
\begin{gathered}\dd{\beta_{i,\mj{k-1}}^{(k-1)}}{\vec{\w ij}}=\eta_{i}^{(k-1)}\left[\epsilon_{i}^{(k-1)}\dd{\exp\left(c_{i,\mj{k-1}}^{(k-1)}\right)}{\vec{\w ij}}\right.\\
\left.-\gamma_{i,\mj{k-1}}^{(k-1)}\dd{\left(\bar{A}_{i}\vec{\boldsymbol{c}_{i}^{(k-1)}}^{\top}\right)}{\vec{\w ij}}\right].
\end{gathered}
\label{eq:deep-quotient-rule-sub}
\end{equation}

Next, we briefly prove (\ref{eq:concat-prop}). For two vectors $A(x),B(x)\dims n$
and input $x\dims m$, let their concatenation be denoted by the vector
$C(x)\dims{2n}$. Then, the partial derivative of $C(x)$ is applied
elementwise. This operation yields
\begin{equation}
\begin{gathered}\dd{}xC(x)=\left[\dd{}xA_{1}(x)^{\top},\ldots,\dd{}xA_{n}(x)^{\top},\right.\\
\left.\dd{}xB_{1}(x)^{\top},\ldots,\dd{}xB_{n}(x)^{\top}\right]^{\top}
\end{gathered}
\label{eq:concat-interim}
\end{equation}
where $\dd{}xC(x)\dims{n\times m}$. Equation (\ref{eq:concat-interim})
is equivalent to
\begin{equation}
\dd{}xC(x)=\begin{bmatrix}\dd{}xA(x)^{\top}, & \dd{}xB(x)^{\top}\end{bmatrix}^{\top}.\label{eq:concatenation}
\end{equation}
Equation (\ref{eq:concatenation}) is the concatenation of the derivative
of $A(x)$ with respect to $x$ and the derivative of $B(x)$ with
respect to $x$. Therefore, 
\[
\dd{}x\left(A(x)\oplus B(x)\right)=\dd{}xC(x)=\left(\dd{}xA(x)\oplus\dd{}xB(x)\right).
\]

Next, we evaluate the partial derivative of $\exp(c_{i,\mj{k-1}}^{(k-1)})$
with respect to $\vec{\w ij}$ and apply (\ref{eq:concat-prop}),
which gives 
\[
\begin{gathered}\dd{\exp\left(b_{i,\mj{k-1}}^{(k-1)}\right)}{\vec{\w ij}}=\gamma_{i,\mj{k-1}}^{(k-1)}a_{i}^{(k-1)\top}\\
\cdot\left(\left(\wt i{k-1}\dd{\phi_{i}^{(k-2)}}{\vec{\w ij}}\right)\oplus\left(\wt i{k-1}\dd{\phi_{\mj{k-1}}^{(k-2)}}{\vec{\w ij}}\right)\right).
\end{gathered}
\]
Then, evaluating the partial derivative of $\bar{A}_{i}\vec{\boldsymbol{c}_{i}^{(k-1)}}^{\top}$with
respect to $\vec{\w ij}$ gives
\[
\begin{gathered}\dd{\left(\bar{A}_{i}\vec{\boldsymbol{c}_{i}^{(k-1)}}^{\top}\right)}{\vec{\w ij}}=\bar{A}_{i}\left(\gamma_{i,\mj{k-1}}^{(k-1)}a_{i}^{(k-1)\top}\left(\wt i{k-1}\right.\right.\\
\left.\left.\dd{\phi_{i}^{(k-2)}}{\vec{\w ij}}\oplus\wt i{k-1}\dd{\phi_{\mj{k-1}}^{(k-2)}}{\vec{\w ij}}\right)\right)_{\bmj{k-1}}^{\top}.
\end{gathered}
\]
For brevity of notation, let $\varkappa_{\mj{\ell+1},\mj{\ell}}^{(\ell)}\dims{1\times\d j(\d{j-1}+1)}$
be defined as
\[
\begin{gathered}\varkappa_{\mj{\ell+1},\mj{\ell}}^{(\ell)}\triangleq\gamma_{\mj{\ell+1},\mj{\ell}}^{(\ell)}a_{\mj{\ell+1}}^{(\ell)}\\
\cdot\left(\left(\wt{\mj{\ell+1}}{\ell}\dd{\phi_{\mj{\ell+1}}^{(\ell-1)}}{\vec{\w ij}}\right)\oplus\left(\wt{\mj{\ell+1}}{\ell}\dd{\phi_{\mj{\ell}}^{(\ell-1)}}{\vec{\w ij}}\right)\right).
\end{gathered}
\]
Then, the partial derivative of $\exp(c_{i,\mj{k-1}}^{(k-1)})$ with
respect to $\vec{\w ij}$ becomes
\begin{equation}
\dd{\exp\left(c_{i,\mj{k-1}}^{(k-1)}\right)}{\vec{\w ij}}=\varkappa_{i,\mj{k-1}}^{(k-1)},\label{eq:deep-quotient-rule-sub-1}
\end{equation}
and the partial derivative of $\bar{A}_{i}\vec{\boldsymbol{c}_{i}^{(k-1)}}^{\top}$with
respect to $\vec{\w ij}$ becomes
\begin{equation}
\dd{\left(\bar{A}_{i}\vec{\boldsymbol{c}_{i}^{(k-1)}}^{\top}\right)}{\vec{\w ij}}=\bar{A}_{i}\left[\varkappa_{i,\mj{k-1}}^{(k-1)}\right]_{\bmj{k-1}}^{\top}.\label{eq:deep-quotient-rule-sub-2}
\end{equation}
Then, substituting (\ref{eq:deep-quotient-rule-sub-1}) and (\ref{eq:deep-quotient-rule-sub-2})
into (\ref{eq:deep-quotient-rule-sub}) gives
\begin{equation}
\begin{gathered}\dd{\beta_{i,\mj{k-1}}^{(k-1)}}{\vec{\w ij}}=\eta_{i}^{(k-1)}\left(\epsilon_{i}^{(k-1)}\varkappa_{i,\mj{k-1}}^{(k-1)}\right.\\
\left.-\gamma_{i,\mj{k-1}}^{(k-1)}\bar{A}_{i}\left[\varkappa_{i,\mj{k-1}}^{(k-1)}\right]_{\bmj{k-1}}^{\top}\right).
\end{gathered}
\label{eq:deep-quotient-rule}
\end{equation}
We continue to analyze the partial derivatives of $\phi_{\mj{\ell+1}}^{(\ell)}$
with respect to $\vec{\w ij}$ from $\ell=k-1,k-2,\ldots,j+1$ until
$\ell=j$. Then, the partial derivative of $\phi_{\mj{\ell+1}}^{(\ell)}$
with respect to $\vec{\w ij}$ is {\small{}
\[
\begin{gathered}\dd{\phi_{\mj{j+1}}^{(j)}}{\vec{\w ij}}=\pi_{\mj{j+1}}^{(j)}\dd{}{\vec{\w ij}}\cdot\wt{\mj{j+1}}j\boldsymbol{B}_{\mj{j+1}}^{(j)}\bar{A}_{\mj{j+1}}^{\top}.\end{gathered}
\]
}We must apply the chain rule because $\boldsymbol{B}_{\mj{j+1},\mj j}^{(j)}$
is also a function of $\w ij$. Therefore,{\small{}
\begin{equation}
\begin{gathered}\dd{\phi_{\mj{j+1}}^{(j)}}{\vec{\w ij}}=\pi_{\mj{j+1}}^{(j)}\left(\dd{\left(\wt{\mj{j+1}}j\boldsymbol{B}_{\mj{j+1}}^{(j)}\bar{A}_{\mj{j+1}}^{\top}\right)}{\vec{\w ij}}\right.\\
\left.+\wt{\mj{j+1}}j\dd{\left(\boldsymbol{B}_{\mj{j+1}}^{(j)}\bar{A}_{\mj{j+1}}^{\top}\right)}{\vec{\w ij}}\right).
\end{gathered}
\label{eq:deep-product-rule-1}
\end{equation}
}Evaluating the partial derivative of $\wt{\mj{j+1}}j\boldsymbol{B}_{\mj{j+1}}^{(j)}\bar{A}_{\mj{j+1}}^{\top}$
with respect to $\vec{\w ij}$ yields 
\[
\dd{\left(\wt{\mj{j+1}}j\boldsymbol{B}_{\mj{j+1}}^{(j)}\bar{A}_{\mj{j+1}}^{\top}\right)}{\vec{\w ij}}=\delta_{i,\mj{j+1}}\iota_{\mj{j+1}}^{(j)}.
\]
and the partial derivative of $\boldsymbol{B}_{\mj{j+1}}^{(j)}\bar{A}_{\mj{j+1}}^{\top}$
with respect to $\vec{\w ij}$ is
\[
\begin{gathered}\dd{\left(\boldsymbol{B}_{\mj{j+1}}^{(j)}\bar{A^{\top}}_{\mj{j+1}}\right)}{\vec{\w ij}}=\tau_{\mj{j+1}}^{(j)}\\
\cdot\left[\left(\phi_{\mj j}^{(j-1)}\dd{\beta_{\mj{j+1},\mj j}^{(j)}}{\vec{\w ij}}\right)^{\top}\right]_{\bmj j}^{\top}.
\end{gathered}
\]
We evaluate the partial derivative of $\beta_{\mj{j+1},\mj j}^{(j)}$
with respect to $\vec{\w ij}$ yielding
\begin{equation}
\begin{gathered}\Upsilon_{m^{(j+1)},m^{(j)}}^{(j)}\triangleq\dd{\beta_{\mj{j+1},\mj j}^{(j)}}{\vec{\w ij}}=\eta_{\mj{j+1}}^{(j)}\epsilon_{\mj{j+1}}^{(j)}\\
\cdot\gamma_{\mj{j+1},\mj j}^{(j)}a_{\mj{j+1}}^{(j)\top}\delta_{i,\mj{j+1}}\left(\xi_{\mj{j+1}}^{(j)}\oplus\xi_{\mj j}^{(j)}\right)\\
-\eta_{\mj{j+1}}^{(j)}\gamma_{\mj{j+1},\mj j}^{(j)}\bar{A}_{\mj{j+1}}\\
\cdot\left[\left(\gamma_{\mj{j+1},\mj j}^{(j)}a_{\mj{j+1}}^{(j)\top}\delta_{i,\mj{j+1}}\right.\right.\\
\left.\left.\cdot\left(\xi_{\mj{j+1}}^{(j)}\oplus\xi_{\mj j}^{(j)}\right)\right)^{\top}\right]_{\boldsymbol{m}^{(j+1)}}^{\top}.
\end{gathered}
\label{eq:deep-quotient-rule-sub-3}
\end{equation}
The following terms capture the recursive nature of the partial derivative
of $\phi_{i}$ with respect to $\vec{\w ij}$. Recall the definitions
of $\Lambda_{\mj{\ell+1}}^{(\ell)}\dims{(\d{\ell}+1)\times\d j(\d{j-1}+1)}$,
$\Upsilon\dims{1\times\d j(\d{j-1}+1)}$, and $\varkappa_{\mj{\ell+1},\mj{\ell}}^{(\ell)}\dims{1\times\d j(\d{j-1}+1)}$
from Table \ref{tab:jacobian-terms}. The term $\Lambda_{\mj{\ell+1}}^{(\ell)}$
contains the structure of the partial derivative of $\phi_{\mj{\ell+1}}^{(\ell)}$
with respect to $\vec{\w ij}$ seen in (\ref{eq:deep-product-rule})
and (\ref{eq:deep-product-rule-1}). The term $\Upsilon_{\mj{\ell+1},\mj{\ell}}^{(\ell)}$
contains the structure of the partial derivative of $\beta_{\mj{\ell+1},\mj{\ell}}^{(\ell)}$
with respect to $\vec{\w ij}$ seen in (\ref{eq:deep-quotient-rule}).
Finally, the term $\varkappa_{\mj{\ell+1},\mj{\ell}}^{(\ell)}$ contains
the structure of the partial derivative of $\exp$$(d_{\mj{\ell+1},\mj{\ell}}^{(\ell)})$
with respect to $\vec{\w ij}$ seen in (\ref{eq:deep-quotient-rule-sub-1})
and (\ref{eq:deep-quotient-rule-sub-3}). Then, the partial derivative
of $\phi_{i}$ with respect to $\vec{\w ij}$ can be written as
\[
\dd{\phi_{i}}{\vec{\w ij}}=\wt ik\Lambda_{i}^{(k-1)},
\]
where $\dd{\phi_{i}}{\vec{\w ij}}\dims{\d k\times\d j\left(\d{j-1}+1\right)}$.
We repeat this process by taking the partial derivative of the outer
layer of the GAT architecture in (\ref{eq:deep-gat-architecture})
with respect to $\vec{\w ij}$ as 
\begin{equation}
\begin{gathered}\dd{\phi_{i}}{a_{i}^{(j)}}=\wt ik\Delta_{i}^{(k-1)}\\
\cdot\left[\left(\dd{}{a_{i}^{(j)}}\phi_{\mj{k-1}}^{(k-2)}\beta_{i,\mj{k-1}}^{(k-1)}\right)^{\top}\right]_{\bmj{k-1}}^{\top}
\end{gathered}
\label{eq:apply-deep-product-rule}
\end{equation}
Both $\phi_{\mj{k-1}}^{(k-2)}$ and $\beta_{i,\mj{k-1}}^{(k-1)}$
are functions of $a_{i}^{(j)}$. Therefore, we apply the product rule
to (\ref{eq:apply-deep-product-rule}) as
\begin{equation}
\begin{gathered}\dd{\phi_{i}}{a_{i}^{(j)}}=\wt ik\Delta_{i}^{(k-1)}\left[\left(\dd{\phi_{\mj{k-1}}^{(k-2)}}{a_{i}^{(j)}}\beta_{i,\mj{k-1}}^{(k-1)}\right.\right.\\
\left.\left.+\phi_{\mj{k-1}}^{(k-2)}\dd{\beta_{i,\mj{k-1}}^{(k-1)}}{a_{i}^{(j)}}\right)^{\top}\right]_{\bmj{k-1}}^{\top}.
\end{gathered}
\label{eq:deep-product-rule-3}
\end{equation}
Evaluating the partial derivative of $\phi_{\mj{k-1}}^{(k-2)}$ with
respect to $a_{i}^{(j)}$ yields 
\[
\begin{gathered}\varsigma_{m^{(k-1)}}^{(k-2)}\triangleq\dd{\phi_{\mj{k-1}}^{(k-2)}}{a_{i}^{(j)}}=\Delta_{\mj{k-1}}^{(k-2)}\left[\left(\dd{\phi_{\mj{k-2}}^{(k-3)}}{a_{i}^{(j)}}\beta_{\mj{k-1},\mj{k-2}}^{(k-2)}\right.\right.\\
\left.\left.+\phi_{\mj{k-2}}^{(k-3)}\dd{\beta_{\mj{k-1},\mj{k-2}}^{(k-2)}}{a_{i}^{(j)}}\right)^{\top}\right]_{\bmj{k-2}}^{\top}.
\end{gathered}
\]
Evaluating the partial derivative of $\beta_{i,\mj{k-1}}^{(k-1)}$
with respect to $a_{i}^{(j)}$ yields
\begin{equation}
\begin{gathered}\dd{\beta_{i,\mj{k-1}}^{(k-1)}}{a_{i}^{(j)}}=\dd{}{a_{i}^{(j)}}\frac{\exp\left(c_{i,\mj{l-1}}^{(k-1)}\right)}{\exp\left(\boldsymbol{c}_{i}^{(k-1)}\right)\bar{A}_{i}^{\top}}.\end{gathered}
\label{eq:apply-quotient-rule-2}
\end{equation}
We apply the quotient rule to (\ref{eq:apply-quotient-rule-2}) and
find
\begin{equation}
\begin{gathered}\dd{\beta_{i,\mj{k-1}}^{(k-1)}}{a_{i}^{(j)}}=\eta_{i}^{(k-1)}\left(\epsilon_{i}^{(k-1)}\dd{\exp\left(c_{i,\mj{k-1}}^{(k-1)}\right)}{a_{i}^{(j)}}\right.\\
\left.-\gamma_{i,\mj{k-1}}^{(k-1)}\dd{\left(\bar{A}_{i}\vec{\boldsymbol{c}_{i}^{(k-1)}}^{\top}\right)}{a_{i}^{(j)}}\right).
\end{gathered}
\label{eq:deep-quotient-rule-2}
\end{equation}
Evaluating the partial derivative of $b_{i,\mj{k-1}}^{(k-1)}$ with
respect to $a_{i}^{(j)}$ gives{\small{}
\begin{equation}
\begin{gathered}\begin{gathered}\dd{\exp\left(c_{i,\mj{k-1}}^{(k-1)}\right)}{a_{i}^{(j)}}=\gamma_{i,\mj{k-1}}^{(k-1)}a_{i}^{(k-1)}\\
\left(\left(\wt i{k-1}\dd{\phi_{i}^{(k-2)}}{a_{i}^{(j)}}\right)\oplus\left(\wt i{k-1}\dd{\phi_{\mj{k-1}}^{(k-2)}}{a_{i}^{(j)}}\right)\right).
\end{gathered}
\end{gathered}
\label{eq:deep-rho}
\end{equation}
}For compactness of notation, let $\varrho_{\mj{\ell+1},\mj{\ell}}^{(\ell)}\dims{1\times2\d j}$
be defined as
\[
\begin{gathered}\varrho_{\mj{\ell+1},\mj{\ell}}^{(\ell)}\triangleq\gamma_{\mj{\ell+1},\mj{\ell}}^{(\ell)}a_{\mj{\ell+1}}^{(\ell)}\left(\left(\wt{\mj{\ell+1}}{\ell}\dd{\phi_{\mj{\ell+1}}^{(\ell-1)}}{a_{i}^{(j)}}\right)\right.\\
\left.\oplus\left(\wt{\mj{\ell+1}}{\ell}\dd{\phi_{\mj{\ell}}^{(\ell-1)}}{a_{i}^{(j)}}\right)\right).
\end{gathered}
\]
Then, (\ref{eq:deep-quotient-rule-2}) becomes 
\[
\begin{gathered}\dd{\beta_{i,\mj{k-1}}^{(k-1)}}{a_{i}^{(j)}}=\eta_{i}^{(k-1)}\left(\epsilon_{i}^{(k-1)}\rho_{i,\mj{k-1}}^{(k-1)}\right.\\
\left.-\gamma_{i,\mj{k-1}}^{(k-1)}\bar{A}_{i}\left[\rho_{i,\mj{k-1}}^{(k-1)}\right]_{\bmj{k-1}}\right).
\end{gathered}
\]
We continue evaluating partial derivatives of $\phi_{\mj{\ell+1}}^{(\ell)}$
from $\ell=k-1,\ldots,j+1$ until $\ell=j$. Then, the partial derivative
of $\phi_{\mj{j+1}}^{(j)}$ with respect to $a_{i}^{(j)}$ is 
\begin{equation}
\begin{gathered}\dd{\phi_{\mj{j+1}}^{(j)}}{a_{i}^{(j)}}=\pi_{\mj{j+1}}^{(j)}\wt{\mj{j+1}}j\tau_{\mj{j+1}}^{(j)}\\
\cdot\left[\left(\phi_{\mj j}^{(j-1)}\dd{}{\vec{a_{i}^{(j)}}}\beta_{\mj{j+1},\mj j}^{(j)}\right)^{\top}\right]_{\bmj j}^{\top}.
\end{gathered}
\label{eq:deep-phi-deriv}
\end{equation}
Then, the partial derivative of $\beta_{\mj{\ell+1},\mj{\ell}}^{(\ell)}$
with respect to $a_{i}^{(j)}$ is
\begin{equation}
\begin{gathered}\vartheta_{m^{(j+1)},m^{(j)}}^{(j)}\triangleq\dd{\beta_{\mj{j+1},\mj j}^{(j)}}{a_{i}^{(j)}}=\\
\eta_{\mj{j+1}}^{(j)}\epsilon_{\mj{j+1}}^{(j)}\dd{\exp\left(c_{\mj{j+1},\mj j}^{(j)}\right)}{a_{i}^{(j)}}\\
-\eta_{\mj{j+1}}^{(j)}\gamma_{\mj{j+1},\mj j}^{(j)}\bar{A}_{\mj{j+1}}\dd{\exp\left(\boldsymbol{c}_{\mj{j+1}}^{(j)}\right)}{a_{i}^{(j)}}.
\end{gathered}
\label{eq:deep-beta-deriv}
\end{equation}
Evaluating the partial derivative of $\exp(c_{\mj{j+1},\mj j}^{(j)})$
with respect to $a_{i}^{(j)}$ yields
\begin{equation}
\begin{gathered}\varrho_{m^{(j+1)},m^{(j)}}^{(j)}\triangleq\dd{\exp\left(c_{\mj{j+1},\mj j}^{(j)}\right)}{a_{i}^{(j)}}=\gamma_{\mj{j+1},\mj j}^{(j)}\\
\cdot\delta_{i,\mj{j+1}}\left(\left(\wt{\mj{j+1}}j\phi_{\mj{j+1}}^{(j-1)}\right)\oplus\left(\wt{\mj j}j\phi_{\mj j}^{(j-1)}\right)\right)^{\top}.
\end{gathered}
\label{eq:deep-att-deriv}
\end{equation}
The following terms capture the recursive nature of the partial derivative
of $\phi_{i}$ with respect to $a_{i}^{(j)}$. Recall the definitions
of $\varsigma_{\mj{\ell+1}}^{(\ell)}\dims{(\d{\ell}+1)\times2\d j}$,
$\vartheta_{\mj{\ell+1},\mj{\ell}}^{(\ell)}\dims{1\times2\d j}$,
and $\varrho_{\mj{\ell+1},\mj{\ell}}^{(\ell)}\dims{1\times2\d j}$
from Table \ref{tab:jacobian-terms}. The term $\varsigma_{\mj{\ell+1}}^{(\ell)}$
contains the structure of the partial derivative of $\phi_{\mj{\ell+1}}^{(\ell)}$
with respect to $a_{i}^{(j)}$ seen in (\ref{eq:deep-product-rule-3})
and (\ref{eq:deep-phi-deriv}). The term $\vartheta_{\mj{\ell+1},\mj{\ell}}^{(\ell)}$
captures the structure of the partial derivative of $\beta_{\mj{\ell+1},\mj{\ell}}^{(\ell)}$
with respect to $a_{i}^{(j)}$ seen in (\ref{eq:deep-quotient-rule-2})
and (\ref{eq:deep-beta-deriv}). The term $\varrho_{\mj{\ell+1},\mj{\ell}}^{(\ell)}$
captures the partial derivative of $\exp(c_{\mj{\ell+1},\mj{\ell}}^{(\ell)})$
with respect to $a_{i}^{(j)}$ seen in (\ref{eq:deep-rho}) and (\ref{eq:deep-att-deriv}).
Then, the partial derivative of $\phi_{i}$ with respect to $a_{i}^{(j)}$
can be written as
\[
\dd{\phi_{i}}{a_{i}^{(j)}}=\wt ik\varsigma_{i}^{(k-1)},
\]
where $\dd{\phi_{i}}{a_{i}^{(j)}}\dims{\d k\times2\d j}$. Therefore,
the first partial derivative of (\ref{eq:deep-gat-architecture})
with respect to (\ref{eq:deep-gat-weights}) is equal to (\ref{eq:deep-gat-jacobian}).
\end{IEEEproof}
\begin{IEEEproof}[Proof of Lemma \ref{lem:lambda-bounds}]
We begin by finding a formula for $\text{\ensuremath{\underbar{\ensuremath{\lambda}}}}_{\mathcal{H}}$
in terms of $N$, where $\text{\ensuremath{\underbar{\ensuremath{\lambda}}}}_{\mathcal{H}}=\lambda_{\min}(\mathcal{H})$,
and $\mathcal{H}=(\mathcal{L}_{G}+\mathcal{B})\otimes I_{n}$. Let
$a(G)$ denote the algebraic connectivity of the graph $G$. Let $P_{N}$
denote the path graph with $N\in\mathbb{Z}_{>0}$ nodes. By \cite[Proposition 1.12]{Belhaiza.DeAbreu.ea2005},
we have that $a(P_{N})\leq a(G)$ for all connected $G$. Therefore,
$\lambda_{\min}(\mathcal{L}_{G}+\mathcal{B})$ is minimized when $\mathcal{L}_{G}=\mathcal{L}_{G}(P_{N})$. 

Let $\mathcal{B}(1,2,\ldots,M)\dims{N\times N}$ denote the matrix
such that $[b_{11}]=[b_{22}]=\ldots=[b_{{\rm MM}}]=\epsilon$, and
all other entries are $0$. We first prove that $\lambda_{\min}(\mathcal{L}_{G}(P_{N})+\mathcal{B}(i))\leq\lambda_{\min}(\mathcal{L}_{G}(P_{N})+\mathcal{B}(1,2,\ldots M))$
for any set $\mathcal{R}\triangleq1,2,\ldots M$ for $\lvert\mathcal{R}\rvert>1$
and any $i\in\mathcal{R}$. We employ the first-order perturbation
expansion for simple eigenvalues, which states that{\small{}
\begin{equation}
\begin{gathered}\lambda_{\min}\left(\mathcal{L}_{G}\left(P_{N}\right)+\mathcal{B}\left(1,2,\ldots,M\right)\right)=\lambda_{\min}\left(\mathcal{L}_{G}\left(P_{N}\right)\right)\\
+\frac{x^{\top}\mathcal{B}\left(1,2,\ldots,M\right)x}{x^{\top}x}+\mathcal{O}\left(\norm{\mathcal{B}\left(1,2,\ldots,M\right)}^{2}\right),
\end{gathered}
\label{eq:lambda-min}
\end{equation}
}where $x$ is a nonzero eigenvector such that $\mathcal{L}_{G}(P_{N})x=\lambda_{\min}(\mathcal{L}_{G}(P_{N}))x$
\cite{Nakatsukasa2017}. Then, let $\Delta_{\lambda}(\mathcal{B}\left(1,2,\ldots,M\right))\triangleq\lambda_{\min}(\mathcal{L}_{G}(P_{N})+\mathcal{B}\left(1,2,\ldots,M\right))-\lambda_{\min}(\mathcal{L}_{G}(P_{N}))$.
Then, (\ref{eq:lambda-min}) becomes
\begin{equation}
\begin{gathered}\Delta_{\lambda}\left(\mathcal{B}\left(1,2,\ldots,M\right)\right)=\frac{x^{\top}\mathcal{B}\left(1,2,\ldots,M\right)x}{x^{\top}x}\\
+\mathcal{O}\left(\norm{\mathcal{B}\left(1,2,\ldots,M\right)}^{2}\right).
\end{gathered}
\label{eq:delta-lambda}
\end{equation}
For a connected graph $G$, the eigenvector corresponding to $\lambda_{\min}(\mathcal{L}_{G})$
is $\mathbbm{1}_N^\top$. Then, (\ref{eq:delta-lambda}) simplifies
to
\[
\Delta_{\lambda}\left(\mathcal{B}\left(1,2,\ldots,M\right)\right)=\frac{1}{N}\sum_{m=0}^{\lvert\mathcal{R}\rvert}\epsilon+\mathcal{O}\left(\epsilon^{2}\right),
\]
where $\lvert\mathcal{R}\rvert$ is the cardinality of $\mathcal{R}$.
Therefore, $\Delta_{\lambda}(\mathcal{B}(i))=\epsilon+\mathcal{O}(\epsilon^{2})$
and $\Delta_{\lambda}(\mathcal{B}(1,2,\ldots,M))=\lvert\mathcal{R}\rvert\epsilon+\mathcal{O}(\epsilon^{2})$.
This shows that $\Delta_{\lambda}(\mathcal{B}(i))<\Delta_{\lambda}(\mathcal{B}(1,2,\ldots,M))$. 

Define $\mathcal{F}(i)\triangleq\mathcal{L}_{G}(P_{N})+\mathcal{B}(i)$
and let $\lambda_{\min}(\mathcal{F}(i)),\lambda_{2}(\mathcal{F}(i)),\ldots,\lambda_{n-1}(\mathcal{F}(i)),\lambda_{\max}(\mathcal{F}(i))$
denote the eigenvalues of $\mathcal{F}(i)$ indexed from smallest
to largest. We will prove that
\[
\lambda_{\min}\left(\mathcal{F}(1)\right)=\lambda_{\min}\left(\mathcal{F}(N)\right)\leq\lambda_{\min}\left(\mathcal{F}(k)\right),
\]
where $1<k<N$. Note that $\mathcal{F}_{(1)}$ and $\mathcal{F}_{(N)}$
are similar matrices that can be obtained from one another by using
a similarity transformation. The similarity transformation from $\mathcal{F}(1)$
to $\mathcal{F}(N)$ is $\mathcal{F}(N)=P\mathcal{F}(1)P^{-1}$, where
$P$ is a permutation matrix that reverses the order of the rows and
columns of $\mathcal{F}(1)$. We see that $\mathcal{F}(1)$ and $\mathcal{F}(N)$
are similar, which shows that they have the same eigenvalues. Therefore,
$\lambda_{\min}(\mathcal{F}(1))=\lambda_{\min}(\mathcal{F}(N))$. 

Next, we consider $\mathcal{F}(k)$ for $1<k<N$. We know that $\mathcal{F}(k)=\mathcal{L}_{G}(P_{N})+\mathcal{B}(k)$.
Additionally, for $\mathcal{F}(1)$ we have $\mathcal{L}_{G}(P_{N})=\mathcal{F}(1)-\mathcal{B}(1)$.
Then,  $\mathcal{F}(k)=\mathcal{F}(1)+\mathcal{B}(k)-\mathcal{B}(1)$.
Weyl's inequality states that for Hermitian matrices $A$ and $B$,
we have $\lambda_{\min}(A)+\lambda_{\max}(B)\leq\lambda_{\min}(A+B).$
Then,
\[
\lambda_{\min}\left(\mathcal{F}(1)\right)+\lambda_{\max}\left(\mathcal{B}(k)-\mathcal{B}(1)\right)\leq\lambda_{\min}\left(\mathcal{F}(k)\right).
\]
Note that $1<k<N$. Then, $\mathcal{B}(k)-\mathcal{B}(1)$ is a matrix
where $[b_{kk}]=1$, $[b_{11}]=-1$, and $[b_{xy}]=0$ for all $x,y\in[N]$
such that $x\neq y$. Therefore, $\lambda_{\max}(\mathcal{B}(k)-\mathcal{B}(1))=1$.
We see that $\lambda_{\min}(\mathcal{F}(1))+1\leq\lambda_{\min}(\mathcal{F}(k))$,
which yields $\lambda_{\min}(\mathcal{F}(1))\leq\lambda_{\min}(\mathcal{F}(k))$.
Therefore, $\mathcal{L}_{G}+\mathcal{B}$ has the smallest minimum
eigenvalue when $\mathcal{L}_{G}+\mathcal{B}=\mathcal{L}_{G}(P_{N})+\mathcal{B}(1)=\mathcal{L}_{G}(P_{N})+\mathcal{B}(N)$. 

Without loss of generality, consider $\mathcal{L}_{G}(P_{N})+\mathcal{B}(1)$.
Next, let \cite[Lemma 1]{Yueh2005} hold. We see that $\mathcal{L}_{G}(P_{N})+\mathcal{B}(1)$
takes the form of a special tridiagonal matrix in \cite{Yueh2005},
where $a=-1,b=2,c=-1,\alpha=0$, and $\beta=\sqrt{ac}=1$. Therefore,
by \cite[Theorem 1]{Yueh2005}, the $k^{\text{th}}$ largest eigenvalue
of $\mathcal{L}_{G}\left(P_{N}\right)+\mathcal{B}(1)$ is given as
\[
\lambda_{k}\triangleq b+2\sqrt{ac}\cos\left(\frac{2(N+1-k)\pi}{2N+1}\right),
\]
where $1\leq k\leq N$. The smallest eigenvalue of $\mathcal{L}_{G}\left(P_{N}\right)+\mathcal{B}(1)$
corresponds to $k=1$. Therefore,
\[
\lambda_{\min}\left(\mathcal{L}_{G}\left(P_{N}\right)+\mathcal{B}(1)\right)=2+2\cos\left(\frac{2N\pi}{2N+1}\right).
\]
Then, recall the definition of $\mathcal{H}$, where $\mathcal{H}=(\mathcal{L}_{G}+\mathcal{B})\otimes I_{n}$.
For two matrices $A\dims{m\times m}$ and $B\dims{n\times n},$ let
$\lambda$ be an eigenvalue of $A$ with corresponding eigenvector
$v$, and let $\mu$ be an eigenvalue of $B$ with corresponding eigenvector
$y$. Then, $\lambda\mu$ is an eigenvalue of $A\otimes B$ with corresponding
eigenvector $v\otimes y$, and any eigenvalue of $A\otimes B$ arises
as a product of eigenvalues of $A$ and $B$ \cite[Theorem 4.2.12]{Horn.Johnson1991}.
All eigenvalues of $I_{n}$ are equal to $1$. Therefore, $\lambda_{\min}(\mathcal{H})=\lambda_{\min}(\mathcal{L}_{G}+\mathcal{B})$,
and $\lambda_{\max}(\mathcal{H})=\lambda_{\max}(\mathcal{L}_{G}+\mathcal{B})$.
This shows that
\[
\text{\ensuremath{\underbar{\ensuremath{\lambda}}}}_{\mathcal{H}}=2+2\cos\left(\frac{2N\pi}{2N+1}\right).
\]

Next, we upper bound $\bar{\lambda}_{\mathcal{H}}$, where $\bar{\lambda}_{\mathcal{H}}=\lambda_{\max}(\mathcal{H})$.
The eigenvalues of $\mathcal{H}$ are maximized when $\mathcal{B}=I_{N}$.
The complete graph with $N$ nodes $K_{N}$ has the largest maximum
eigenvalue of any connected graph. For any square matrix $A$ with
eigenvalue $\lambda$ and corresponding eigenvector $v$, where $Av=\lambda v$,
$\lambda+1$ is an eigenvalue of $A+I$ with the corresponding eigenvector
$v$, where $I$ is the identity matrix. The largest eigenvalue of
the graph Laplacian of the complete graph is $\lambda_{\max}(\mathcal{L}_{G}(K_{N}))=N$
\cite[Lemma 2]{Anderson.Morley1985}. Therefore, $\lambda_{\max}(\mathcal{L}_{G}(K_{N})+I_{N})\leq N+1$.
Finally, we invoke \cite[Theorem 4.2.12]{Horn.Johnson1991} which
shows that $\lambda_{\max}(\mathcal{H})=\lambda_{\max}(\mathcal{L}_{G}+I_{N})$.
Therefore, $\bar{\lambda}_{\mathcal{H}}\leq N+1$.
\end{IEEEproof}
\begin{IEEEproof}[Proof of Lemma \ref{lem:lagrange-remainder-gnn}]
The $z^{\text{th}}$ component of the GNN's first Lagrange remainder
$T_{y,z}$ is expressed as $T_{y,z}=[T_{y,z}(s_{1}),\ldots,T_{y,z}(s_{n})]^{\top},$
where $T_{y,z}(s_{m})=\tilde{\theta}_{y,z}^{\top}\nabla_{\hat{\theta}_{y,z}}^{2}\phi_{y,i}(\kappa_{i},\kappa_{j:j\in\mathcal{N}_{i}^{k}},\hat{\theta}_{y,z}+s_{m}(\tilde{\theta}_{y,z})\tilde{\theta}_{y,z},\hat{\theta}_{y,j:j\in\mathcal{\bar{N}}_{i}^{k-1}\backslash z})\tilde{\theta}_{y,z}$
for all $m\in[n]$. The use of the projection operator in (\ref{eq:update-law-1})
and (\ref{eq:update-law-2}) ensure that $\hat{\theta}_{y,z}(t)\in\mathbb{f}_{y,z}$
for all $t\in\mathbb{R}_{\geq0}$, for $y=1,2$, and all $i\in V$,
where $\mathbb{f}_{y,z}\triangleq\{\theta\in\mathbb{R}^{p}:\lVert\theta\rVert\leq\bar{\theta}\}$.
Additionally, due to the facts that $\mathbb{f}_{y,z}$ is a convex
set, $s_{m}(\hat{\theta})\in[0,1]$ for all $m\in[n]$, and $\theta_{y,z}^{\ast},\hat{\theta}_{y,z}\in\mathbb{f}_{y,z}$,
the term $\overset{\circ}{\theta}_{y,z}\in\mathbb{f}_{y,z}$ for all
$z\in V$. Due to the fact that $s_{m}(\tilde{\theta}_{y,z})$ is
unknown, we conduct this analysis considering the worst-case bound
in which $\lVert\hat{\theta}_{y,z}+s_{m}(\tilde{\theta}_{y,z})\tilde{\theta}_{y,z}\rVert\leq\bar{\theta}$
for all $m\in[n]$. Thus, we consider arbitrary values of $\theta_{y,z}\in\mathbb{f}_{y,z}$
for all $z\in V$. Additionally, in this bound, we consider the complete
communication graph as it represents the maximal inter-agent aggregation
that can occur at each GNN layer.

Then, we consider $T_{y,z}(s_{m})=\tilde{\theta}_{y,z}^{\top}\nabla_{\theta_{y,z}}^{2}\phi_{y,i}(\kappa_{i},\kappa_{j:j\in\mathcal{N}_{i}^{k}},\theta_{y,j:j\in\mathcal{\bar{N}}_{i}^{k-1}})\tilde{\theta}_{y,z}$.
For simplicity of notation, we denote $\phi_{y,i}(\kappa_{i},\kappa_{j:j\in\mathcal{N}_{i}^{k}},\theta_{y,j:j\in\mathcal{\bar{N}}_{i}^{k-1}})$
as $\phi_{y,i}$. For $\kappa_{i}\in\mathbb{R}^{n}$, let $\kappa=[\kappa_{i}^{\top}]_{i\in V}^{\top}\dims{nN}$.
Then, the norm of $T_{y,z}$ can be upper-bounded as
\begin{equation}
\lVert T_{y,z}\rVert\leq\lVert\tilde{\theta}_{y,z}\rVert^{2}\norm{\nabla_{\theta_{y,z}}^{2}\phi_{y,i}}_{F},\label{eq:lagrange-bound-here}
\end{equation}
where $\lVert\cdot\rVert_{F}$ denotes the Frobenius norm. The tensor
of the second partial derivative of $\phi_{y,i}$ with respect to
$\text{vec}(\w zp)$ and $\text{vec}(\w zq)$ is composed of blocks
of the form $\frac{\partial^{2}\phi_{y,i}}{\partial\text{vec}(\w zp)\text{vec}(\w zq)}$
for $p,q\in[k]$. To bound the norm of the second partial derivatives
of $\phi_{y,i}$ with respect to $\text{vec}(\w zp)$ and $\text{vec}(\w zq)$,
we must first bound the norm of the first partial derivative of $\phi_{y,i}$
with respect to $\theta_{y,z}$ in terms of the norm of the ensemble
input $\kappa$. By the triangle inequality, 
\begin{equation}
\norm{\nabla_{\theta_{y,z}}\phi_{y,i}}_{F}\leq\sum_{q=0}^{k}\norm{\frac{\partial\phi_{y,i}}{\partial\text{vec}(\w zq)}}_{F}.\label{eq:jacobian-bound}
\end{equation}
The use of bounded activation function ensures that the GNN input
$\kappa$ appears in (\ref{eq:jacobian-bound}) only for the term
with $q=0$. This is a direct result of the formula for $\varphi_{m^{(\ell+1)}}^{(\ell)}$
in Table \ref{tab:deep-jacobians}, in which $\kappa$ only explicitly
appears for $j=0$. Otherwise, the base case for $\varphi_{m^{(\ell+1)}}^{(\ell)}$
when $\ell=j$ contains $\boldsymbol{\phi}^{(\ell-1)}$, whose Frobenius
norm is upper-bounded by $N(\sqrt{\d{\ell-1}}\bar{\sigma}^{(\ell-1)}+1)$.
Thus, we examine how the norm of the partial derivative of $\phi_{y,i}$
with respect to $\text{vec}(\w zq)$ scales according to the norm
of the ensemble GNN input $\kappa$ when $q=0$. 

Let $\boldsymbol{\rho}^{(w)}$ denote the set of $w^{\text{th}}$-order
strictly increasing polynomials with coefficients ${\tt c}_{w},{\tt c}_{w-1},\ldots,{\tt c}_{0}\in\mathbb{R}_{\geq0}$,
where at least one coefficient ${\tt c}_{i}$ is strictly positive
for $i\in\{1,\ldots,w\}$. We find that $\lVert\pi_{m^{(\ell+1)}}^{(\ell)}\rVert_{F}\leq\sqrt{\d{\ell}}\bar{\sigma}^{\prime(\ell)}$,
where $\bar{\sigma}^{\prime(\ell)}$ denotes the upper bound of the
derivative of the activation function at the $\ell^{\text{th}}$ layer.
Then, when $q=0$, $\lVert\varphi_{\mj 1}^{(0)}\rVert_{F}\in\boldsymbol{\rho}^{(1)}$
for all $m^{(1)}\in V$, where $\boldsymbol{\rho}^{(1)}$ is the set
of strictly increasing linear functions of the norm of the ensemble
GNN input. The term $\lVert\varphi_{\mj 1}^{(0)}\rVert_{F}$ is upper-bounded
by a first-order polynomial of the GNN input which has leading coefficient
upper-bounded by $d^{(0)}\sqrt{N}\bar{\sigma}^{\prime(0)}$

We find that $\lVert\w z{\ell}\rVert_{F}\leq\bar{\theta}$ and $\lVert\Delta_{m^{(\ell)}}^{(\ell-1)}\rVert_{F}\leq\bar{\theta}\bar{\sigma}^{\prime(\ell)}\sqrt{N(\d{\ell})(\d{\ell-1}+1)}$
for all $\ell\in[k-1]$ and all $m^{(\ell)}\in V$. Then, $\lVert\varphi_{\mj 2}^{(1)}\rVert_{F}\in\boldsymbol{\rho}^{(1)}$
holds for all $m^{(2)}\in V$. We bound the term $\lVert\varphi_{\mj{\ell+1}}^{(\ell)}\rVert_{F}$
for each layer of the GNN, and find that $\lVert\varphi_{\mj{\ell+1}}^{(\ell)}\rVert_{F}\in\boldsymbol{\rho}^{(1)}$
holds for all $\ell\in[k-1]$ and all $m^{(\ell+1)}\in V$. For brevity
of notation, let ${\tt g}^{(\ell)}\triangleq\bar{\theta}^{\ell}N^{\frac{1}{2}+\frac{3\ell}{2}}(d^{(0)})^{\frac{1}{2}}\prod_{i=0}^{\ell}(\bar{\sigma}^{\prime(i)}(d^{(i)})^{\frac{1}{2}})\prod_{i=0}^{\ell-2}(d^{(i)}+1)^{\frac{1}{2}}$.
The term $\lVert\varphi_{\mj{\ell+1}}^{(\ell)}\rVert_{F}$ is upper-bounded
by a first-order polynomial of the GNN input which has leading coefficient
upper-bounded by ${\tt g}^{(\ell)}$. We reach the highest level of
the partial derivative of $\phi_{y,i}$ with respect to $\text{vec}(\w zq)$,
where 
\[
\norm{\frac{\partial\phi_{y,i}}{\partial\text{vec}(\w zq)}}_{F}\leq\norm{\w zk}_{F}\norm{\Delta_{m^{(k)}}^{(k-1)}}_{F}\sum_{\boldsymbol{m}^{(k-1)}}\norm{\varphi_{m^{(k-1)}}^{(k-2)}}_{F}.
\]
By the use of bounded activation functions with bounded first derivatives
and the use of a bounded search space in (\ref{eq:theta-star-1})
and (\ref{eq:theta-star-2}), we find that
\begin{equation}
\norm{\frac{\partial\phi_{y,i}}{\partial\text{vec}(\w zq)}}_{F}\in\boldsymbol{\rho}^{(1)}.\label{eq:gnn-gradient-bound}
\end{equation}
Then, the first-order polynomial of the GNN input which upper-bounds
(\ref{eq:gnn-gradient-bound}) has a leading coefficient upper-bounded
by $\bar{\theta}{\tt g}^{(k-1)}$. Next, using (\ref{eq:jacobian-bound})
and (\ref{eq:gnn-gradient-bound}) yields $\lVert\nabla_{\theta_{y,z}}\phi_{y,i}\rVert_{F}\in\boldsymbol{\rho}^{(1)}$.
Therefore, the norm of the first partial derivative of the GNN $\phi_{y,i}$
with respect to the weights $\theta_{y,z}$ is linear in the ensemble
input. 

We return to bounding the blocks of the tensor of the second partial
derivative of $\phi_{y,i}$ with respect to $\theta_{y,z}$, where
by the submultiplicity of the Frobenius norm and the triangle inequality,
it follows that
\begin{equation}
\norm{\nabla_{\theta_{y,z}}^{2}\phi_{y,i}}_{F}\leq\sum_{p=0}^{k}\sum_{q=0}^{k}\norm{\frac{\partial\phi_{y,i}}{\partial\text{vec}(\w zp)\text{vec}(\w zq)}}_{F}.\label{eq:hessian-bound}
\end{equation}
By the use of bounded activation functions with bounded first and
second derivatives, the GNN input only appears in (\ref{eq:hessian-bound})
when $p=0$ or $q=0$. Once again, this is a result of the formula
for $\varphi_{m^{(\ell+1)}}^{(\ell)}$ in Table \ref{tab:deep-jacobians},
in which $\kappa$ only explicitly appears for $j=0$. Thus, we will
consider the double derivative tensor block in which $p=q=0$. The
norm of all other blocks of the second partial derivative of $\phi_{y,i}$
with respect to $\theta_{y,z}$ will scale in terms of $\lVert\kappa\rVert$
at an order less than or equal to the case when $p=q=0$. By the use
of bounded activation functions with bounded first and second derivatives
and the use of a bounded search space in (\ref{eq:theta-star-1})
and (\ref{eq:theta-star-2}), we find that
\begin{equation}
\norm{\frac{\partial\pi_{m^{(\ell)}}^{(\ell-1)}}{\partial\text{vec}(\w z0)}}_{F}\in\boldsymbol{\rho}^{(1)},\label{eq:dpi-dw-bound}
\end{equation}
holds for all $\ell\in[k-1]$ and $m^{(\ell)}\in V$. Then, using
(\ref{eq:dpi-dw-bound}), we find that
\begin{equation}
\norm{\frac{\partial\varphi_{\mj 1}^{(0)}}{\partial\text{vec}(\w z0)}}_{F}\in\boldsymbol{\rho}^{(2)},\label{eq:dvartheta-dvecw-bound}
\end{equation}
holds for all $m^{(1)}\in V$. Next, we examine
\[
\norm{\frac{\partial\varphi_{\mj 2}^{(1)}}{\partial\text{vec}(\w z0)}}_{F}=\norm{\frac{\partial}{\partial\text{vec}(\w z0)}\Delta_{\mj 2}^{(1)}\left[\left(\varphi_{\mj 1}^{(0)}\right)^{\top}\right]_{\boldsymbol{m}^{(1)}}^{\top}}_{F}.
\]
By the use of bounded activation functions with bounded second derivatives,
(\ref{eq:dpi-dw-bound}), (\ref{eq:gnn-gradient-bound}), and the
use of a bounded search space in (\ref{eq:theta-star-1}) and (\ref{eq:theta-star-2}),
we find that
\begin{equation}
\norm{\frac{\partial\Delta_{\mj{\ell+1}}^{(\ell)}}{\partial\text{vec}(\w z0)}}_{F}\in\boldsymbol{\rho}^{(1)},\label{eq:first-gnn-delta-bound}
\end{equation}
holds for all $\ell\in[k-1]$ and $m^{(\ell+1)}\in V$. Then, we find
that
\begin{equation}
\norm{\frac{\partial\varphi_{\mj 2}^{(1)}}{\partial\text{vec}(\w z0)}}_{F}\in\boldsymbol{\rho}^{(2)},\label{eq:first-gnn-varphi-bound}
\end{equation}
holds for all $m^{(2)}\in V$. We repeat this analysis from (\ref{eq:first-gnn-delta-bound})
and (\ref{eq:first-gnn-varphi-bound}) for each layer of the GNN,
showing that 
\begin{equation}
\norm{\frac{\partial\varphi_{\mj{\ell+1}}^{(\ell)}}{\partial\text{vec}(\w z0)}}_{F}\in\boldsymbol{\rho}^{(2)},\label{eq:varphi-bound}
\end{equation}
holds for $\ell=0,1,\ldots,k-1$ and for all $m^{(\ell+1)}\in V$.
Then, we reach the outermost term in the second partial derivative
of the GNN with respect to its layer weights, where
\[
\begin{gathered}\norm{\frac{\partial^{2}\phi_{y,i}}{\partial\text{vec}(\w z0)\partial\text{vec}(\w z0)}}_{F}=\left\Vert \frac{\partial}{\partial\text{vec}(\w zp)}\w zk\Delta_{m^{(k)}}^{(k-1)}\right.\\
\left.\left[\left(\varphi_{m^{(k-1)}}^{(k-2)}\right)^{\top}\right]_{\boldsymbol{m}^{(k-1)}}^{\top}\right\Vert _{F}.
\end{gathered}
\]

Based on the use of bounded activation functions with bounded first
and second derivatives, (\ref{eq:first-gnn-delta-bound}), (\ref{eq:varphi-bound}),
and the use of a bounded search space in (\ref{eq:theta-star-1})
and (\ref{eq:theta-star-2}), we find that
\begin{equation}
\norm{\frac{\partial^{2}\phi_{y,i}}{\partial\text{vec}(\w z0)\partial\text{vec}(\w z0)}}_{F}\in\boldsymbol{\rho}^{(2)}.\label{eq:dphi-dweights-bound}
\end{equation}
Since the norm of all other blocks of the second partial derivative
of $\phi_{y,i}$ with respect to $\text{vec}(\w z0)$ and $\text{vec}(\w z0)$
will scale in terms of $\lVert\kappa\rVert$ at an order less than
or equal to the case when $p=q=0$, we use (\ref{eq:dphi-dweights-bound})
to conclude that
\[
\norm{\nabla_{\theta_{y,z}}^{2}\phi_{y,i}}_{F}\in\boldsymbol{\rho}^{(2)}.
\]
Additionally, since $\lVert T_{y,z}\rVert\leq\lVert\tilde{\theta}_{y,z}\rVert^{2}\lVert\nabla_{\theta_{y,z}}^{2}\phi_{y,i}\rVert_{F}$,
then
\[
\norm{T_{y,z}}\leq\rho(\lVert\kappa\rVert)\norm{\tilde{\theta}_{y,z}}^{2},
\]
where $\rho(\lVert\kappa\rVert)$ is a strictly increasing, quadratic
polynomial in the norm of the ensemble GNN input $\lVert\kappa\rVert$. 
\end{IEEEproof}
\begin{IEEEproof}[Proof of Lemma \ref{lem:lagrange-remainder-gat}]
This analysis is conducted considering the worst-case bound in which
$\lVert\hat{\theta}_{y,z}+s_{m}(\tilde{\theta}_{y,z})\tilde{\theta}_{y,z}\rVert\leq\bar{\theta}$
for all $m\in[n]$. Thus, we consider arbitrary values of $\theta_{y,z}\in\mathbb{f}_{y,z}$
for all $z\in V$. In this bound, we consider the complete communication
graph as it represents the maximal inter-agent aggregation that can
occur at each GAT layer. The norm of $T_{y,z}$ can be upper-bounded
as
\begin{equation}
\lVert T_{y,z}\rVert\leq\lVert\tilde{\theta}_{y,z}\rVert^{2}\norm{\nabla_{\theta_{y,z}}^{2}\phi_{y,i}}_{F}.\label{eq:lagrange-bound-here-1}
\end{equation}

The tensor of the second partial derivative of $\phi_{y,i}$ with
respect to $\theta_{y,z}$ is composed of blocks of the form $\frac{\partial^{2}\phi_{y,i}}{\partial\text{vec}(\w zp)\text{vec}(\w zq)}$
for $p,q\in[k]$, $\frac{\partial^{2}\phi_{y,i}}{\partial\text{vec}(a_{z}^{(p)})\text{vec}(\w zq)}$
for $p\in[k-1],q\in[k]$, $\frac{\partial^{2}\phi_{y,i}}{\partial\text{vec}(a_{z}^{(p)})\text{vec}(a_{z}^{(q)})}$
for $p,q\in[k-1]$, and $\frac{\partial^{2}\phi_{y,i}}{\partial\text{vec}(W_{z}^{(p)})\text{vec}(a_{z}^{(q)})}$
for $p\in[k]$ and $q\in[k-1]$. To bound the norm of the second partial
derivatives of $\phi_{y,i}$, we must bound the norm of the first
partial derivatives of $\phi_{y,i}$ with respect to its layer weights
and its attention weights, respectively. By the triangle inequality,
\begin{equation}
\begin{gathered}\norm{\nabla_{\theta_{y,z}}\phi_{y,i}}_{F}\leq\sum_{q=0}^{k}\norm{\frac{\partial\phi_{y,i}}{\partial\text{vec}(\w zq)}}_{F}\\
+\sum_{q=0}^{k-1}\norm{\frac{\partial\phi_{y,i}}{\partial\text{vec}(a_{z}^{(q)})}}_{F}.
\end{gathered}
\label{eq:gat-jacobian-sum}
\end{equation}

Once again, the input of the GNN $\kappa$ appears explicitly in (\ref{eq:gat-jacobian-sum})
only for the term with $q=0$, and all other terms will scale at an
order less than or equal to the case when $q=0$. Thus, we begin by
quantifying how the Frobenius norm of the first derivative of the
GAT with respect to its base layer weights scales according to the
norm of the ensemble GAT input. Based on the fact that $\lVert\beta_{x,y}^{(z)}\rVert_{F}\leq1$
for all $x,y\in V$ and all $z\in[k-1]$ and the use of a bounded
search space in (\ref{eq:theta-star-1}) and (\ref{eq:theta-star-2}),
we find that 
\begin{equation}
\norm{\frac{\partial\beta_{m^{(1)},m^{(0)}}^{(0)}}{\partial\text{vec}(\w z0)}}_{F}\in\boldsymbol{\rho}^{(1)},\label{eq:base-upsilon-bound}
\end{equation}
holds for all $m^{(0)},m^{(1)}\in V$. Then, by the use of activation
functions with bounded first derivatives, (\ref{eq:base-upsilon-bound}),
and the use of a bounded search space in (\ref{eq:theta-star-1})
and (\ref{eq:theta-star-2}),
\begin{equation}
\norm{\frac{\partial\phi_{m^{(1)}}^{(0)}}{\partial\text{vec}(\w z0)}}_{F}\in\boldsymbol{\rho}^{(2)},\label{eq:base-lambda-bound}
\end{equation}
holds for all $m^{(1)}\in V$. 

Next, we examine the Frobenius norm of the partial derivative of $\beta_{m^{(2)},m^{(0)}}^{(1)}$
with respect to $\text{vec}(\w z0)$. Based on the fact that $\lVert\beta_{x,y}^{(z)}\rVert_{F}\leq1$
for all $x,y\in V$ and all $z\in[k-1]$, (\ref{eq:base-lambda-bound}),
and the use of a bounded search space in (\ref{eq:theta-star-1})
and (\ref{eq:theta-star-2}), we find that 
\begin{equation}
\norm{\frac{\partial\beta_{m^{(2)},m^{(1)}}^{(1)}}{\partial\text{vec}(\w z0)}}_{F}\in\boldsymbol{\rho}^{(2)},\label{eq:first-upsilon-bound}
\end{equation}
holds for all $m^{(1)},m^{(2)}\in V$. Then, for the Frobenius norm
of the corresponding partial derivative of $\phi_{m^{(2)}}^{(1)}$
with respect to $\text{vec}(\w z0)$, we have 
\begin{equation}
\norm{\frac{\partial\phi_{m^{(2)}}^{(1)}}{\partial\text{vec}(\w z0)}}_{F}\in\boldsymbol{\rho}^{(3)},\label{eq:first-lambda-bound}
\end{equation}
which holds for all holds for all $m^{(2)}\in V$. We repeat the analysis
from (\ref{eq:first-upsilon-bound}) and (\ref{eq:first-lambda-bound})
for each layer of the GAT, finding that $\lVert\frac{\partial\beta_{m^{(\ell+1)},m^{(\ell)}}^{(\ell)}}{\partial\text{vec}(\w z0)}\rVert_{F}\in\boldsymbol{\rho}^{(\ell+1)}(\lVert\kappa\rVert)$
and $\lVert\frac{\partial\phi_{m^{(\ell+1)}}^{(\ell)}}{\partial\text{vec}(\w z0)}\rVert_{F}\in\boldsymbol{\rho}^{(\ell+2)}$
holds for $\ell=0,1,\ldots,k-1$ and all $m^{(\ell)},m^{(\ell+1)}\in V$.
Then, we reach the outermost term, where
\[
\begin{gathered}\norm{\frac{\partial\phi_{y,z}}{\partial\text{vec}(\w z0)}}_{F}=\left\Vert \frac{\partial}{\partial\text{vec}(\w z0)}\w zk\Delta_{\mj k}^{(k-1)}\right.\\
\left.\left[\left(\Lambda_{\mj{k-1}}^{(k-2)}\beta_{\mj k,\mj{k-1}}^{(k-1)}+\phi_{\mj{k-1}}^{(k-2)}\Upsilon_{\mj k,\mj{k-1}}^{(k-1)}\right)^{\top}\right]_{\boldsymbol{m}^{(k-1)}}^{\top}\right\Vert _{F}.
\end{gathered}
\]

Based on the use of bounded activation functions with bounded derivatives,
the fact that $\lVert\beta_{x,y}^{(z)}\rVert_{F}\leq1$ for all $x,y\in V$
and all $z\in[k-1]$, and the use of a bounded search space in (\ref{eq:theta-star-1})
and (\ref{eq:theta-star-2}), we find that 
\begin{equation}
\norm{\frac{\partial\phi_{y,z}}{\partial\text{vec}(\w z0)}}_{F}\in\boldsymbol{\rho}^{(k)},\label{eq:dgat-dweights-bound}
\end{equation}
A structurally identical argument can be conducted to show that $\lVert\frac{\partial\phi_{y,z}}{\partial\text{vec}(a_{z}^{(0)})}\rVert_{F}\in\boldsymbol{\rho}^{(k)}$and
thus, $\lVert\nabla_{\theta_{y,z}}\phi_{y,i}\rVert_{F}\in\boldsymbol{\rho}^{(k)}$.

We now return to the scaling analysis of the second partial derivative
of $\phi_{y,i}$ with respect to $\theta_{y,z}$. By the submultiplicity
of the Frobenius norm and the triangle inequality,
\[
\begin{gathered}\norm{\nabla_{\theta_{y,z}}^{2}\phi_{y,i}}_{F}\leq\sum_{p=0}^{k}\sum_{q=0}^{k}\norm{\frac{\partial^{2}\phi_{y,i}}{\partial\text{vec}(\w zp)\text{vec}(\w zq)}}_{F}\\
+\sum_{p=0}^{k-1}\sum_{q=0}^{k}\norm{\frac{\partial^{2}\phi_{y,i}}{\partial\text{vec}(a_{z}^{(p)})\text{vec}(\w zq)}}_{F}\\
+\sum_{p=0}^{k-1}\sum_{q=0}^{k-1}\norm{\frac{\partial^{2}\phi_{y,i}}{\partial\text{vec}(a_{z}^{(p)})\text{vec}(a_{z}^{(q)})}}_{F}\\
+\sum_{p=0}^{k}\sum_{q=0}^{k-1}\norm{\frac{\partial^{2}\phi_{y,i}}{\partial\text{vec}(W_{z}^{(p)})\text{vec}(a_{z}^{(q)})}}_{F}.
\end{gathered}
\]
We first compute how the norm of the second partial derivative of
$\phi_{y,i}$ with respect to $\text{vec}(\w zp)$ and $\text{vec}(\w zq)$
scales according to the norm of the ensemble GAT input $\lVert\kappa\rVert$.
Once again, we consider the case in which $p=q=0$, as all other cases
will scale at a rate less than or equal to that of $p=q=0$ with respect
to $\lVert\kappa\rVert$. Based on (\ref{eq:dgat-dweights-bound})
and the use of a bounded search space in (\ref{eq:theta-star-1})
and (\ref{eq:theta-star-2}), we find that 
\begin{equation}
\norm{\frac{\partial^{2}\beta_{m^{(1)},m^{(0)}}^{(0)}}{\partial\vec{\w z0}\partial\vec{\w z0}}}_{F}\in\boldsymbol{\rho}^{(k+1)},\label{eq:base-gat-upsilon-bound}
\end{equation}
holds for all $m^{(0)},m^{(1)}\in V$. Then, based on the use of bounded
activation functions with bounded first and second derivatives, (\ref{eq:dgat-dweights-bound}),
(\ref{eq:base-gat-upsilon-bound}), and the use of a bounded search
space in (\ref{eq:theta-star-1}) and (\ref{eq:theta-star-2}), 
\begin{equation}
\norm{\frac{\partial^{2}\phi_{m^{(1)}}^{(0)}}{\partial\text{vec}(\w z0)\partial\text{vec}(\w z0)}}_{F}\in\boldsymbol{\rho}^{(2k)},\label{eq:base-gat-lambda-bound}
\end{equation}
holds for all $m^{(1)}\in V$. 

Next, we examine the norm of the partial derivative of $\Upsilon_{m^{(2)},m^{(1)}}^{(1)}$
with respect to $\text{vec}(\w z0)$. Based on the use of bounded
activation functions with bounded first and second derivatives, (\ref{eq:dgat-dweights-bound}),
(\ref{eq:base-gat-lambda-bound}), and the use of a bounded search
space in (\ref{eq:theta-star-1}) and (\ref{eq:theta-star-2}), 
\begin{equation}
\norm{\frac{\partial^{2}\beta_{m^{(2)},m^{(1)}}^{(1)}}{\partial\text{vec}(\w z0)\partial\text{vec}(\w z0)}}_{F}\in\boldsymbol{\rho}^{(2k)},\label{eq:first-gat-upsilon-bound}
\end{equation}
and
\begin{equation}
\norm{\frac{\partial^{2}\phi_{m^{(2)}}^{(1)}}{\partial\text{vec}(\w z0)\partial\text{vec}(\w z0)}}_{F}\in\boldsymbol{\rho}^{(2k)},\label{eq:first-gat-lambda-bound}
\end{equation}
holds for all $m^{(1)},m^{(2)}\in V$. We repeat the analysis from
(\ref{eq:first-gat-upsilon-bound}) and (\ref{eq:first-gat-lambda-bound})
for each layer of the GAT, finding that
\[
\norm{\frac{\partial^{2}\beta_{m^{(\ell+1)},m^{(\ell)}}^{(\ell)}}{\partial\text{vec}(\w z0)\partial\text{vec}(\w z0)}}_{F}\in\boldsymbol{\rho}^{(2k)},
\]
 and
\[
\norm{\frac{\partial^{2}\phi_{m^{(\ell+1)}}^{(\ell)}}{\partial\text{vec}(\w z0)\partial\text{vec}(\w z0)}}_{F}\in\boldsymbol{\rho}^{(2k)}
\]
 holds for $\ell=0,1,\ldots,k-1$ and for all $m^{(\ell)},m^{(\ell+1)}\in V$.
Then, we reach the outermost term, where
\[
\begin{gathered}\norm{\frac{\partial^{2}\phi_{y,i}}{\partial\text{vec}(\w z0)\text{vec}(\w z0)}}_{F}=\left\Vert \frac{\partial}{\partial\text{vec}(\w z0)}\wt ik\Delta_{\mj k}^{(k-1)}\right.\\
\left.\left[\left(\Lambda_{\mj{k-1}}^{(k-2)}\beta_{\mj k,\mj{k-1}}^{(k-1)}+\phi_{\mj{k-1}}^{(k-2)}\Upsilon_{\mj k,\mj{k-1}}^{(k-1)}\right)^{\top}\right]_{\mathbf{m}^{(k-1)}}^{\top}\right\Vert _{F}.
\end{gathered}
\]
Based on the use of bounded activation functions with bounded first
and second derivatives, (\ref{eq:dgat-dweights-bound}), and the use
of a bounded search space in (\ref{eq:theta-star-1}) and (\ref{eq:theta-star-2}),
\[
\norm{\frac{\partial^{2}\phi_{y,i}}{\partial\text{vec}(\w z0)\text{vec}(\w z0)}}_{F}\in\boldsymbol{\rho}^{(2k)}.
\]

Structurally identical arguments can be carried out to prove that
$\norm{\frac{\partial^{2}\phi_{y,i}}{\partial\text{vec}(a_{z}^{(0)})\text{vec}(\w z0)}}_{F}\in\boldsymbol{\rho}^{(2k)},\norm{\frac{\partial^{2}\phi_{y,i}}{\partial\text{vec}(a_{z}^{(0)})\text{vec}(a_{z}^{(0)})}}_{F}\in\boldsymbol{\rho}^{(2k)},and\norm{\frac{\partial^{2}\phi_{y,i}}{\partial\text{vec}(\w z0)\text{vec}(a_{z}^{(0)})}}_{F}\in\boldsymbol{\rho}^{(2k)}$,
and thus,
\begin{equation}
\norm{\nabla_{\theta_{y,z}}\phi_{y,i}}_{F}\in\boldsymbol{\rho}^{(2k)}.\label{eq:gat-hessian-bound}
\end{equation}
Additionally, since $\lVert T_{y,z}\rVert\leq\lVert\tilde{\theta}_{y,z}\rVert^{2}\lVert\nabla_{\theta_{y,z}}^{2}\phi_{y,i}\rVert_{F}$,
then
\[
\norm{T_{y,z}}\leq\rho(\lVert\kappa\rVert)\norm{\tilde{\theta}_{y,z}}^{2},
\]
where $\rho(\lVert\kappa\rVert)$ is a strictly increasing, polynomial
of degree $2k$ in the norm of the ensemble GNN input $\lVert\kappa\rVert$,
where $k$ is the number of GAT layers. 
\end{IEEEproof}
\bibliographystyle{IEEEtran}
\bibliography{bib/encr,bib/lb-gnn,bib/main,bib/ncr}

\vspace{-10mm}
\begin{IEEEbiography}[{\includegraphics[width=1\columnwidth]{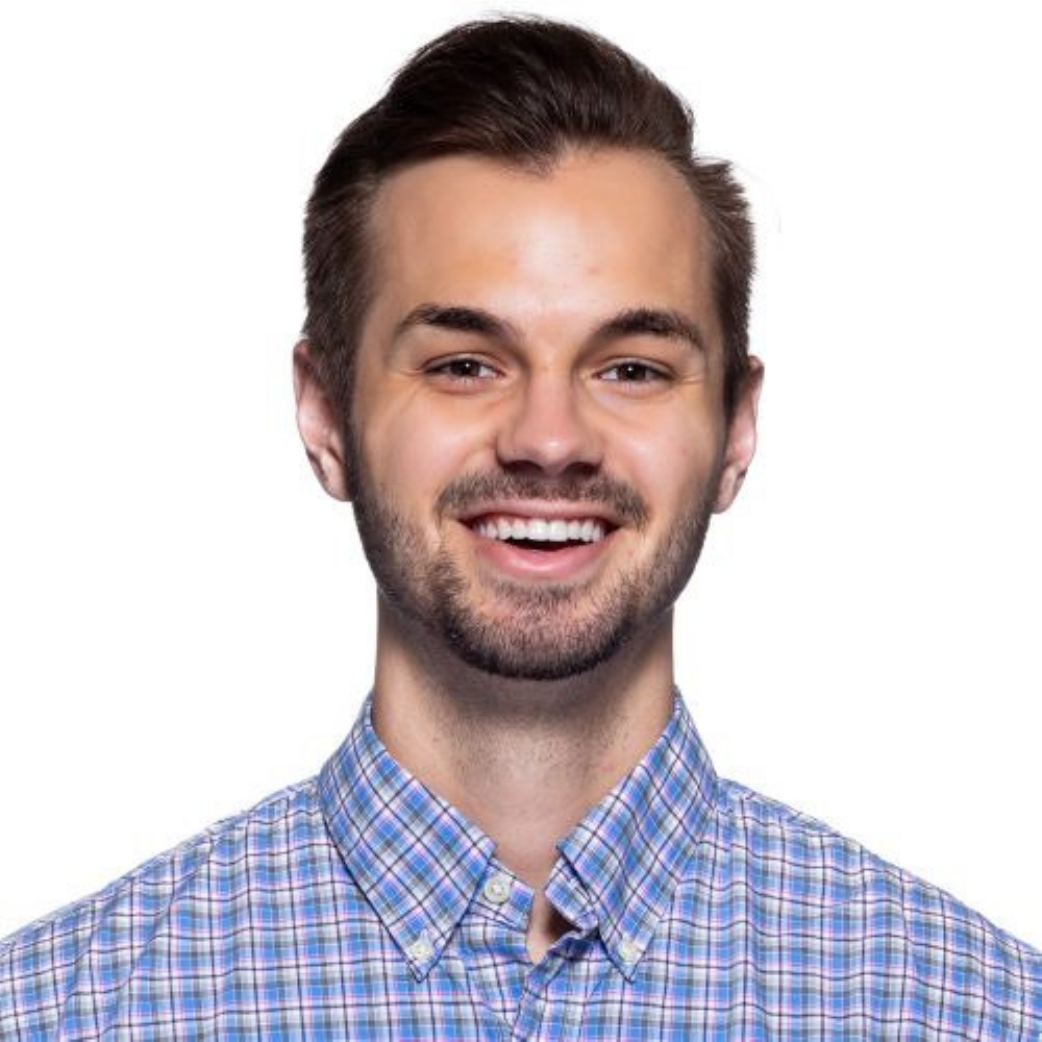}}]{Brandon Fallin}
 received his Bachelor of Science degree in Aerospace Engineering
at the University of Florida in May 2022. He subsequently earned his
Master of Science degree in Aerospace Engineering, also from the University
of Florida. He is currently pursuing a Ph.D. in Aerospace Engineering
at the University of Florida under the supervision of Dr. Warren Dixon.
His research interests include privacy and obfuscation with application
to nonlinear control systems.
\end{IEEEbiography}

\vspace{-12mm}
\begin{IEEEbiography}[{\includegraphics[width=1\columnwidth]{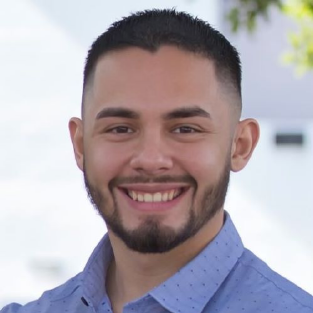}}]{Cristian F. Nino}
 received the B.S. degrees in Mathematics and Mechanical Engineering
from the University of Florida, Gainesville, FL, USA. He subsequently
earned the M.S. degree in Mechanical Engineering, also from the University
of Florida, with a focus on control systems. He is currently pursuing
the Ph.D. degree in Mechanical Engineering at the University of Florida
under the guidance of Dr. Warren Dixon. Mr. Nino is a recipient of
the SMART Scholarship, the NSF Fellowship, and the Machen Florida
Opportunity Scholarship. His research interests include robust adaptive
nonlinear control, multi-agent target tracking, distributed state
estimation, reinforcement learning, Lyapunov-based deep learning,
and geometric mechanics and control.
\end{IEEEbiography}

\vspace{-12mm}
\begin{IEEEbiography}[{\includegraphics[width=1\columnwidth]{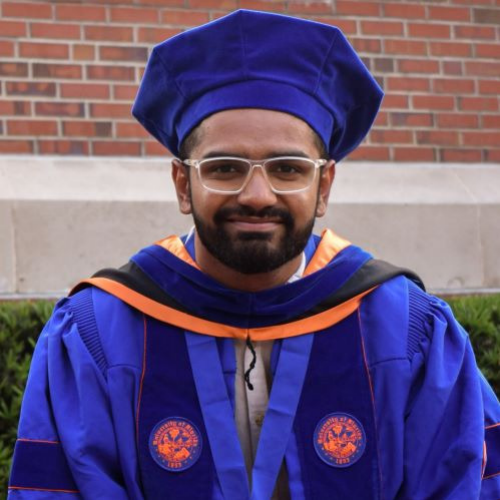}}]{Omkar Sudhir Patil}
 received his Bachelor of Technology (B.Tech.) degree in production
and industrial engineering from Indian Institute of Technology (IIT)
Delhi in 2018, where he was honored with the BOSS award for his outstanding
bachelor's thesis project. In 2019, he joined the Nonlinear Controls
and Robotics (NCR) Laboratory at the University of Florida under the
guidance of Dr. Warren Dixon to pursue his doctoral studies. Omkar
received his Master of Science (M.S.) degree in mechanical engineering
in 2022 and Ph.D. in mechanical engineering in 2023 from the University
of Florida. During his Ph.D. studies, he was awarded the Graduate
Student Research Award for outstanding research. In 2023, he started
working as a postdoctoral research associate at NCR Laboratory, University
of Florida. His research focuses on the development and application
of innovative Lyapunov-based nonlinear, robust, and adaptive control
techniques. 
\end{IEEEbiography}

\pagebreak{}

\vspace{-12mm}
\begin{IEEEbiography}[{\includegraphics[width=1\columnwidth]{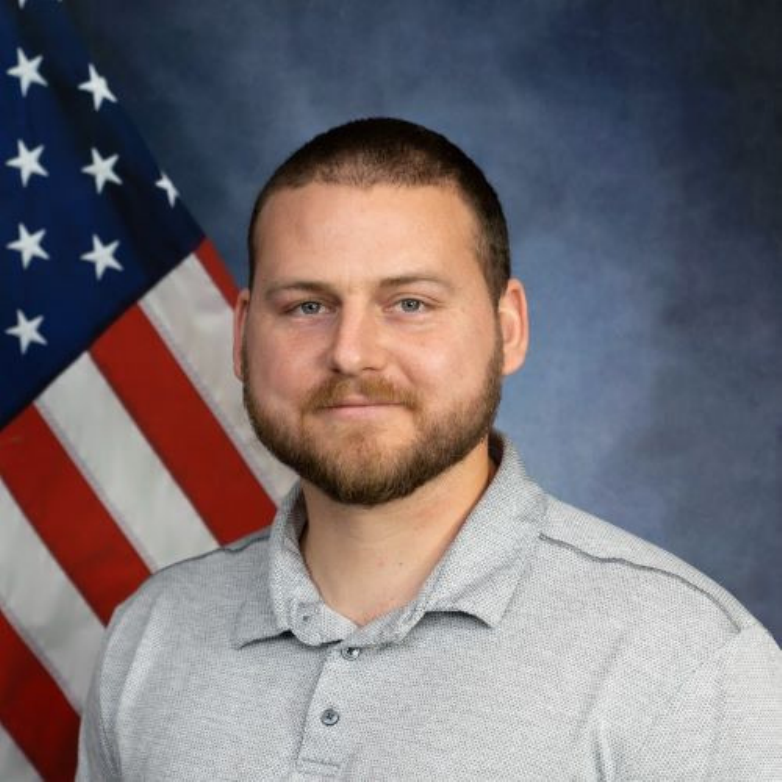}}]{Zachary I. Bell}
 received his Ph.D. from the University of Florida in 2019 and is
a researcher for the Air Force Research Lab. His research focuses
on cooperative guidance and control, computer vision, adaptive control,
and reinforcement learning.
\end{IEEEbiography}

\vspace{-12mm}
\begin{IEEEbiography}[{\includegraphics[width=1\columnwidth]{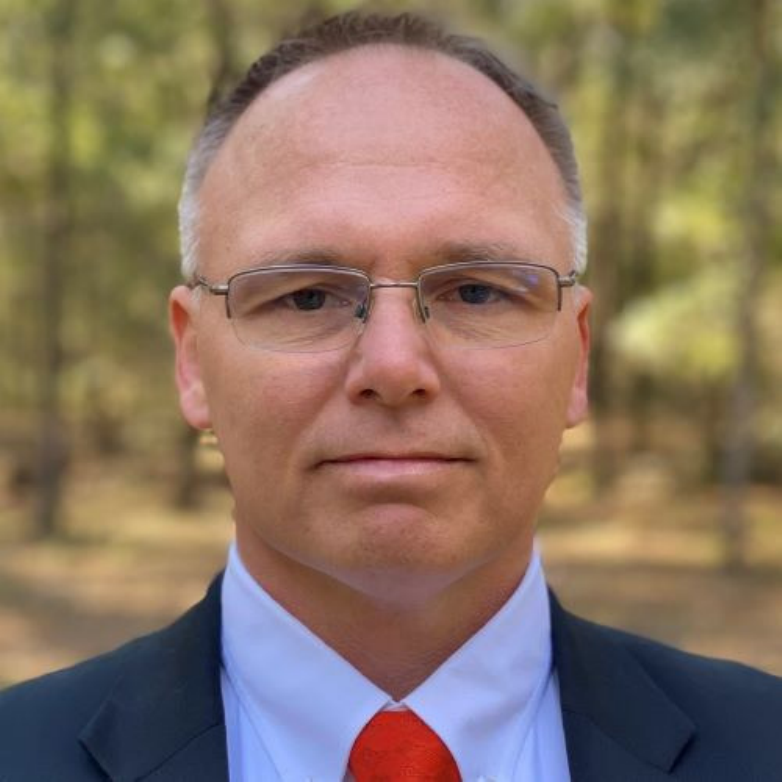}}]{Prof. Warren Dixon}
 received his Ph.D. in 2000 from the Department of Electrical and
Computer Engineering from Clemson University. He worked as a research
staff member and Eugene P. Wigner Fellow at Oak Ridge National Laboratory
(ORNL) until 2004, when he joined the University of Florida in the
Mechanical and Aerospace Engineering Department. His main research
interest has been the development and application of Lyapunov-based
control techniques for uncertain nonlinear systems. His work has been
recognized by the 2019 IEEE Control Systems Technology Award, (2017-2018
\& 2012-2013) University of Florida College of Engineering Doctoral
Dissertation Mentoring Award, 2015 \& 2009 American Automatic Control
Council (AACC) O. Hugo Schuck (Best Paper) Award, the 2013 Fred Ellersick
Award for Best Overall MILCOM Paper, the 2011 American Society of
Mechanical Engineers (ASME) Dynamics Systems and Control Division
Outstanding Young Investigator Award, the 2006 IEEE Robotics and Automation
Society (RAS) Early Academic Career Award, an NSF CAREER Award (2006-2011),
the 2004 Department of Energy Outstanding Mentor Award, and the 2001
ORNL Early Career Award for Engineering Achievement. He is an ASME
Fellow (2016) and IEEE Fellow (2016), was an IEEE Control Systems
Society (CSS) Distinguished Lecturer (2013-2018), served as the Director
of Operations for the Executive Committee of the IEEE CSS Board of
Governors (BOG) (2012-2015), and served as an elected member of the
IEEE CSS BOG (2019-2020). His technical contributions and service
to the IEEE CSS were recognized by the IEEE CSS Distinguished Member
Award (2020). He was awarded the Air Force Commander's Public Service
Award (2016) for his contributions to the U.S. Air Force Science Advisory
Board. He is currently or formerly an associate editor for ASME Journal
of Journal of Dynamic Systems, Measurement and Control, Automatica,
IEEE Control Systems, IEEE Transactions on Systems Man and Cybernetics:
Part B Cybernetics, and the International Journal of Robust and Nonlinear
Control.

\enlargethispage{-9.5cm}
\end{IEEEbiography}

\end{document}